\documentclass[fleqn,usenatbib]{mnras}

\usepackage{newtxtext,newtxmath}
\usepackage{longtable}
\usepackage[flushleft]{threeparttable}
\usepackage{xcolor}

\definecolor{rvdb}{rgb}{0.9,0.45,0}

\usepackage[T1]{fontenc}

\DeclareRobustCommand{\VAN}[3]{#2}
\let\VANthebibliography\thebibliography
\def\thebibliography{\DeclareRobustCommand{\VAN}[3]{##3}\VANthebibliography}

\usepackage{graphicx}	
\usepackage{amsmath}	
\usepackage{amssymb}	
\usepackage{xcolor}

\title[The GOGREEN and GCLASS Surveys Data Release]{The GOGREEN and GCLASS Surveys: First Data Release}
\author[Balogh et al.]{
\newauthor
Michael L. Balogh$^{1,2}$\thanks{mbalogh\@@uwaterloo.ca}, 
Remco F. J. van der Burg$^{3}$,
Adam Muzzin$^{4}$, 
Gregory Rudnick$^{5}$, 
\newauthor 
Gillian Wilson$^{6}$,
Kristi Webb$^{1,2}$, 
Andrea Biviano$^{7,8}$,
Kevin Boak$^{1}$,
Pierluigi Cerulo$^{9}$,
\newauthor
Jeffrey Chan$^{6}$,
M. C. Cooper$^{10}$, 
David G. Gilbank$^{11,12}$, 
Stephen Gwyn$^{13}$,
Chris Lidman$^{14,15}$,
\newauthor
Jasleen Matharu$^{16,17}$,
Sean L. McGee$^{18}$, 
Lyndsay Old$^{19}$, 
Irene Pintos-Castro$^{20}$, 
\newauthor
Andrew M. M. Reeves$^{1,2}$,
Heath Shipley$^{21}$,
Benedetta Vulcani$^{22}$,
Howard K.C. Yee$^{20}$, 
\newauthor
M. Victoria Alonso$^{23}$,
Callum Bellhouse$^{18,22}$,
Kevin C. Cooke$^{5}$,
Anna Davidson$^{5}$,
\newauthor
Gabriella De Lucia$^{7}$, 
Ricardo Demarco$^{9}$, 
Nicole Drakos$^{1,2,24}$,
Sean P. Fillingham$^{10,25}$,
\newauthor
Alexis Finoguenov$^{26}$,
Ben Forrest$^{6}$,
Caelan Golledge$^{5}$,
Pascale Jablonka$^{27}$,
\newauthor
Diego Lambas Garcia$^{23}$,
Karen McNab$^{1,2}$,
Hernan Muriel$^{23}$,
Julie B. Nantais$^{28}$, 
\newauthor
Allison Noble$^{29}$,
Laura C. Parker$^{30}$,
Grayson Petter$^{5}$,
Bianca M. Poggianti$^{22}$,
\newauthor
Melinda Townsend$^{5}$,
Carlos Valotto$^{23}$,
Tracy Webb$^{21}$,
Dennis Zaritsky$^{31}$
\\Author affiliations are listed at the end of the paper
}

\date{\today}

\pubyear{2020}

\begin{document}
\label{firstpage}
\pagerange{\pageref{firstpage}--\pageref{lastpage}}
\maketitle

\begin{abstract}
We present the first public data release of the GOGREEN and GCLASS surveys of galaxies in dense environments, spanning a redshift range $0.8<z<1.5$.  The surveys consist of deep, multiwavelength photometry and extensive Gemini GMOS spectroscopy of galaxies in 26 overdense systems ranging in halo mass from small groups to the most massive clusters.  The objective of both projects was primarily to understand how the evolution of galaxies is affected by their environment, and to determine the physical processes that lead to the quenching of star formation.  There was an emphasis on obtaining unbiased spectroscopy over a wide stellar mass range ($M\gtrsim 2\times 10^{10}~\mathrm{M}_\odot$), throughout and beyond the cluster virialized regions.   The final spectroscopic sample includes 2771 unique objects, of which 2257 have reliable spectroscopic redshifts.  Of these, 1704 have redshifts in the range $0.8<z<1.5$, and nearly 800 are confirmed cluster members.  Imaging spans the full optical and near-infrared wavelength range, at depths comparable to the UltraVISTA survey, and includes \textit{HST}/WFC3 F160W (GOGREEN) and F140W (GCLASS).  This data release includes fully reduced images and spectra, with catalogues of advanced data products including redshifts, line strengths, star formation rates, stellar masses and rest-frame colours.  Here we present an overview of the data, including an analysis of the spectroscopic completeness and redshift quality.
\end{abstract}




\section{Introduction}
Distant galaxy clusters have proven to be a rich source of information about our Universe.  Because of their high spatial density of galaxies, all at nearly equal distance from the observer, clusters provide an efficient way to observe large samples of galaxies and to uncover fundamental relationships between them.  Homogeneous spectroscopic and photometric surveys of such systems have provided much insight into galaxy evolution in general, and the role played by large scale structure in particular.

The early work of \citet{BO78a,BO78b} was among the first to demonstrate that galaxies evolve, as their observations of modestly distant clusters revealed a population significantly bluer than that of local clusters.  This evolution turned out not to be specific to clusters, but characteristic of galaxy evolution in general \citep[e.g.][]{Autofib_lf,L96}.  Pioneering work by \citet{YEC} optimized the use of multiobject spectroscopy, employing band-limiting filters and on-the-fly mask design, to execute the first large and homogeneous study of galaxy clusters at $z>0.2$.  Though the main science driver of this survey (CNOC) was to measure the average matter density of the Universe, $\Omega_m$, from the dynamical masses and stellar light content of the clusters \citep{CNOC1}, it was also well-suited to studies of the galaxy population itself \citep[e.g.][]{A2390,B+97,B+98,PSG,Erica}.  A contemporaneous redshift survey of massive clusters at a similar redshift \citep{D+97,D+99} took advantage of {\it Hubble Space Telescope} imaging to consider also the morphological evolution of galaxies.  Again, the highly multiplexed spectroscopy led to numerous advances in our understanding of galaxy evolution \citep[e.g.][]{P+99}.

With the advent of truly wide-field, highly multiplexed spectroscopic surveys \citep{SDSS_tech,2dF_colless,Owings,OwingsII} the statistical properties of the nearby cluster population relative to the surrounding field became much better defined \citep[e.g.][]{deP_clus,deP_BO,2dF-sfr,Sloan_sfr,2015MNRAS.450.2749G,2017ApJ...838..148P}.  At the same time, the power of large programs on 8-m class telescopes was being exploited to push targeted cluster surveys to redshifts $z>0.5$ \citep{Halliday+04,Ediscs-survey,2008A&A...482..419M,IMACSI,XXLXXII,ACTI,ACTII,SPTI,SPT-GMOS}.
The benefits of a large and homogeneous sample again provided a wealth of insight into galaxy evolution, now extending back more than six billion years in lookback time \citep[e.g.][]{2005ApJ...630..206F,Pogg05,2007MNRAS.374..809D,2007ApJ...660.1151D,Pogg+08,Ediscs_psb2,Ediscs_psb,Rudnick+09,V+10, V+11,IMACSII,XXLXXXVII}.

Our knowledge of galaxy evolution in the general field population has continued to advance to earlier and earlier times, thanks to both photometric and sparsely-sampled spectroscopic surveys \cite[e.g.][]{GDDS,VVDS1,zCOSMOS_10K,2010A&A...523A..74V,DEEP2_data,Ultravista_Muzzin,zFOURGE,VUDS,VIPERS}.  As these surveys tend to be over relatively narrow fields, they contain few, if any, massive galaxy clusters.   Primarily photometric studies of clusters \citep[e.g.][]{RCS,2008ApJ...684..905E,SpARCS09_Muzzin,SpARCS09_Wilson,RCS2,2017ApJ...851..139L,2019ApJ...876...40P} have been used to learn about the  statistical properties of the galaxy population, including the evolution of the stellar mass function.  Ambitious efforts have been undertaken to obtain relatively sparse follow-up spectroscopy of bright targets on large samples of clusters \citep[e.g.][]{2013ApJ...779..138B,Madcows,2019ApJ...870....7K,2020ApJS..247...25B}, often motivated by interests in cosmology.  But highly complete, homogeneous spectroscopy of the faint, quiescent population that dominates local clusters, is challenging and expensive for clusters at $z>0.8$, with near-infrared spectroscopy required to probe beyond $z\sim 1.5$ \citep[e.g.][]{SpARCS_Nantais,SpARCS_Nantais2,2017ApJ...843..126D}.  

The GMOS spectrographs \citep{GMOS} on the twin Gemini telescopes are well suited to the study of galaxy clusters at these higher redshifts, as the field of view contains the full virialized cluster volume, and the red sensitivity and high multiplex capability makes it feasible to obtain redshifts for hundreds of faint, quiescent galaxies in a relatively short time.  The Gemini CLuster Astrophysics Spectroscopic Survey \citep[GCLASS;][]{GCLASS12} and Galaxy Environment Evolution Collaboration 2 \citep[GEEC2;][]{GEEC2-survey} were independent, but contemporaneous, GMOS surveys of high- and low-mass clusters, respectively, over the redshift range $0.8<z<1.3$.  These surveys provided a new perspective on galaxy evolution within dense environments  \citep{GEEC2-1,Noble+13,Mok13,Mok14,Muzzin14,Foltz15,Noble+16,Foltz+18}, the stellar mass content and dynamics of clusters \citep{GEEC2-Hou,vdB+13,GCLASS_vdB,GCLASS_dynamics}, and the growth of the brightest cluster galaxies \citep{Lidman_bcg,Lidman_bcg2}.  Among other things, they revealed that the fraction of quenched galaxies in clusters was already very high by $z=1$, but that it has been established in a fundamentally different way from that in local clusters \citep{Balogh+16}.

GEEC2 and GCLASS pushed the limits of what could be achieved with the detectors available at the time, and with the challenges of coordinating multipartner, multisemester time requests through normal PI-mode observing.  The introduction of red-sensitive, Hamamatsu CCDs on both Gemini telescopes \citep{GMOS-Ham}, together with a new Large and Long Program (LLP) proposal category, opened up an opportunity for an ambitious cluster survey that would approach $z=1.5$, the practical limit for ground-based, optical spectroscopy.  

The Gemini Observations of Galaxies in Rich Early Environments (GOGREEN) survey was launched in 2014, in the first round of Gemini LLPs.  The last of the spectroscopic data were obtained in July, 2019.  The goal was to take advantage of upgrades to the red-sensitive Hamamatsu detectors and  obtain spectroscopy for ${\sim} 1000$ galaxies, representative of all galaxy types with stellar masses $M\gtrsim10^{10.3}~\mathrm{M}_\odot$, over the full virialized regions of 21 galaxy systems at $1<z<1.5$ and with halo masses spanning $10^{13}\lesssim M/\mathrm{M}_\odot\lesssim 10^{15}$.  By covering a wide parameter range in redshift, halo mass and stellar mass, the objective was to provide the best available constraints on cluster galaxy evolution at this important epoch.  The survey was introduced and described in \cite[][Paper~I]{gogreen-survey}.  The first results have already proved surprising.  While the excess fraction of quenched cluster galaxies, relative to the surrounding field at the same epoch, is just as high as we observe at $z=0$, the shape of the stellar mass function of quiescent galaxies is independent of environment \citep{gogreen-HLF,gogreen-smf}.  Similar to low-redshift clusters, the distribution of star formation rates among the star--forming population shows little or no environmental dependence \citep{gogreen-sfr}.  More unexpected, perhaps, is that the mass-weighted ages of the quiescent galaxies are only slightly older among the cluster population than in the field (Webb et al., submitted).  Together these observations are ruling out models where the majority of the cluster population was quenched upon accretion; instead, they must have ceased forming stars long before, and be already quenched when they reached the cluster's virialized region.  Ongoing work includes measuring the halo mass dependence of this effect (Reeves et al., in prep), the abundance of recently quenched and post-starburst galaxies (McNab et al. in prep), the morphological differences between cluster and field galaxies (Chan et al. in prep), and the dynamical mass profiles of these clusters (Biviano et al. in prep). 

In August 2020, the data for both GOGREEN and GCLASS are being made publicly available\footnote{GEEC2 data were released with \citet{GEEC2-survey}.}.  This data release includes images, spectra, catalogues, and a wide range of advanced data products.  While many of the survey details for GOGREEN and GCLASS were provided in Paper~I and \citet{GCLASS12}, respectively, here we summarize the most important features (\S~\ref{sec-survey}) and provide updates to the data processing steps where required (\S~\ref{sec-obs}).  We describe how we derive advanced products from these data in \S~\ref{sec-ddp}.   An overview of the galaxy sample contained in these two surveys, including the spectroscopic completeness, is given in \S~\ref{sec-ssample}, and a summary of the data release contents is provided in \S~\ref{sec-DR}.

All GCLASS and GOGREEN results use the following standard systems and assumptions.  Magnitudes are on the AB system.  Unless otherwise specified, we assume a \citet{Chab} initial mass function and a flat cosmology with $\Omega_m=0.3$ and $H_0=70$~km/s/Mpc.  

\section{Survey Design, cluster sample, and spectroscopic target selection}\label{sec-survey}
The design of the GCLASS and GOGREEN surveys are described in \citet{GCLASS12} and Paper~I, respectively.  Both surveys are founded on extensive multiobject spectroscopy of galaxy clusters with the Gemini telescopes: GCLASS covering a redshift range $0.8<z<1.3$ and GOGREEN spanning an overlapping $1<z<1.5$.  There are five clusters in common between the two surveys, with the additional GOGREEN spectroscopy extending the sample to lower stellar masses.  In total the two surveys target $26$ unique galaxy groups and clusters.

\subsection{The cluster sample}
The GCLASS sample of ten massive clusters, and nine of the clusters in GOGREEN, are drawn from the Spitzer Adaptation of the Red  Cluster Sequence (SpARCS) Survey \citep{SpARCS09_Wilson,SpARCS09_Muzzin,Demarco_Sparcs}. 
All of these clusters were discovered from shallow $z^{\prime}$ and IRAC 3.6\,$\mu$m images, via their overdensity of "red-sequence" galaxies \citep[e.g.][]{GY00}.  
Three of the GOGREEN clusters (SPT0205, SPT0546 and SPT2106) were selected from the South Pole Telescope (SPT) survey.  These systems were detected via their Sunyaev-Zeldovich (SZ) signature and subsequently spectroscopically confirmed \citep{SPT0546,SPT2106,SPT0205}.
To extend the GOGREEN sample to lower halo masses, nine galaxy groups in the COSMOS and Subaru-XMM Deep Survey (SXDS) were included.  These were selected based on spectroscopically confirmed detections at X-ray wavelengths, drawn from updated versions of the catalogues described in \citet{alexis_sdf,alexis_cosmos} and \citet{George+11}.  In some cases our spectroscopy revealed numerous structures along the line of sight, and resulted in either additional groups in the sample (SXDF76) or a significantly revised redshift estimate for the group (SXDF64).  

The coordinates and redshifts of all 26 galaxy clusters\footnote{In the rest of this paper we will avoid making an arbitrary distinction between "clusters" and "groups", and refer to all of our targeted, overdense systems as clusters.} included in our final sample are given in Table~\ref{tab-cluster_sample}.  For the GOGREEN SpARCS and SPT clusters, the centre is chosen to be the position of the brightest cluster galaxy, as described in \citet[][GOGREEN]{gogreen-smf} and \citet[][GCLASS]{Lidman_bcg}.    For the clusters in the COSMOS/SXDF fields, which are deliberately selected to be low-richness, the position and velocity centre can be less robustly defined than for the massive clusters.  We use all available public redshifts, in addition to GOGREEN, to identify spectroscopic members (see \S~\ref{sec-vdisp}) and take the spatial centres to be the unweighted average positions of those members within a 1\,Mpc area centered on the X-ray detection.  The full procedure and results are described in more detail, in Reeves et al. (in prep).  

\begin{figure*}{}
\includegraphics[clip=true,trim=0mm 0mm 0mm 0mm,width=3.4in,angle=0]{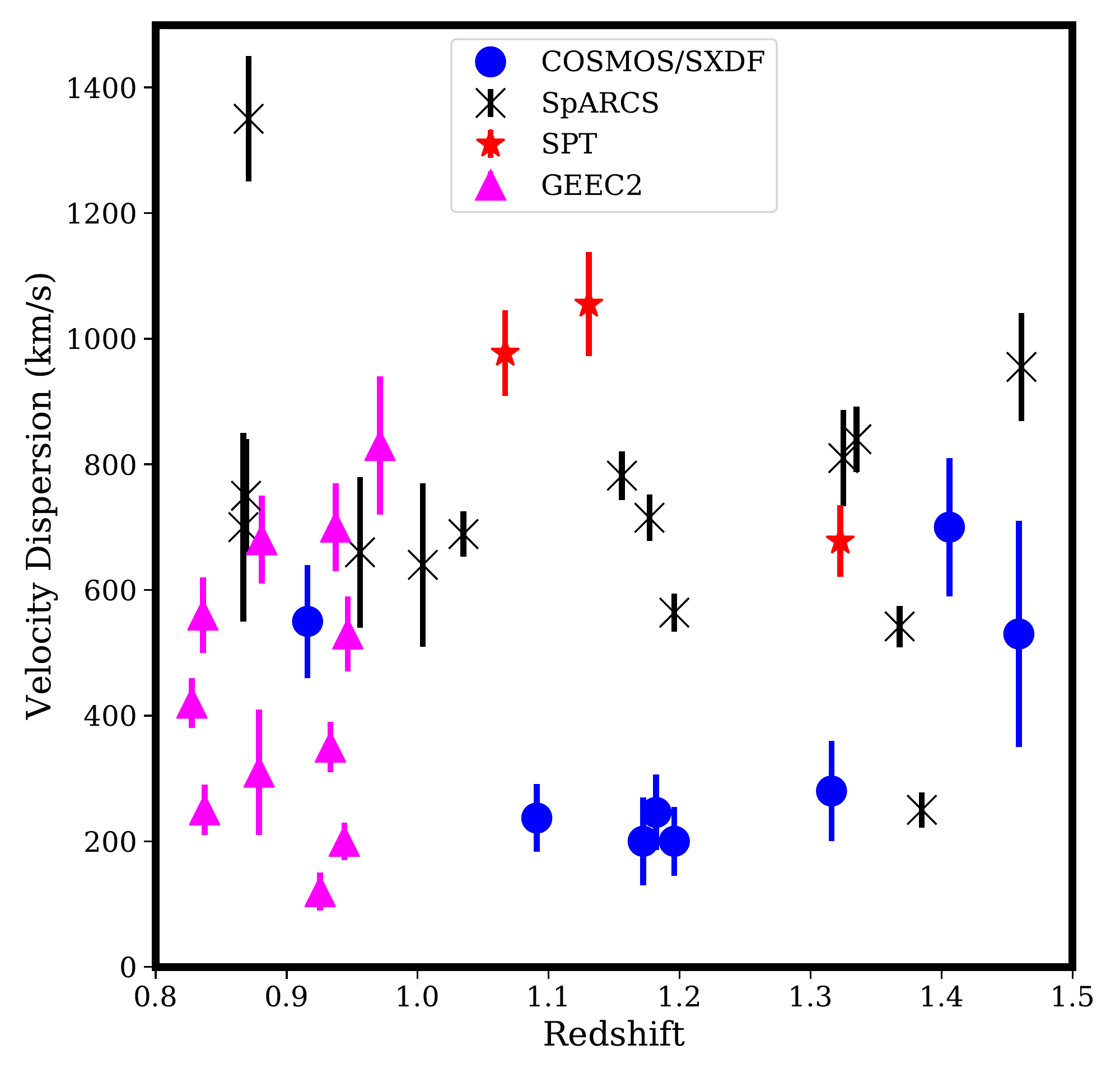}\includegraphics[clip=true,trim=0mm 0mm 0mm 0mm,width=3.4in,angle=0]{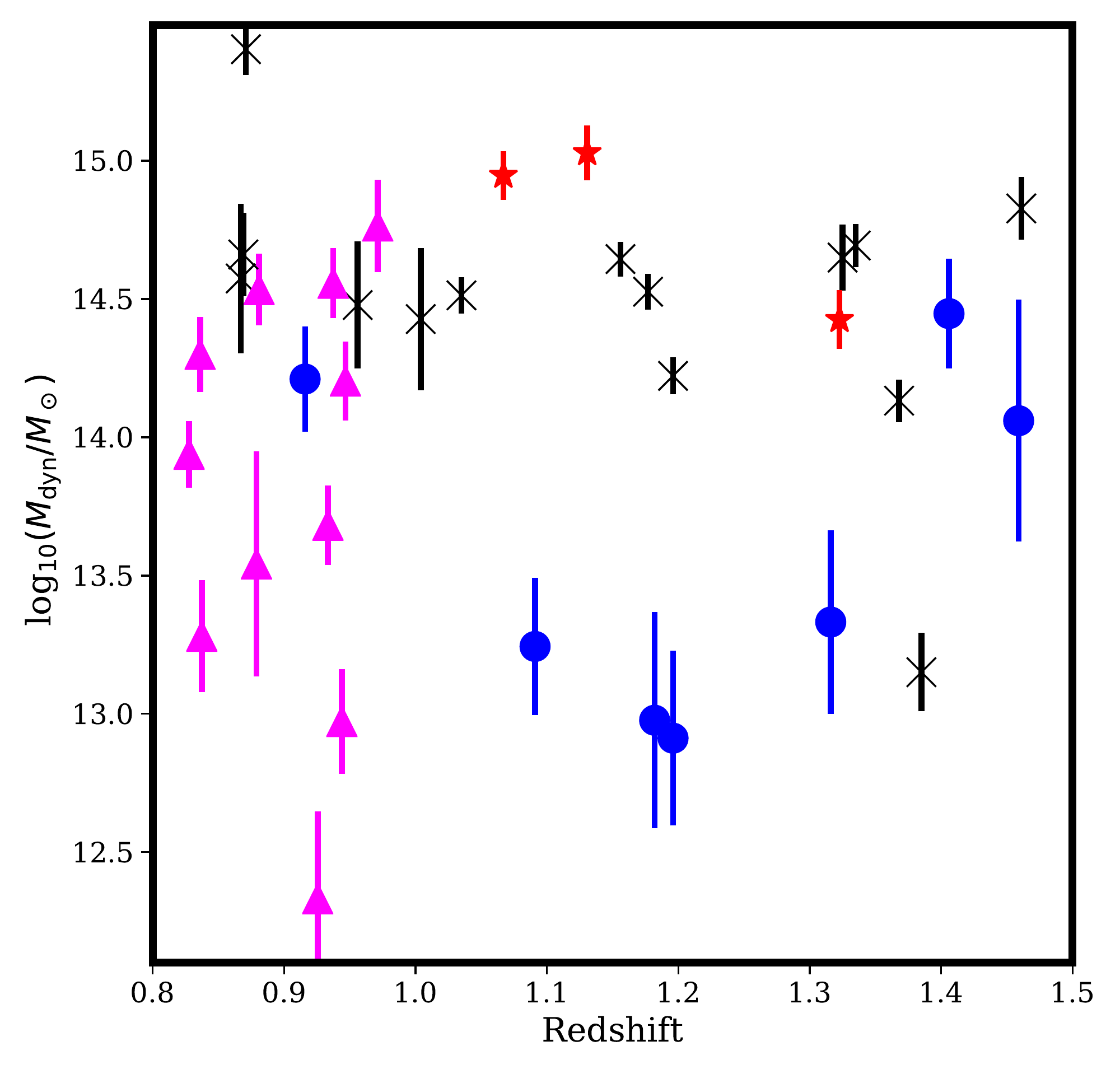}
	\caption{{\it Left: }The intrinsic velocity dispersion for each system in the GOGREEN and GCLASS samples is shown as a function of redshift.  We also compare with the previously released GEEC2 sample \citep{GEEC2-survey}.  {\it Right:} Velocity dispersions converted to dynamical mass estimates, based on the relation of \citet{Saro}.
	\label{fig-sigmaz}}
\end{figure*}

Figure~\ref{fig-sigmaz} shows the velocity dispersions and redshifts of our sample.  One cluster (COSMOS-125) is excluded, as a robust velocity dispersion could not be measured.  Velocity dispersion calculations are described in \S~\ref{sec-vdisp}, and include redshifts from the literature.  The three different selection classes are highlighted: SZ-selected (SPT), galaxy overdensity selected (SpARCS), and X-ray selected (COSMOS/SXDF).  While these three selections were expected to correspond to very massive, typical, and low-mass clusters, respectively, in practice there is significant overlap between them.  We also include the GEEC2 clusters \citep{GEEC2-survey} on this figure; these fill in the low-dynamical mass region at $0.8<z<1$.  Here we estimate total dynamical masses from the velocity dispersion for all these systems using the simulation--derived scaling of \citet{Saro}, and show these in the right panel (again omitting COSMOS-125).  Dynamical masses span more than two orders of magnitude, from poor groups to very massive clusters.

\begin{table*}
  \begin{threeparttable}
	\begin{tabular}{lrrlrrrl}
		Name & RA & Dec & $z$ &N$_z$&N$_{\rm mem}$&$\sigma_{\rm int}$&References\\
		&\multicolumn{2}{c}{~~~~~~~~~~~(J2000)}&&&&km/s&(Literature members)\\
		\hline
		\multicolumn{8}{c}{SPT Clusters}\\
		SPT-CL J0546-5345 &86.6562	&-53.7580	&1.068&63&34 (67)&$980\pm70$&\citet{ACTII}\\
		SPT-CL J2106-5844 &316.5191 &-58.7411	&1.126&67&37 (50)&$1055\pm85$&\citet{SPT2106}\\ 
		SPT-CL J0205-5829 &31.4390	&-58.4829	&1.323&65&24 (28)&$680\pm60$&\citet{SPT0205}\\
		\hline
		\multicolumn{8}{c}{SpARCS Clusters}\\
		SpARCS0034-4307	  &8.6751	&-43.1315	&0.867&126&44&$700\pm150$&\\
		SpARCS0036-4410	  &9.1875	&-44.1805	&0.869&114&48&$750\pm90$&\\
		SpARCS1613+5649	  &243.3110 &56.8250	&0.871&152&94&$1350\pm100$&\\ 
		SpARCS1047+5741	  &161.8890 &57.6871	&0.956&137&30&$660\pm120$&\\
		SpARCS0215-0343	  &33.8500 &-3.7256	&1.004&110&48&$640\pm130$&\\
		SpARCS1051+5818	  &162.7968 &58.3009	&1.034&176&42&$690\pm40$&\\
		SpARCS1616+5545	  &244.1718 &55.7571	&1.157&195&60&$780\pm40$&\\ 
		SpARCS1634+4021	  &248.6475 &40.3643	&1.177&176&69&$715\pm40$&\\
		SpARCS1638+4038	  &249.7152 &40.6452	&1.194&161&56&$565\pm30$&\\
		SpARCS0219-0531	  &34.9316 &-5.5249	&1.328&56&9&$810\pm80$&\\
		SpARCS0035-4312	  &8.9571	&-43.2068	&1.335&121&29&$840\pm50$&\\
		SpARCS0335-2929   &53.7649	&-29.4822	&1.368&66&12 (27)&$540\pm30$& J. Nantais (priv. comm)\\
		SpARCS1034+5818	  &158.70560 &58.3092	&1.388&40&11&$250\pm30$&\\
		SpARCS1033+5753	  &158.3565 &57.8900	&1.460&61&9&$955\pm90$&\\
		\hline
		\multicolumn{7}{c}{COSMOS/SXDF Clusters}& See  \S~\ref{sec-vdisp}\\
		SXDF64XGG	      &34.3319  &-5.2067	&0.916&17&1 (8)&$530\pm80$&\\
		SXDF49XGG	      &34.4996	&-5.0649	&1.091&101$^{1}$&6 (14)&$255\pm50$&\\
		COSMOS-63	      &150.3590 &1.9352	    &1.1722&26&5 (8)&N/A& \\
		SXDF76bXGG	      &34.7474	&-5.3235	&1.182&80$^{2}$&7&$210\pm65$&\\
		COSMOS-221	      &150.5620 &2.5031	    &1.196&54&9&$200\pm50$&\\
		COSMOS-28	      &149.4692 &1.6685	    &1.316&54&10&$285\pm75$&\\ 
		COSMOS-125	      &150.6208 &2.1675	    &1.404&39&7 (9)&N/A&\\
		SXDF87XGG	      &34.5360	&-5.0630	&1.406&101$^{1}$&8&$700\pm110$&\\ 
		SXDF76aXGG	      &34.7461	&-5.3041	&1.459&80$^{2}$&6&$520\pm180$&\\
		\hline
	\end{tabular}
    \begin{tablenotes}
      \small
      \item $^{1}$ SXDF49XGG and SXDF87XGG share a single GMOS field.  This number represents the total number of spectra in that field, so is the same for both groups.
      \item $^{2}$ SXDF76aXGG and SXDF76bXGG share a single GMOS field.  This number represents the total number of spectra in that field, so is the same for both groups.\\
    \end{tablenotes}
	\caption{The table presents the 26 galaxy clusters and groups in the GOGREEN and GCLASS samples, ordered by redshift within three approximate halo mass classes.  Redshifts are given in column (4), and the total number of GOGREEN and/or GCLASS spectra yielding good redshifts is given in column (5).  The number of spectroscopically confirmed cluster members from GOGREEN and GCLASS combinedare shown in column (6).   A following number in parentheses indicates the total number of members used to calculate the velocity dispersion, including values taken from the literature referenced in column (8).  
	The intrinsic velocity dispersion of cluster members is given in column (7). 
	\label{tab-cluster_sample}}
  \end{threeparttable}
\end{table*}
\subsection{The GCLASS survey}
The goal of GCLASS was to obtain at least ${\sim} 50$ spectroscopic cluster members in each of ten clusters at $0.8<z<1.3$, for the main purpose of understanding galaxy evolution in dense environments.  To this end, priority was given to bright galaxies near the core of the cluster and close to the cluster red sequence in $z^{\prime}-[3.6\,\mu\mathrm{m}]$ colour.  Nod-and-shuffle spectroscopy, combined with a strategy that used two offset GMOS pointings, allowed efficient sampling of the dense cluster cores as well as substantial radial coverage.   For more details we refer the reader to \citet{GCLASS12}.

\subsection{The GOGREEN survey}
The GOGREEN survey is built upon a Gemini Large Program that ran from 2014 to 2019, using both telescopes over ten semesters to execute 432 hours of telescope time.  Strong support from Gemini, through program extensions and Director's Discretionary time, ensured that the project executed nearly 100 per cent of the originally allocated time (438h).  The majority of this time was spent on multiobject spectroscopy; about 45h were used to obtain deep $z^\prime$ imaging with GMOS for target selection and mask design.  Observations were acquired primarily in Priority Visitor mode, with the remainder executed in queue mode.  Details of the observing runs are given in Appendix~\ref{sec-speclog}.

Spectroscopic targets for the twelve SpARCS and SPT clusters were selected from deep IRAC and GMOS $z^{\prime}$ imaging, as described in Paper~I.  These were magnitude limited at $z^{\prime}<24.5$ and $[3.6\,\mu\mathrm{m}]<22.5$, with broad, magnitude-dependent colour cuts to exclude galaxies at $z<0.7$ and $z>1.5$ (see \S~\ref{sec-comp} for an illustration).  For the nine clusters in COSMOS and SXDF, target selection was based on available\footnote{An unpublished, updated catalogue for SXDS was kindly provided by R. Quadri.} photometric redshifts \citep{W+09,Quadri+12,Ultravista_Muzzin}.    

Each cluster was observed with up to six slitmasks over the course of the survey, all centred at the same position and at the same position angle\footnote{The position angle is normally 90 degrees, to minimize atmospheric dispersion.  However, guide star constraints often necessitated a different choice.  In any case, the angle selected was independent of any feature of the target itself, including its orientation on the sky.}.  The faintest targets were observed on multiple masks, to increase the signal-to-noise of the spectrum.  Existing data were examined when designing new masks, and objects were removed or reobserved in part based on the signal accumulated to that point.  More detail about the mask design strategy is given in Paper~I.

Imaging data were obtained through separate applications to multiple facilities following approval of the Large Program.  While the minimum requirement for these data were to image the full GMOS field of view, most of the images cover a much wider field, allowing future, photometric studies of the environments surrounding these clusters out to many times the virial radius.

\section{Observations and Data Reduction}\label{sec-obs}
\subsection{Multiwavelength Imaging}
\subsubsection{Spitzer IRAC images}
All SpARCS clusters have shallow ($5$-$\sigma$ depth of 7\,$\mu$Jy) imaging from the SWIRE survey \citep{SWIRE}, from which the clusters were identified.  Deeper channel 1 and 2 data ($5$-$\sigma$ depth in IRAC channel 1 of at least 2\,$\mu$Jy, or AB=23.1) exist for most of the clusters, from SERVS \citep{SERVS}, S-COSMOS \citep{COSMOS_Spitzer}, SpUDS 
\citep[PI J.  Dunlop, as described in][]{CANDELS} and GTO programs 40033 and 50161 (PI G. Fazio).
The remaining clusters were observed as PI programs (PI Brodwin, from program ID 70053 and 60099 and McGee, program 13046).  Some MIPS data from SWIRE and GTO programs 40033 and 50161 (PI G. Fazio) are available for the SpARCS clusters, but these are not used in the analysis nor included in this release.

There were small astrometric offsets (typically $\sim 1$\arcsec\ between the stacks from these different programs.  These were corrected with respect to the USNO-B1 catalogue \citep{USNO-B}, and all reduced data from different programs were stacked together.  The stacks cover 10$\arcmin$ on a side, with a pixel scale of 0.2\,arcsec/pixel, and have an AB magnitude zeropoint of 21.58.

\subsubsection{GMOS $z^{\prime}$-band images}\label{sec-zim} 
The deep GMOS imaging in $z^{\prime}$ obtained as part of GOGREEN is described in Paper~I.  The images were reduced by Gemini staff using their pipeline for producing mask-design preimages.  These are scientifically useful, though not optimized for photometry of the faintest sources.  The sky subtraction in particular is not optimal, as effects from bright sources remain in the sky frames, leading to non-Poisson noise in the background.  However, for most systems we have wider field $z-$band coverage from other facilities that supersedes the GMOS imaging \citep{gogreen-smf}.  

GCLASS $z^\prime$ imaging was obtained from CTIO and CFHT, as described in \citet{SpARCS09_Wilson}, \citet{SpARCS09_Muzzin} and \citet{GCLASS12}.  

\subsubsection{Ground-based optical and infrared imaging}\label{sec-groundbasedimaging}
A significant effort was undertaken to obtain deep imaging for all GOGREEN systems outside the COSMOS and SXDF fields.  New and archival data were obtained from Subaru, VLT, Magellan, Blanco and CFHT, spanning the full observable optical/infrared wavelength range from $u$ to $K$.  These data and basic processing steps are described in \citet{gogreen-smf}.  

Near-infrared data were processed with custom {\sc pyraf} scripts, based closely on the procedure described in \citet{2008A&A...489..981L}.  
All images are astrometrically registered, to within $0\farcs1$ precision, to the USNO-B1 catalogue \citep{USNO-B}.  

The Subaru Suprime-Cam data were processed at the CADC using a dedicated
pipeline (Gwyn 2020, ASP Conference Series, in press). The archival raw
images are detrended using
SDFRed\footnote{\url{https://www.subarutelescope.org/Observing/Instruments/SCam/sdfred/sdfred2.html.en}},
astrometrically calibrated using {\it Gaia} DR2 as a reference, and 
photometrically calibrated using Pan-STARRS photometry converted into
the Suprime-Cam photometric system. Image defects (bad columns, cosmic
rays) are masked and then the calibrated, detrended images are
resampled and combined using
SWarp\footnote{\url{https://www.astromatic.net/software/swarp}}.  The more recent, HypersuprimeCam data were fully processed using the {\sc hscPipe} software\footnote{\url{https://hsc.mtk.nao.ac.jp/pipedoc/pipedoc_6_e/}} (v6.0).

For GOGREEN, the $J-$ and $K_s$-band photometric zero points were calibrated with respect to 2MASS \citep{2MASS}.  Calibration of all the other filters was done based on the universality of the stellar locus \citep[cf.][]{high2009stellarlocus}.  Stellar spectra were obtained from the library of \citet{Pickles98}, supplemented with \citet{Ivanov04} in the near-infrared, and \citet{Kelly2014}. These spectra were convolved with the response function for each telescope/filter combination used, and colour-colour diagrams were inspected to measure shifts relative to the stars identified in our own data (for examples, see  Appendix~\ref{app-photcal}).

Colour measurements are based on aperture photometry performed on PSF-homogenised image stacks, constructed by convolving the individual stacks with kernels created with \texttt{PSFEx} \citep{psfex}. For more details we refer to \citet{gogreen-smf}.
Image depths are reported in \citet{gogreen-smf}, and calculated from the median 5$\sigma$ limits within 2\arcsec\ apertures measured on the PSF-homogenized images, corrected for Galactic extinction.  Depths for most images are comparable to those of the UltraVISTA \citep{2012A&A...544A.156M,Ultravista_Muzzin} survey.  

For GCLASS, optical $ugriz$ data for the northern clusters were obtained with MegaCam at CFHT.  The southern clusters were imaged in $ugri$ with IMACS on Magellan, while the $z^\prime$ band imaging came from the MOSAIC-II camera on the Blanco telescope at CTIO.  These data are described in \citet{vdB+13}.  Near-infrared imaging of the GCLASS systems from CFHT, Blanco and VLT are described in \citet{Lidman_bcg}. 

\subsubsection{Hubble Space Telescope Imaging}
New \textit{HST}/WFC3 F160W imaging for the twelve GOGREEN clusters was obtained in a Cycle 25 program (GO-15294: PI Wilson). Each cluster was targeted with a $1 \times 2$ mosaic of WFC3 pointings centered on the cluster, which covered a region of $136'' \times  233''$.  At the redshift of GOGREEN clusters this corresponds to a $\sim 1.1 \times 1.9$ Mpc rectangular region on the sky.  Each pointing has 1-orbit depth. The orientation of the \textit{HST} pointings (\texttt{ORIENT}) are constrained to within $20^{\circ}$ of the GMOS mask orientation to maximize the overlap between the imaging and the GOGREEN spectroscopic observations (see \S~\ref{sec-gogreen} below for details).

The data are reduced and combined using \textsc{Astrodrizzle} \citep[version 2.1.22,][]{Gonzagaetal2012}.  The data reduction steps are described in Chan et al. (in prep).  For the final drizzling, a pixel scale of $0\farcs06$ pixel$^{-1}$, a square kernel, and a~\texttt{pixfrac} of 0.8 are adopted.  Two sets of weight maps are produced using inverse variance map (\texttt{IVM}) and error map (\texttt{ERR}) weighting, respectively.  The characteristic PSF of each cluster is constructed by median-stacking 5--22 bright, unsaturated stars. The full-width-half-maximum (FWHM) of the PSFs are $\sim0\farcs17 - 0\farcs18$.

We also provide the WFC3 F140W imaging obtained for the ten GCLASS clusters, described in \cite{Matharu2019}.  The pixel scale and \texttt{pixfrac} values are the same as for the F160W images, and the stellar FWHM is $\sim 0\farcs23$.  


\subsection{Spectroscopy}\label{sec-spectroscopy}
\subsubsection{GCLASS}
The GCLASS spectroscopy is described in \citet{GCLASS12}.  All the data were obtained with GMOS-S and GMOS-N, which cover a 5.5$\times$5.5\arcmin\ field of view.  Slits were 1\arcsec\ wide and 3\arcsec\ long.  With the R150 grating this results in a spectral resolution of $R=\lambda/\Delta\lambda_{\rm FWHM}=440$
 (see \S~\ref{sec-gogreen}).  Two overlapping pointings per cluster were used, so as to increase the coverage in the dense core of the clusters and somewhat extend the radial coverage beyond that of the GMOS field of view.  Wavelength calibration was done exclusively using sky lines, following \citet{GDDS}.  

\subsubsection{GOGREEN}\label{sec-gogreen}
A full log of the GOGREEN spectroscopic observations is given in Table~\ref{tab-specobs}.
Like GCLASS, spectroscopy was obtained with the GMOS-S and GMOS-N instruments, but with a single pointing per cluster.  All observations on GMOS-S were obtained with the Hamamatsu detector array, which consists of three chips.  Two of these have enhanced red response, while the chip at the blue end has enhanced blue response, relative to the older EEV detectors used for GCLASS.  
On GMOS-N, observations prior to 2017 were obtained with an array of EEV deep depletion detectors.  
In 2017, the GMOS-N detector was replaced with a Hamamatsu array identical to the one on GMOS-S.

GOGREEN used the same grating and slit size as GCLASS.  We measure the spectral resolution for both surveys by fitting a Gaussian to the [O\,II] emission lines as described in \citet[][see \S~\ref{sec-indices}]{gogreen-sfr}.  
We find $R=440\pm 60$, or $\Delta\lambda_{\rm FWHM}\sim19.3$ at $\lambda=8500$\AA.  

The spectroscopic data reduction was based on the {\sc iraf}\footnote{ "IRAF is distributed by the National Optical Astronomy Observatories,
    which are operated by the Association of Universities for Research
    in Astronomy, Inc., under cooperative agreement with the National
    Science Foundation."} tools provided by Gemini, via the {\it Ureka} distribution.  A variance (VAR) and data quality (DQ) plane were propagated through all the reduction steps.   
Wavelength calibration was done using CuAr arc lamps, usually taken after a night's observing.  At our low resolution, this lamp provides $\sim 10$ useful lines over the wavelength range $6200\,\mbox{\AA}<\lambda<10700$\,\AA.  The typical {\it rms} of the wavelength solution is $\sim 0.5$\,\AA.
All spectra (regardless of detector) are linearized and rebinned to 3.91\,\AA\ per pixel.
  
While GOGREEN used an order-sorting filter that blocked light bluer than 5150\AA, in practice the wavelength calibration is not robust for $\lambda\lesssim6000$\,\AA\ due to the lack of good arc lines at this resolution. 
To account for simple shifts in the zeropoint due to instrument flexure, we cross-correlate each sky spectrum with that of a reference slit, ideally chosen to have an accurate wavelength solution.  The median shift for each mask is computed, and applied to the wavelength solution of that mask.  Shifts are typically $<0.5$ pixels, though on occasion can be two or three times larger.  

An important correction must be made for light that extends beyond the slit edges, an effect that is strongest at $\lambda>8500$\AA.  This presents a problem for microshuffle nod-and-shuffle masks, where light from one spectrum contaminates the sky region of its shuffled counterpart, or other surrounding slits, in a way that does not subtract off.  This effect was described by \citet{GDDS} and interpreted as charge diffusion.  However, we observe the same effect using very different CCDs, and expect the origin is related to the instrument optics.
A simple, average empirical model was developed in \citet{GDDS}, to provide an average, approximate correction to all objects on a mask; this was applied to the GCLASS spectra.  For GOGREEN the problem was acute, as we rely on the data at $>8500$\AA\ for redshift and absorption line studies at $z\gtrsim1.2$.  We therefore developed a more sophisticated, empirical model for the effect.  This model and its application to our data is described in Appendix~\ref{sec-cdc}. We treat this as a form of scattered light, and refer to it as such throughout this paper. 
We fully reduce each mask, including wavelength calibration, sky subtraction and scattered light correction.  All images of a given slit (sometimes taken on multiple nights or, rarely, even multiple semesters) are then median combined, rejecting the lowest and highest pixels.  The one-dimensional (1D) spectra are extracted with a weighted Gaussian profile, as described in Paper~I.

Telluric features in our spectra beyond $\lambda=9000$\AA\ are strong, and variable on short timescales.  Initial attempts to correct for this absorption based on the baseline calibration of a single standard star taken each semester proved inadequate.  We therefore derived a correction for each mask, using bright objects on the mask itself\footnote{This still is not ideal, as data for a given mask will have been obtained over several hours or even several nights.  A frame-by-frame correction might do better, though without a dedicated bright star the signal would generally be too low for a good correction.  In the end, this correction proves adequate for most of our data.}.  Midway through the project, when it was realized that a better procedure was needed, we added a single bright star to each mask for this purpose.  As these stars were chosen to minimize impact on existing science slits they were not usually good telluric standards in the normal sense (they are often late-type stars), but they proved suitable for the procedure described in Appendix~\ref{sec-telluric}.  For masks where a star is not available we used either a bright galaxy spectrum, or a stacked spectrum of the 5--15 brightest objects.  

The wavelength-dependent response of the observing system was calibrated using standard star observations, generally taken once per semester as part of Gemini baseline calibrations, not directly associated with this program.  As described in \citet{GCLASS12} and \citet{gogreen-survey}, we find that the shape of this response is stable over time, and we apply a single correction to all spectra (though separately for GCLASS and GOGREEN).  We expect that this calibration is typically good to about 10 per cent, based on a comparison with SED templates fit to the photometry.  An absolute flux calibration is determined where possible, by comparing to the available $i-$band photometry as described in Appendix~\ref{sec-fluxcal}.  The main exception is for the spectra associated with SpARCS1033, for which K-band imaging (and, hence, photometric catalogues) are not yet available.  Some galaxies with spectra do not have a match in the photometric catalogues because they lie in a masked region; no absolute calibration is applied in these cases.

\section{Derived Data Products}\label{sec-ddp}
\begin{figure}{}
\includegraphics[clip=true,trim=65mm 0mm 80mm 0mm,width=3.4in,angle=0]{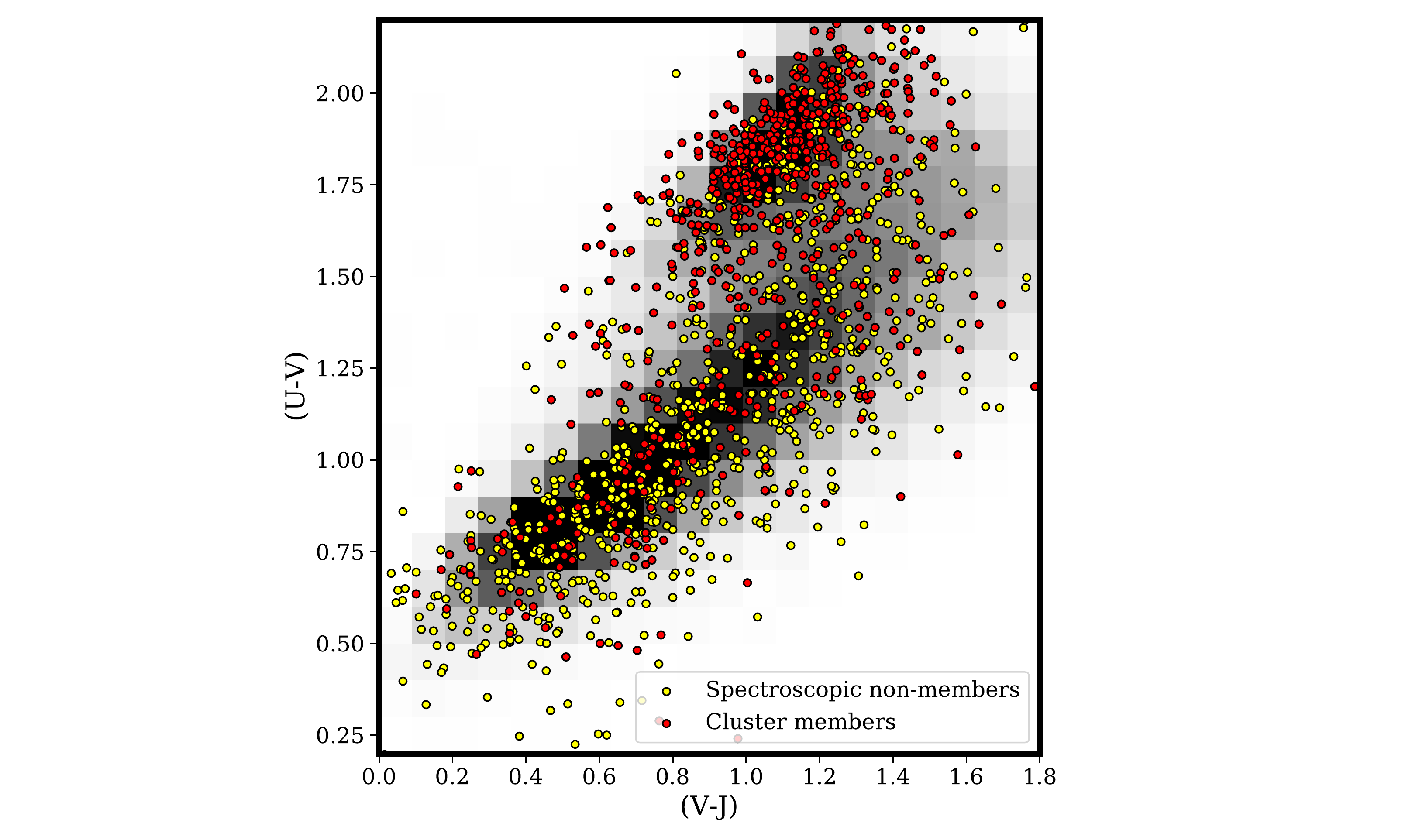}
	\caption{The rest-frame (U-V) and (V-J) colours are shown for all galaxies with spectroscopic redshifts in GOGREEN and GCLASS.  Field galaxies with spectroscopic redshifts $0.8<z<1.5$ are shown as yellow points.  Red points are those galaxies identified as spectroscopic cluster members.  The linear greyscale corresponds to the density of points in the full photometric sample (including Ultravista and SPLASH), within the same redshift limits and $10^9<M/\mathrm{M}_\odot<10^{12}$.\label{fig-uvj}}
\end{figure}
\begin{figure}{}
\includegraphics[clip=true,trim=0mm 0mm 0mm 0mm,width=3in,angle=0]{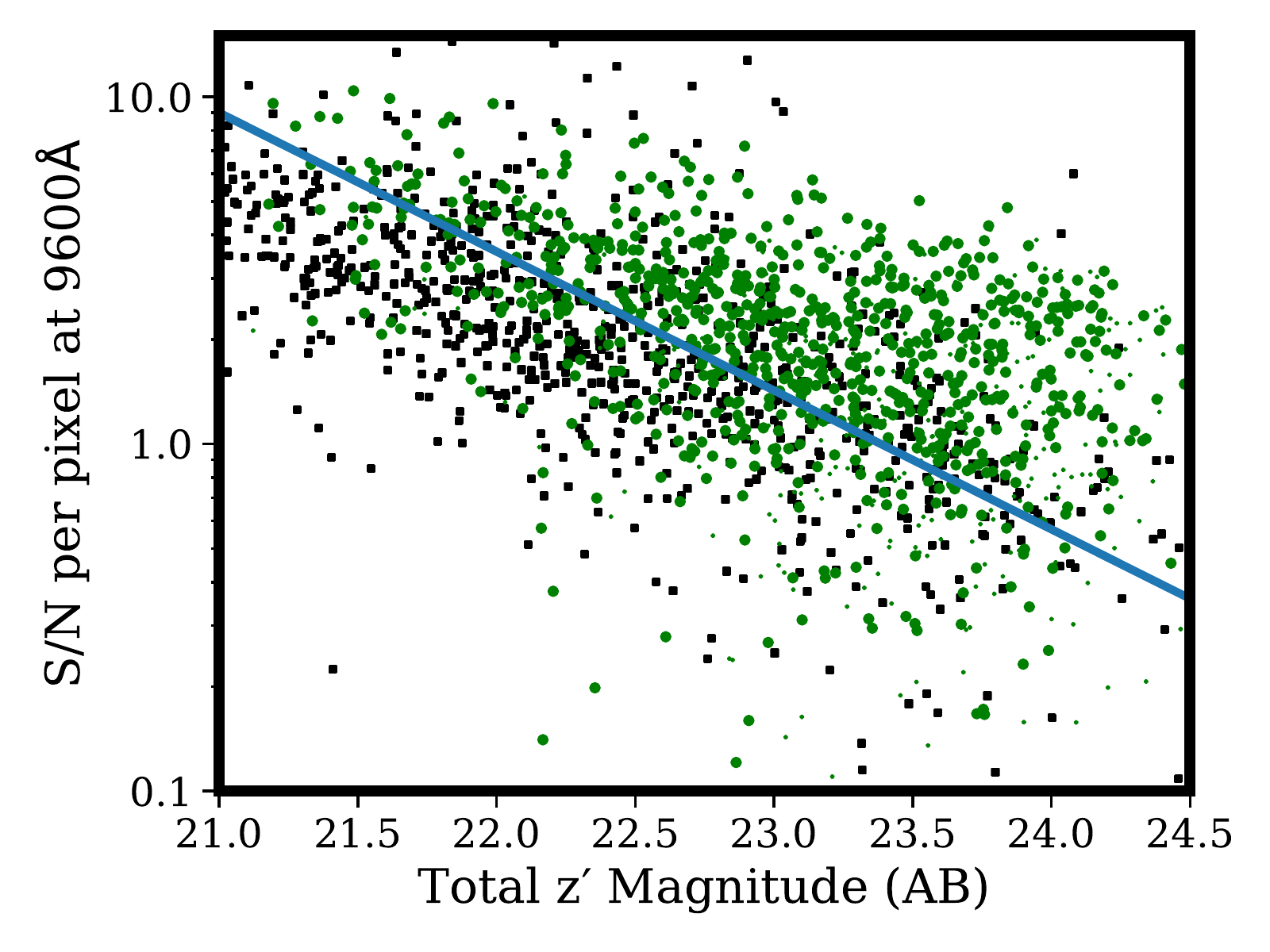}
	\caption{The S/N per 3.9\AA\ pixel, measured at a wavelength of 9600\AA, is shown as a function of total $z^\prime$-band magnitude, for all GCLASS spectra (black squares) and GOGREEN (green dots) primary targets.  Heavier symbols correspond to spectra that yield a reliable redshift (see \S~\ref{sec-redshifts}).  The straight line is shown as a reference; it represents the dependence of S/N on magnitude that would be expected if all spectra were obtained in the same conditions and with the same exposure time (arbitrary normalizaion).  This demonstrates how the longer exposure times on the faintest galaxies in GOGREEN successfully ensures that most spectra obtain $\mathrm{S/N}>1$ per pixel, independent of magnitude. \label{fig-snrmag}}
\end{figure}

\subsection{Photometric Products}\label{sec-photoz}
\subsubsection{Photometric Redshifts, Stellar Masses and Rest-Frame Colours}
Photometric redshifts for GOGREEN are determined using {\sc EAZY} \citep[][version May 2015]{EAZY}, as described in \citet{gogreen-smf}.  Small residuals relative to the spectroscopic redshifts (see \S~\ref{sec-redshifts}) were corrected by fitting and applying a low order polynomial correction.  These corrected redshifts are what are provided in the data release, and used throughout this paper.

Stellar masses and rest-frame colours are measured by fitting stellar population synthesis models of \citet{BC03} with the {\sc FAST} \citep{FAST} code, assuming a \citet{Chab} initial mass function.  Details are again provided in \citet{gogreen-smf}.  These models assume simple star formation histories parametrised with a declining exponential function.  This is known to underestimate the stellar mass by up to 0.3\,dex compared with nonparametric (binned) star formation histories \citep{2019ApJ...876....3L}.    

\begin{figure*}{}
\includegraphics[clip=true,trim=0mm 0mm 0mm 0mm,width=2.3in,angle=0]{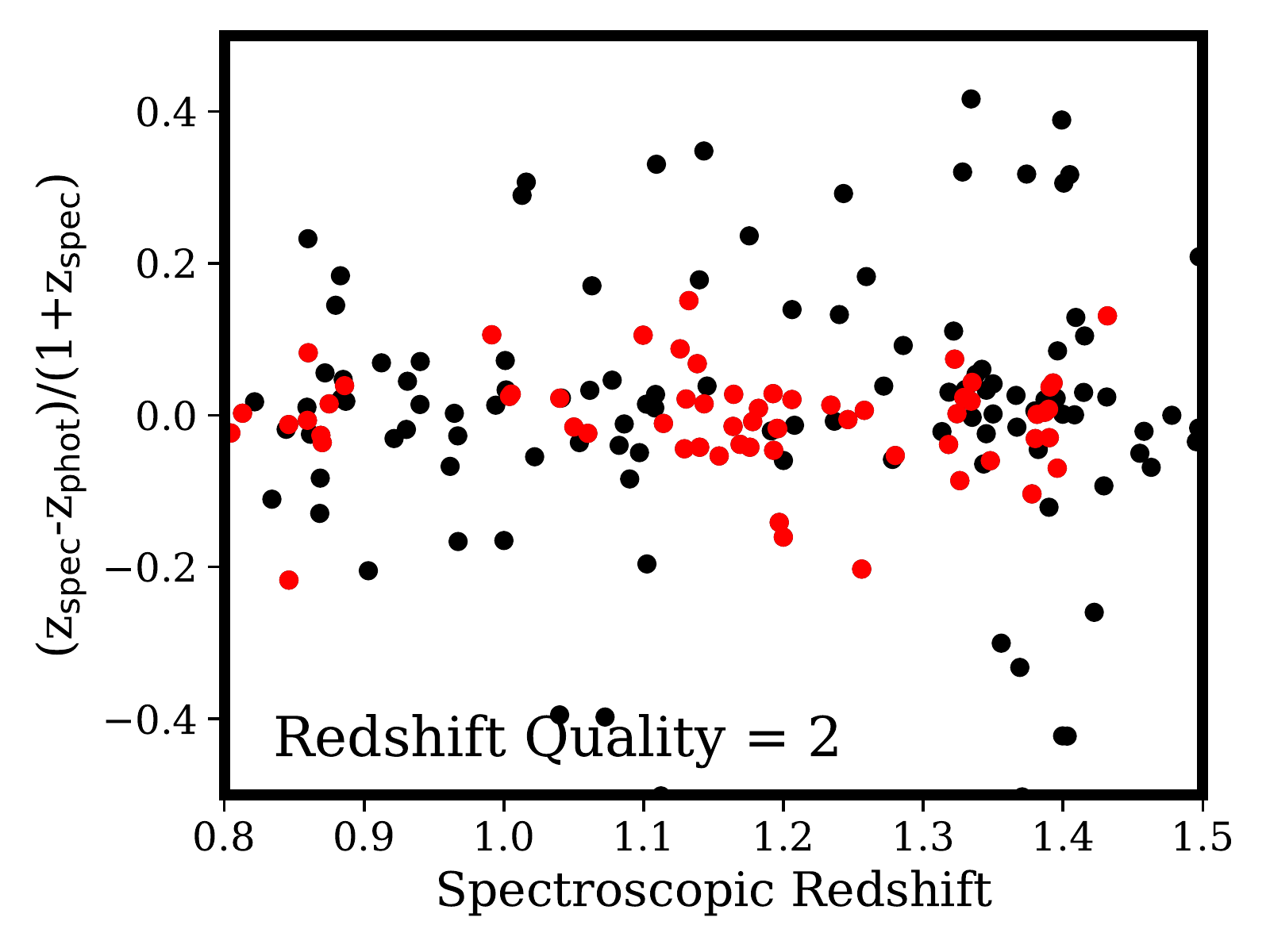}
\includegraphics[clip=true,trim=0mm 0mm 0mm 0mm,width=2.3in,angle=0]{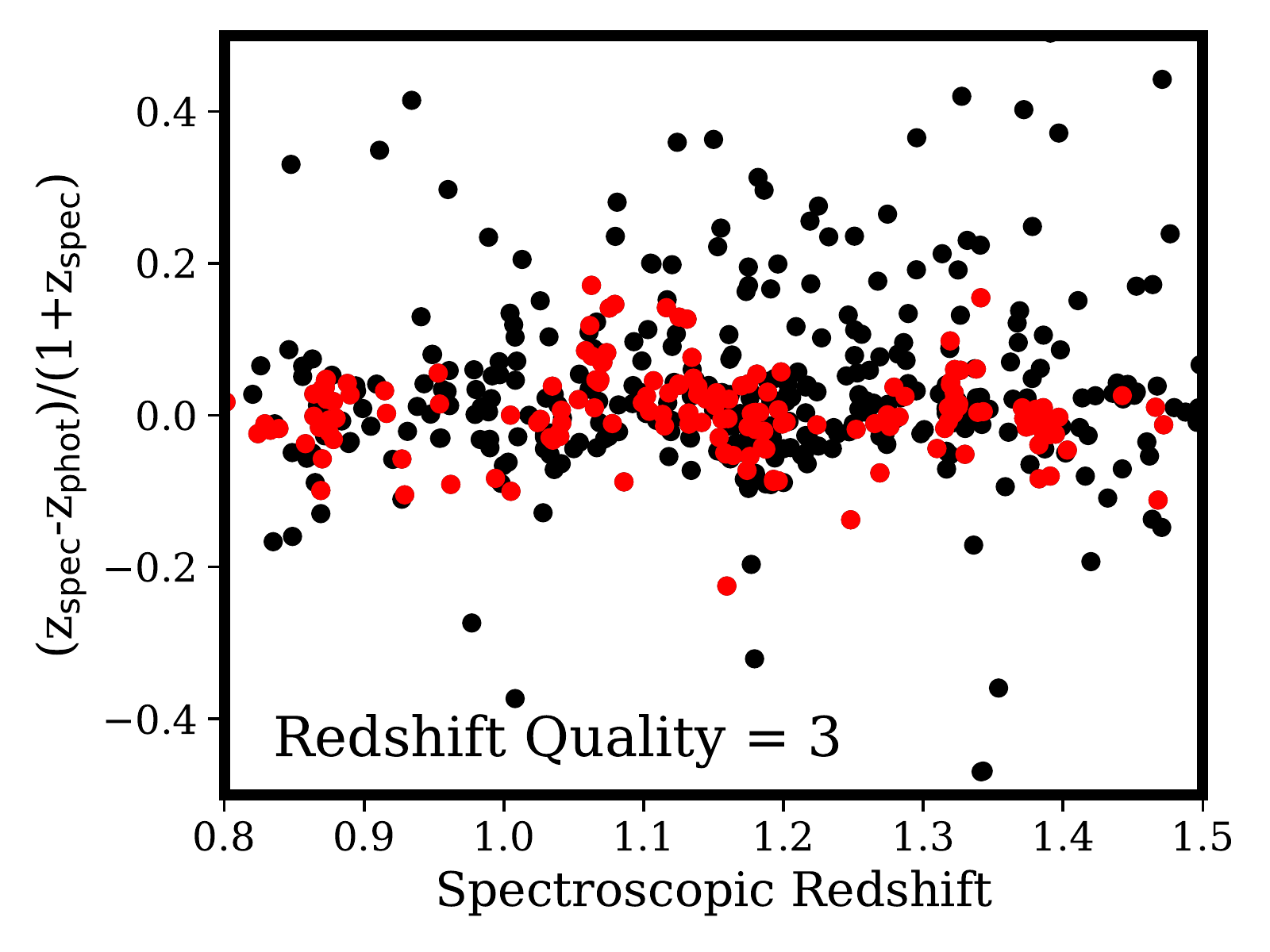}
\includegraphics[clip=true,trim=0mm 0mm 0mm 0mm,width=2.3in,angle=0]{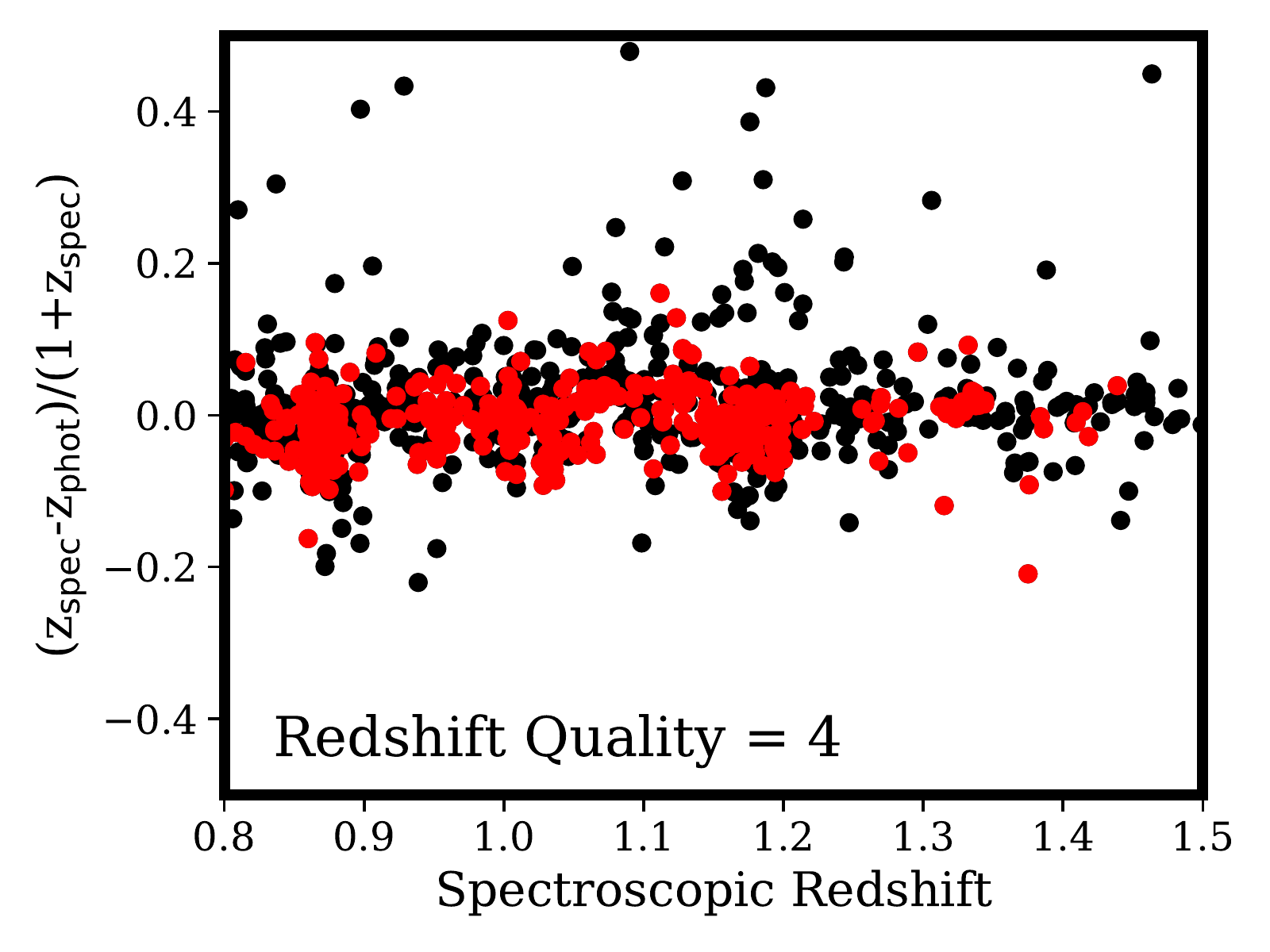}
	\caption{The difference between spectroscopic and photometric redshifts is shown for the full sample (except SpARCS1033, for which photometric catalogues are not available), for galaxies with  spectroscopic redshifts $0.8<z<1.5$.  The three panels correspond to increasing redshift quality, as indicated.   Note that only quality 3 and 4 are considered acceptable for most analyses.  A small number of outliers lie outside the bounds of the plot.  Quiescent galaxies are indicated with red symbols.  The standard deviation of $\Delta z/(1+z)$ is 0.05, with 7 per cent outliers, for galaxies with quality flag 3 or 4. \label{fig-zspeczphot}}
\end{figure*}

The rest-frame (U-V) and (V-J) colours are shown in Figure~\ref{fig-uvj}, for all galaxies in the GCLASS and GOGREEN samples with secure redshifts (see \S~\ref{sec-redshifts}) within $0.8<z<1.5$.  Galaxies identified as likely cluster members (see \S~\ref{sec-vdisp}) are shown in red.  For comparison, the full photometric sample, including the Ultravista \citep{Ultravista_Muzzin} and SPLASH \citep{SPLASH} catalogues\footnote{We compute UVJ colours for all galaxies in the SPLASH catalogue using the same procedure as for the clusters, and these are provided in the data release as described in \S~\ref{sec-DR}.  For Ultravista, we use the UVJ colours in the catalogue, which were computed in the same way.}, is shown in greyscale.  The bimodality of the colour distribution is evident, as is the dominance of the quiescent population among cluster members.  We use these colours to define the quiescent galaxy population through this work, as 
\begin{equation}
 (\mathrm{U}-\mathrm{V}) > 1.3 \,\,\cap\,\, (\mathrm{V}-\mathrm{J})<1.5 \,\,\cap\,\, (\mathrm{U}-\mathrm{V}) > 0.88\,(\mathrm{V}-\mathrm{J}) + 0.59 
 \end{equation}
\noindent as defined in \citet{Ultravista_Muzzin} for $1{<}z{<}4$, adapted from \citet{W+09}.

\subsection{Spectroscopic Products}
In Figure~\ref{fig-snrmag}, we show the correlation between S/N per pixel in each final spectrum and the total $z^{\prime}$-band magnitude of the galaxy.  The S/N ratio here is the average S/N within $9500<\lambda<9700$\AA.  There is significant scatter, reflecting the fact that spectra were obtained with a range of exposure times, detectors, and in a variety of conditions over several years.  Some scatter is also likely caused by different galaxy sizes, which affect the amount of light going down the slits of fixed width.  Notable is the flattening of the relation for GOGREEN spectra at magnitudes ${>}23.25$, relative to the declining S/N that would be expected for fixed exposure time (solid line).  This is because galaxies fainter than this magnitude are observed on multiple masks, to build up the total integration time.  As a result, $\sim 80$ per cent of the spectra have  $\mathrm{S/N}>1$ per pixel.

\subsubsection{Redshifts}\label{sec-redshifts}
Spectroscopic redshifts for GCLASS were determined by comparison with templates within the iGDDS software \citep{GDDS}.  For GOGREEN we used the Manual and Automatic Redshifting Software \citep[{\sc marz},][]{hinton2016}, with the default templates supplemented by stacked spectra of red and blue $z>1.5$ galaxies from GMASS \citep{GMASS}. Redshifts were determined interactively, using custom software to show the image, one- and two-dimensional spectra, and photometric redshift information (with 68 per cent confidence limits) for each galaxy in addition to the {\sc marz} cross-correlation results. Quality flags are assigned based on a largely subjective assessment of the redshift likelihood, and considering the photometric redshift.  These flags are described in \S~\ref{sec-zcat}, and included in the data release as Redshift\_Quality (see Table~\ref{tab-zcat.fits}).  The GCLASS redshift quality flags have been translated onto this system, so all spectra in this release have a uniform classification.  We consider galaxies with Reshift\_Quality$>2$ to be "good", and these are used in all the analysis here and in accompanying science papers, unless explicitly stated otherwise.

To quantify the accuracy and precision of the photometric redshifts, we compare these with the spectroscopic redshifts in Figure~\ref{fig-zspeczphot}\footnote{Recall from \S~\ref{sec-photoz} that the photometric redshifts have had a small empirical correction applied to remove offsets relative to the spectroscopic redshifts.}.  For this comparison we consider only the subsample of galaxies with good quality spectroscopic redshifts in the range $0.8<z<1.5$.  Quiescent galaxies are indicated with red circles. 
Considering all galaxies with good spectroscopic redshifts (Redshift\_Qualit$>2$), we find excellent agreement with the photometric redshifts, with an outlier fraction\footnote{Following \citet{gogreen-smf}, we define outliers as those for which $|\Delta z|/(1+z)$>0.15.} of 7.3 per cent, and a standard deviation (after rejecting outliers) of $0.05$ in $\Delta z/(1+z)$.  Restricted to the quiescent galaxy subsample the standard deviation is similar, at 5.8 per cent, but the outlier fraction is much smaller, only 1.1 per cent.    We use this as an indicator of our photometric redshift accuracy.

However, we also recognize that some of the outliers may be due to incorrect spectroscopic redshifts.  We see in fact that the outlier fraction increases from 4.5 per cent for the quality class 4 galaxies, to 13 per cent for quality class 3, and 22 per cent for quality class 2 (which generally should not be used for science).  The redshift quality flag is not independent of the photometric redshift, since good agreement with the latter is considered when assigning confidence to the spectroscopic redshift.  It is therefore not possible from this analysis to robustly determine what fraction of spectroscopic redshifts might be incorrect, but it is likely $<10$ per cent of those with quality flag 3, and significantly more for those with quality flag 2.

\begin{figure}{}
\includegraphics[clip=true,trim=0mm 0mm 0mm 0mm,width=3in,angle=0]{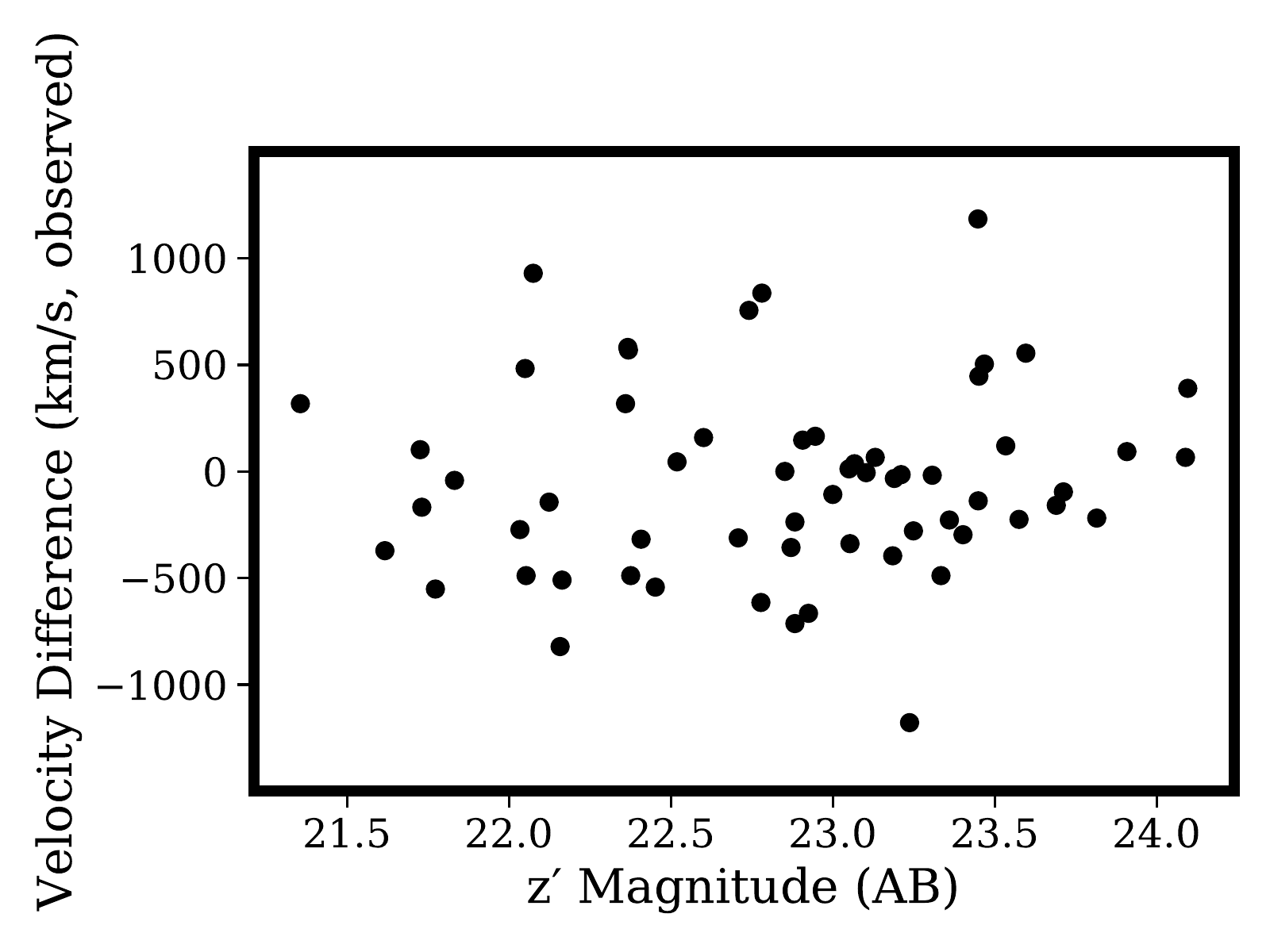}
	\caption{The difference between redshifts (expressed as a velocity $cz$) measured in GOGREEN and GCLASS, for 61 galaxies in common, with good redshifts.  The difference is in the sense GOGREEN$-$GCLASS, so negative values indicate larger redshifts in GCLASS.  One galaxy, with a redshift larger in GCLASS by $0.034$, is outside the plotted area.  The median offset is $-95$\,km/s, and the standard deviation (excluding the outlier) is $278$\,km/s.\label{fig-zcomp}}
\end{figure}

To assess the accuracy and precision of the GOGREEN spectroscopic redshifts, we compare the spectroscopic redshifts of the 61 objects with good redshifts in common with GCLASS.  The observed-frame velocity difference of these spectra are shown in Figure~\ref{fig-zcomp}, as a function of magnitude in the $z^{\prime}$-band.  Of these 61 objects, only one (not shown on the plot), is a significant outlier, with a velocity difference of $>1000$\,km/s.  The standard deviation of the rest of the sample is $393.5$\,km/s, with no significant dependence on magnitude, redshift or S/N in the spectrum.  Assuming GOGREEN and GCLASS have comparable, normally distributed redshift uncertainties, this means the typical uncertainty on an individual redshift is $278$\,km/s.  In the rest-frame of the clusters (all at $z>0.8$), this corresponds to an uncertainty of $<154$km/s.  There is also an unresolved bias, such that GOGREEN redshifts are smaller than those measured in GCLASS, by $95$\,km/s (observed frame).  As this is small relative to the typical uncertainty, and to the velocity dispersions of the clusters, we do not apply any correction.  For the three SPT clusters, we have 22 GOGREEN redshifts that overlap with published redshifts in \citet{SPT0205}, \citet{SPT0546} and \citet{SPT2106}.  The median and standard deviation of the redshift difference is 25km/s and 550km/s, respectively, and there are no significant outliers.  

About 15 per cent of the GOGREEN spectra obtained were severely compromised by a) contamination from bright, nearby objects; b) excess flux from the target or neighbours in the sky sampled region; c) poor sky subtraction or telluric correction; or d) large and critical regions of wavelength space affected by uncorrected scattered light or detector issues.  We flag these spectra and exclude them from the redshift completeness statistics here.  As there is no possibility to obtain a redshift from them, regardless of the intrinsic source characteristics, it is appropriate to treat them as if they were never observed.

\begin{figure}
\includegraphics[clip=true,trim=0mm 0mm 0mm 0mm,width=3in,angle=0]{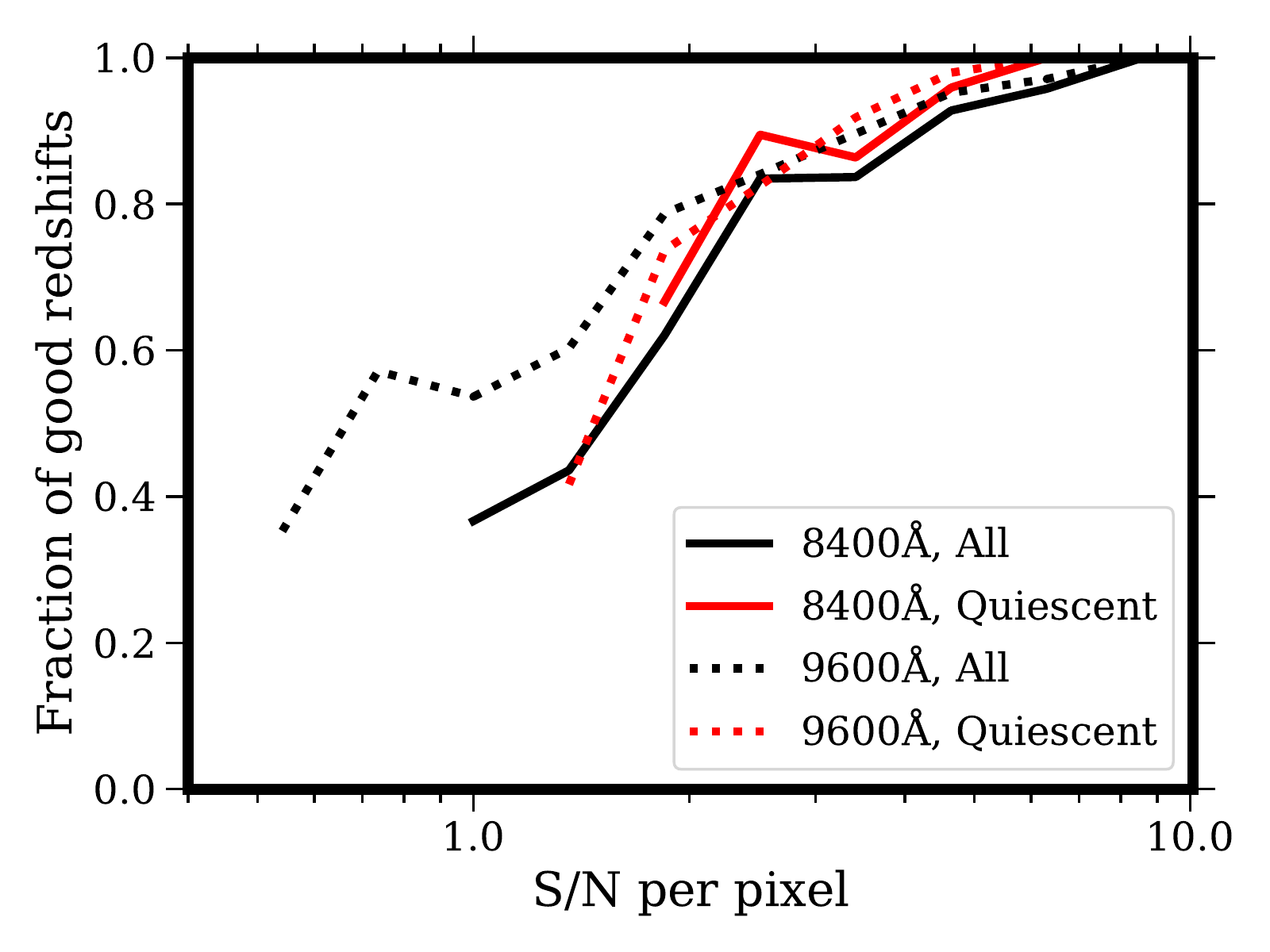}
	\caption{The fraction of GOGREEN primary targets for which a reliable redshift was obtained is shown as a function of signal-to-noise ratio per pixel.  The S/N is measured at two different wavelengths, as shown in the legend.  Black lines represent the success rate for galaxies of all types, while red lines are restricted to quiescent galaxies, as identified by their $UVJ$ colours. \label{fig-zsuccess_snr}}
\end{figure}

In Figure~\ref{fig-zsuccess_snr} we show the fraction of GOGREEN spectra for which we obtained a confident redshift, as a function of S/N per pixel.  Only primary targets -- those that satisfy our colour and magnitude selection boundaries -- that are not flagged as described above are considered here.  The S/N is considered at two different wavelengths --- 8400\AA\ and 9600\AA\ --- as shown.  We obtain redshifts for about 50 per cent of these targets at S/N$\sim 1$ per pixel at 9600\AA.  The success rate is $>80$ per cent at $\mathrm{S/N}>2$.  For $\mathrm{S/N}>2$, the success rate is independent of galaxy type.  Lower S/N redshifts primarily come from emission line galaxies.  

\begin{figure}
\includegraphics[clip=true,trim=0mm 0mm 0mm 0mm,width=3in,angle=0]{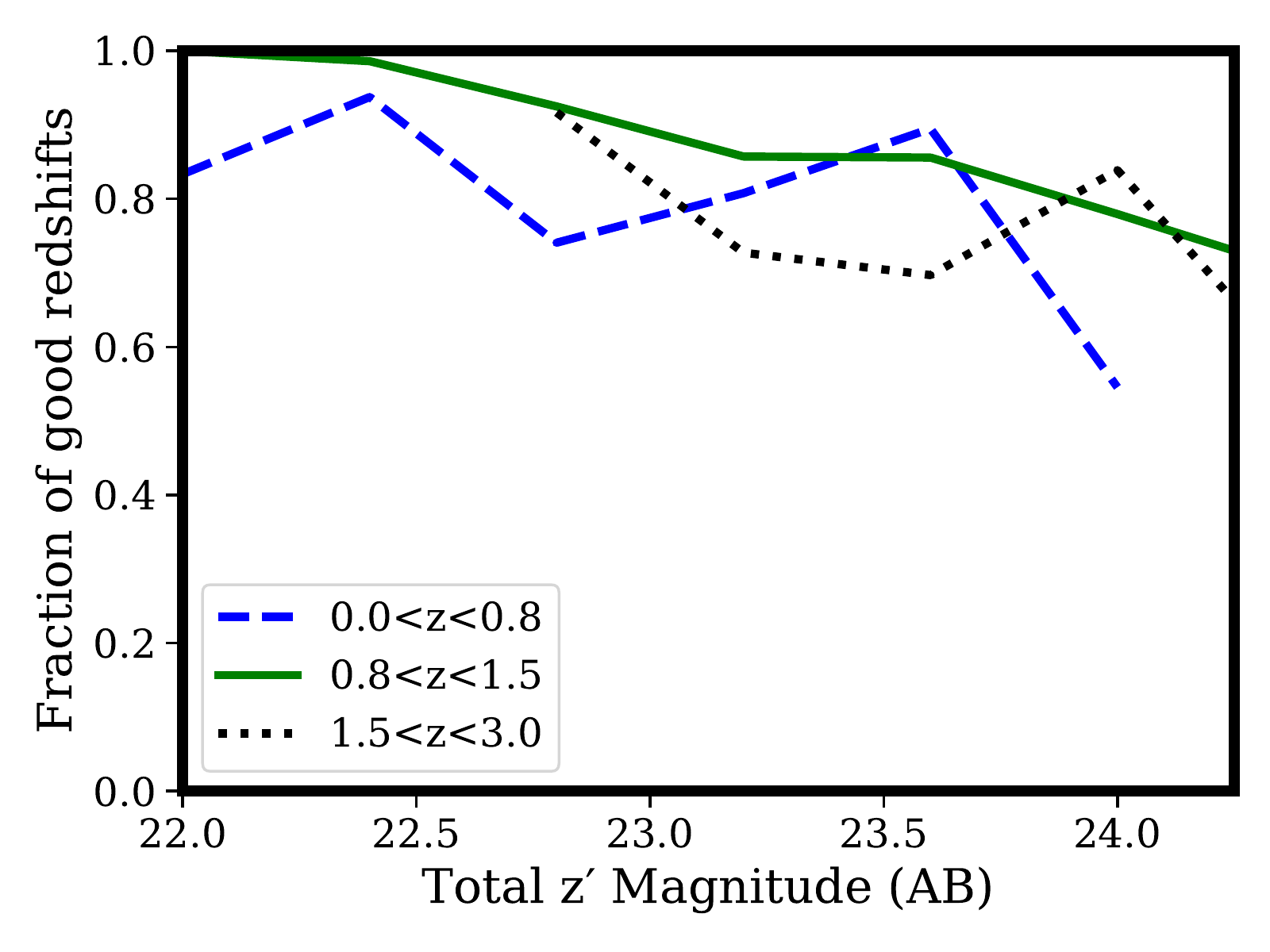}
	\caption{The fraction of GOGREEN targets for which a reliable redshift was obtained is shown as a function of total $z^\prime$-band magnitude.  Targets are binned by their photometric redshift, as shown in the legend.  Only targets for which the spectrum is unflagged and has S/N>1 per pixel at 9600\AA\ are included.  The success rate is highest for galaxies that have photometric redshifts within our target range of $0.8<z<1.5$; it drops 
	for higher and lower redshift galaxies for which key spectral features fall outside our wavelength range. \label{fig-zsuccess_mag}}
\end{figure}

In Figure~\ref{fig-zsuccess_mag} we show how the GOGREEN redshift success rate depends on magnitude and the photometric redshift of the target.  We only consider unflagged galaxies with $\mathrm{S/N}>1$ per pixel here.  As expected, our redshift success rate is somewhat lower for galaxies with photometric redshifts $z<0.8$ or $z>1.5$, for which our spectral range does not cover key spectral features.  For galaxies in our target range of $0.8<z<1.5$, the success rate is $>80$ per cent at all magnitudes.  It is plausible that many of the galaxies for which we do not obtain a good redshift, especially at the faintest magnitudes, are actually at redshifts that lie well outside our targeted range despite what is indicated by their most probable photometric redshift.
\subsubsection{Line Indices and [O\,II]-based star formation rates}\label{sec-indices}
Several spectroscopic indices are computed and provided as part of this data release.  The [O\,II]$\lambda$3727 emission line was measured as described in \citet{gogreen-sfr}.  Briefly, a model consisting of a Gaussian component superposed on a linear continuum was fit to every spectrum, and compared with a continuum-only model.  The Bayesian Information Criterion (BIC) is then used to identify objects for which the fit with the Gaussian component is preferred.  Objects with $\Delta{\rm BIC}>10$ are likely to be secure detections, while those with $\Delta{\rm BIC} < 10$ are likely secure non-detections.  The catalogue provides equivalent widths and line fluxes, as well as the $\Delta{\rm BIC}$ value.  The width of the Gaussian component is also left as a free parameter (as is the redshift, with a tight prior), and the values were used to determine the spectral resolution reported in \S~\ref{sec-spectroscopy}.

\begin{figure*}{}
\includegraphics[clip=true,trim=0mm 0mm 0mm 0mm,width=7.0in,angle=0]{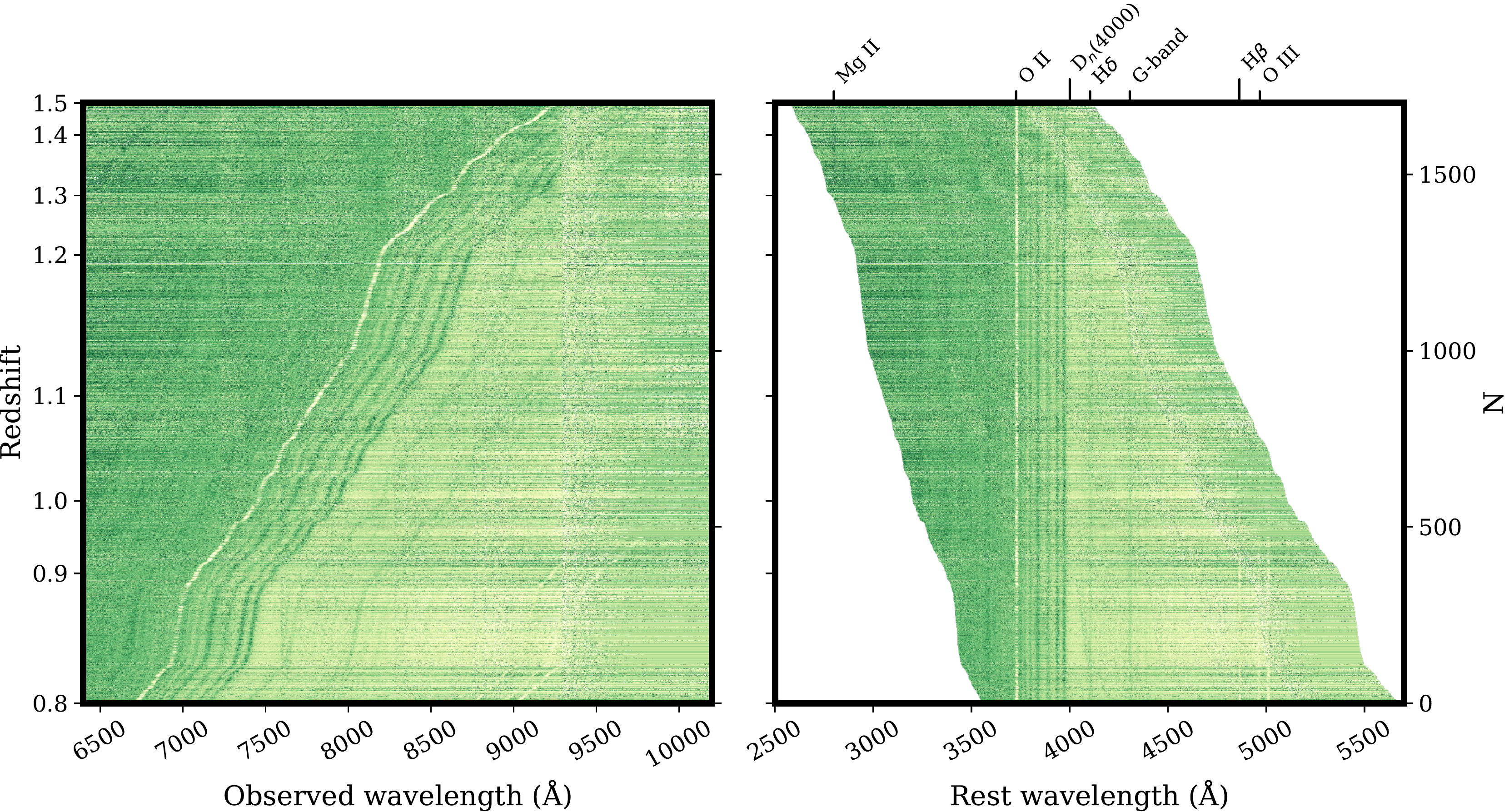}
	\caption{These figures show all GOGREEN and GCLASS spectra with robust redshifts between $0.8<z<1.5$.  The spectra are arranged in order of increasing redshift, from bottom to top as shown on the y-axis. The left panel shows the spectra in observed wavelength, while on the right they are in rest-frame coordinates.  The [O\,II] emission line, 4000\AA\ break and several absorption lines (indicated at the top of the right panel) are clearly visible.  \label{fig-spectra}}
\end{figure*}

Star formation rates (SFRs) are estimated from the [O\,II] fluxes, using the calibration of \citet{2010MNRAS.405.2594G}.  This is a stellar-mass dependent calibration relative to H$\alpha$-derived SFRs at $z=0$.  Because [O\,II] is strongly dependent on dust, ionization, and metallicity, there is significant scatter in this relation.  It is therefore most useful for determining the average SFR of a population, rather than as an indicator for individual galaxies.

We also provide measurements of the 4000\,\AA\ break, and the equivalent widths of the [O\,II] emission and H$\delta$ absorption lines, following the definitions in \citet{PSG}.  The D$_{n}$4000 index is an age-sensitive colour, based on the average value of $f_\nu$ in a narrow wavelength range on either side of the break.  Individual measurements of H$\delta$ typically have large uncertainties, and should be used with caution.  The [O\,II] equivalent widths are provided as an alternative to the Gaussian-modeling approach described above.

\subsubsection{Cluster Velocity Dispersions and membership}\label{sec-vdisp}
Cluster members for GCLASS were selected based on a simple $\Delta v\leq 1500$\,km/s criterion,  
as described in \citet{GCLASS12}.  Velocity dispersions presented in \citet{GCLASS_dynamics} were determined using standard methods \citep[e.g.][]{Beers}.  As these clusters are generally very well defined in velocity space, these simple methods work adequately for most purposes.

For the clusters in GOGREEN, velocity dispersions are computed including all available redshifts from the literature.  The total number of members used is reported in parentheses in column (6) of Table~\ref{tab-cluster_sample}, with the references provided in the final column.  The membership and velocity dispersion for the SPT and SpARCS clusters are derived as described in Biviano et al. (in prep).  First, the main redshift peak is identified following \citet{Beers91}, and any significant substructure identified using the KMM algorithm  \citep{KMM}.  A small amount of substructure is only identified in two clusters (SpARCS1051 and SpARCS1616).  Membership is then assigned based on their location in projected phase space.  Two different algorithms are used.  CLEAN \citep{CLEAN} is an established method, theoretically motivated based on the results of simulations.  We also consider a new algorithm, CLUMPS, that is less model dependent.  The difference in membership resulting from the two algorithms is small (usually within 10 per cent), and we defer the details Biviano et al. (in prep).  The number of galaxies identified as a member by either method is what is listed in Table~\ref{tab-cluster_sample}.


The COSMOS and SXDF systems in GOGREEN are more challenging because they have relatively few spectroscopic members, often without a well-defined central concentration.  Furthermore, the location of the X-ray detection provides an important prior on the central location, though the centroid of the low X-ray fluxes are often uncertain, as well.  We use an iterative clipping algorithm, again making use of all available public spectroscopy as well as that from GOGREEN, to measure the velocity dispersion.  For COSMOS, the literature redshifts are drawn from a compilation provided by M. Salvato (priv comm).  SXDF redshifts are taken from 
UDSz \citep{bradshaw2013high, mclure2012sizes}, XMM-LSS \citep{2013AA...557A..81M, 2013MNRAS.429.1652C} and
VANDELS \citep{2018arXiv181105298P}. Cluster members are defined as those within $1$\,Mpc and $2.5\sigma$ of the iteratively-determined spatial and velocity centre, respectively. In each iteration, the centre itself is the unweighted average position of those members within a 1\,Mpc area centered on the X-ray detection. In some cases, there are many galaxies within the velocity dispersion of the cluster that lie outside the $1$\,Mpc limit; in that sense, membership with the larger scale structure defining these systems can be larger than what is reported in Table~\ref{tab-cluster_sample}.  Uncertainties on the velocity dispersion are computed by bootstrapping the sample of selected cluster members.  Full details will be available in Reeves et al. (in preparation).

\section{Spectroscopic sample characteristics}\label{sec-ssample}
The final spectroscopic catalogue for GCLASS and GOGREEN consists of $2771$ unique objects, of which $2257$ have good redshifts:  $1529$ from GOGREEN, and $728$ from GCLASS\footnote{Where there is an observation from both surveys, we retain only the GOGREEN spectrum and redshift for the final catalogues and analysis.}.
An overview of our spectroscopic sample in the range $0.8<z<1.5$ is shown in Figure~\ref{fig-spectra}.  All one-dimensional spectra with good redshifts, from both surveys, are displayed in order of redshift.  Key emission and absorption features are clearly visible throughout the redshift range, as is the excellent quality of the sky subtraction at the red end of the spectra.

\begin{figure}
\includegraphics[clip=true,trim=0mm 0mm 0mm 0mm,width=3in,angle=0]{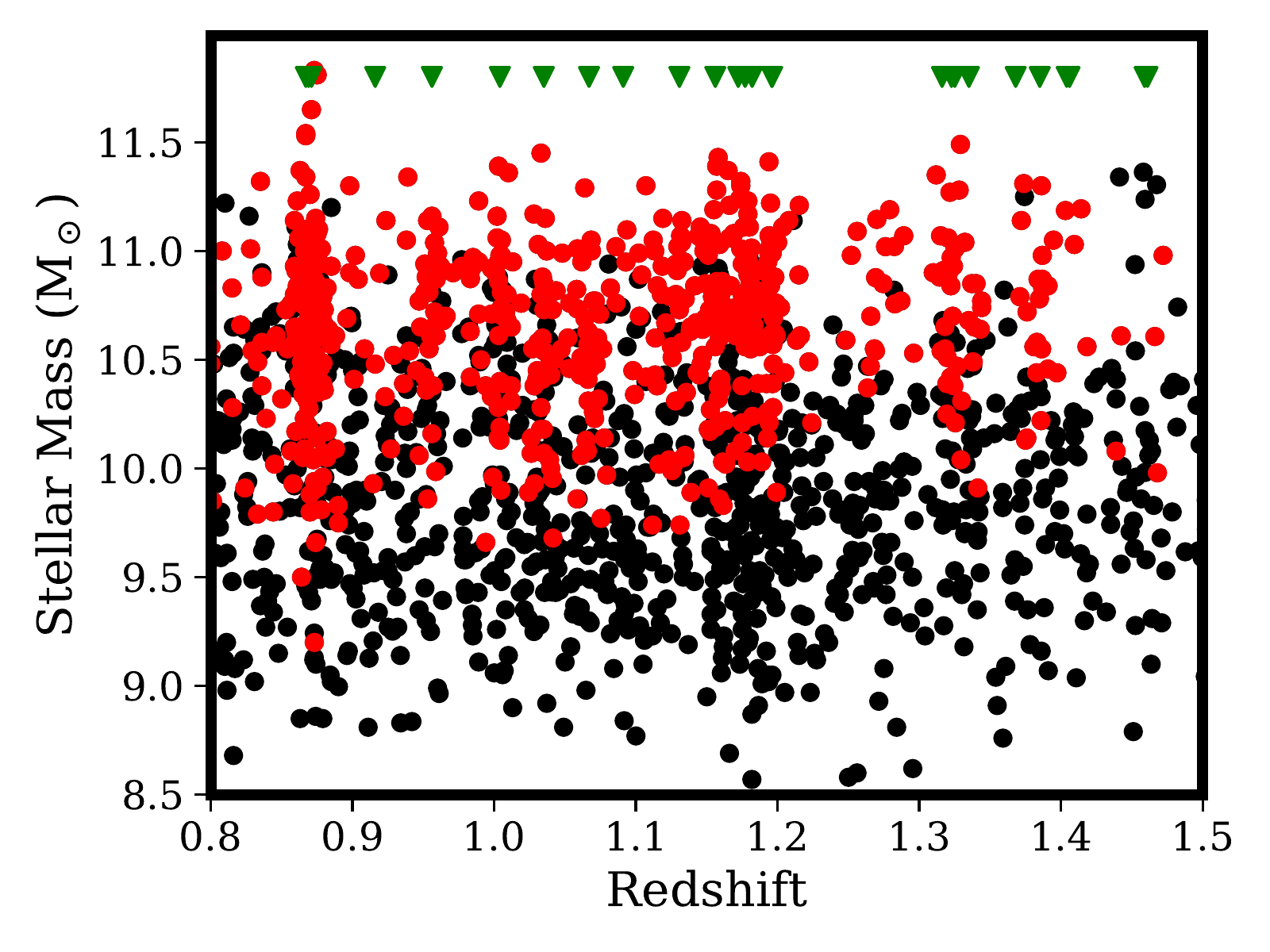}
\caption{The stellar masses and redshifts are shown for the spectroscopic sample within $0.8<z<1.5$.  Galaxies classified as quiescent from their UVJ colours are indicated in red.  The mean redshifts of each system are shown with inverted, green triangles at the top of the plot.  Note that data for SpARCS1033 is not represented in this plot, because photometric catalogues are not available at this time due to the lack of $K$ band imaging.  \label{fig-zmass}}
\end{figure}

In Figure~\ref{fig-zmass} we show the distribution in redshift and stellar mass, for most of the spectroscopic sample within $0.8<z<1.5$; SpARCS1033 is omitted from the plot since $K$-band selected catalogues are not available at this time.  Quiescent galaxies are indicated with red circles.    The clustering in redshift space is readily apparent, and corresponds to the locations of the targeted systems, indicated with green triangles at the top of the Figure.  

\subsection{GOGREEN Completeness}\label{sec-comp}
We now consider how the spectroscopic completeness in GOGREEN depends on colour, magnitude, stellar mass, spectral type, and position.   This can vary significantly from cluster to cluster.  The geometric sampling in particular is non-trivial to characterize because the GOGREEN spectroscopy does not uniformly sample the area around the cluster, but is preferentially aligned along one (arbitrary) axis, perpendicular to the dispersion direction.  

\begin{figure*}
\includegraphics[clip=true,trim=0mm 0mm 0mm 0mm,width=7.5in,angle=0]{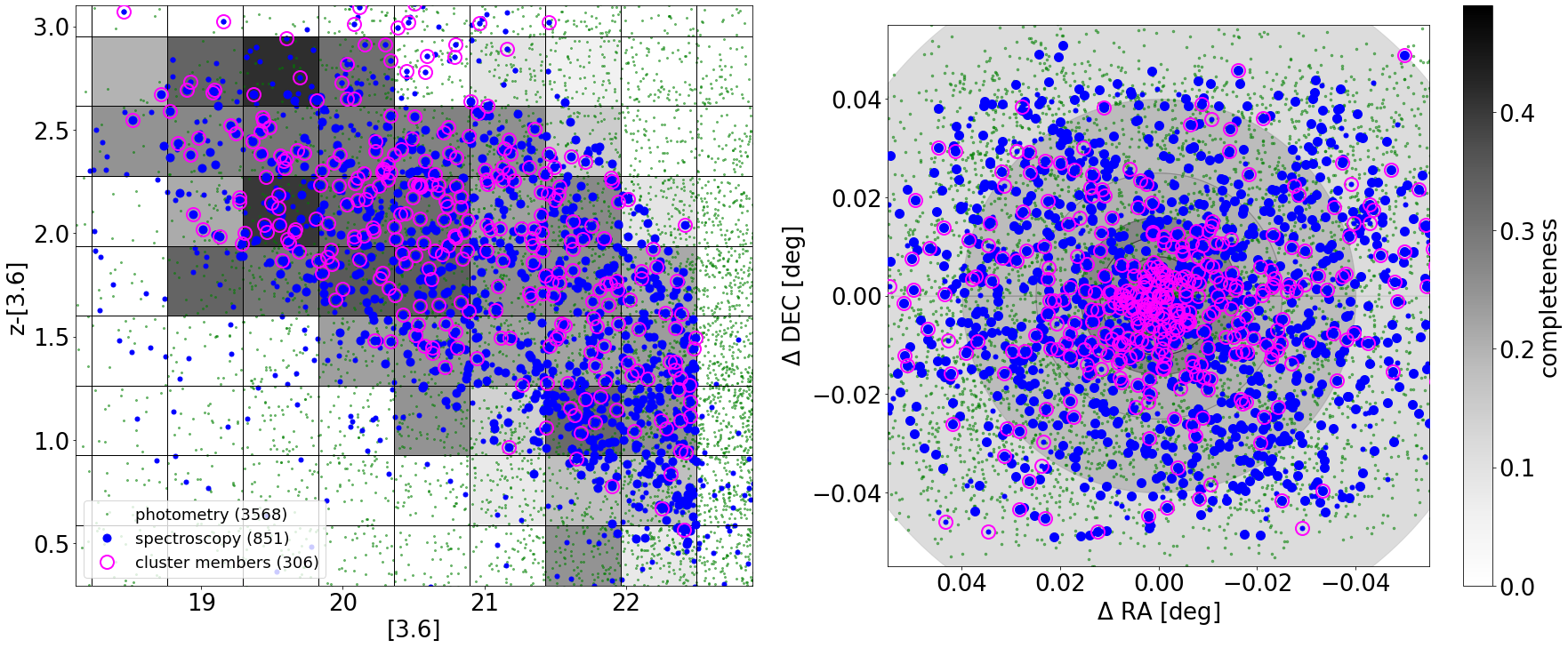}
\caption{These images show the spectroscopic sampling rate in the twelve SpARCS and SPT clusters observed by GOGREEN, relative to directly observable selection criteria.  On the {\it left} we show the distribution in colour-magnitude space.  Most of the objects with spectroscopy (blue points) are confined to a band that reflects the selection criteria.  Magenta circles show galaxies that were ultimately identified as cluster members.  The greyscale grid shows the completeness, relative to the full photometric sample in each bin.  The numbers in the legend correspond to the number of objects that lie within the main colour selection boundaries, and that are included in the GOGREEN preimaging selection catalogue.  This excludes, for example, some GCLASS objects.  On the {\it right}, we show the same points distributed spatially, relative to the cluster centre.  Again the greyscale represents the average completeness, within each of three broad, radial bins. \label{fig-zIRACcompleteness}}
\end{figure*}

The completeness with respect to the observed colour and magnitude selection in GOGREEN is described in detail, and presented for the twelve SpARCS and SPT clusters included in the GOGREEN survey (including SpARCS1033, as the $K$-band catalogues are not needed for this analysis), in Appendix~\ref{sec-obscomp}.  Here we provide a summary of the average completeness for these twelve systems, in  Figure~\ref{fig-zIRACcompleteness}.   Targets were selected with primary limits [3.6$\mu$m]<22.5 and $z^\prime$<24.25, with colour cuts to exclude foreground and background galaxies.  While the blue cut is the same for all fields, the red selection limit was adjusted for each cluster, depending on its redshift. This defines the band within which most of our spectroscopy lies in the left panel of Figure~\ref{fig-zIRACcompleteness}; the few objects outside those bounds are "mask-filler" objects.  The greyscale corresponds to the completeness within each colour-magnitude cell;  overall, spectroscopic sampling within the colour boundaries is about 25 per cent, dropping somewhat with magnitude but not strongly dependent on colour.  The panel on the right shows the spatial distribution of the spectroscopy for the same twelve clusters, with greyscale again representing completeness in broad radial bins.  While there was no explicit geometric selection criteria in GOGREEN, the single GMOS pointing in each cluster effectively limits the spectroscopy to a broad stripe no more than 5\arcmin\ in length.  

\begin{figure*}{
		\includegraphics[clip=true,trim=0mm 0mm 0mm 0mm,width=7.5in,angle=0]{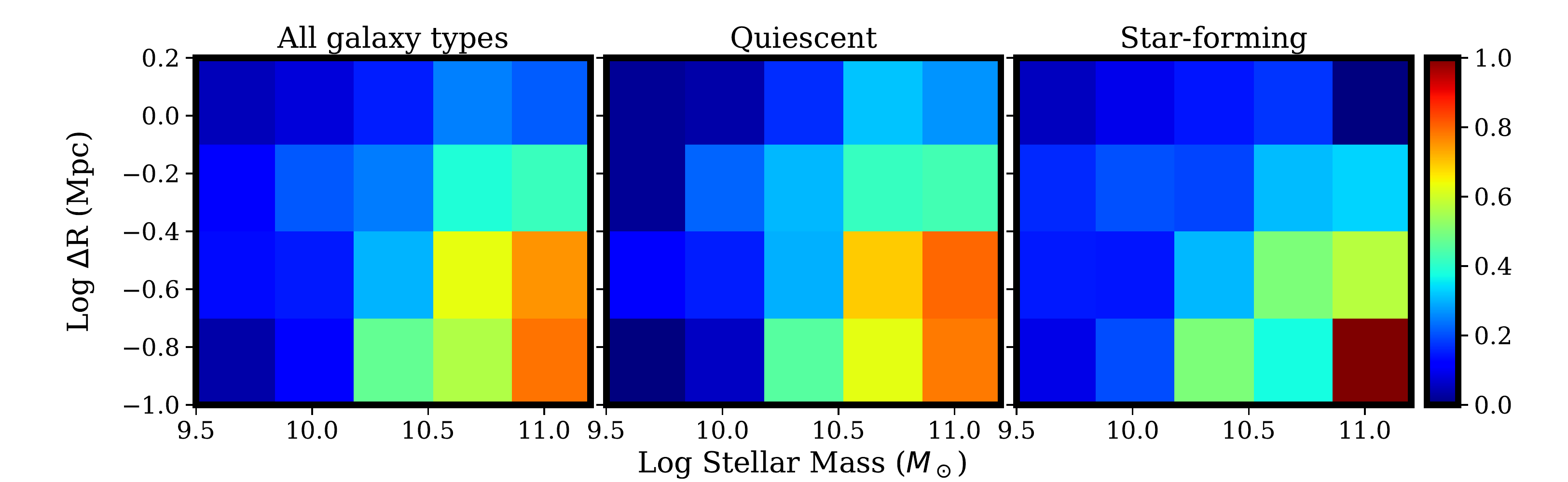}}
	\caption{The spectroscopic completeness of the sample is estimated as a function of distance from the cluster centre and stellar mass.  The parent sample is taken to be all galaxies in the photometric sample for which the cluster redshift lies within the 68 per cent confidence limits of the photometric redshift. \label{fig-MRC-all}}
\end{figure*}

For most analyses, what is likely to be most relevant is how complete the spectroscopic sample is relative to the population of likely  cluster members, in terms of physical parameters like stellar mass, clustercentric distance and galaxy type.  We can evaluate this using the photometric catalogues, available for all our systems except SpARCS1033.  Here, we consider the parent cluster population to be all galaxies for which the cluster redshift lies within the 68 per cent confidence limits of the photometric redshift.  We then count how many of those galaxies have a robust redshift, regardless of whether or not that redshift is consistent with that of the cluster.    The ratio of these numbers is shown as the completeness in Figure~\ref{fig-MRC-all}, as a function of clustercentric distance and stellar mass.  Above a mass of $\sim 10^{10.2}~\mathrm{M}_\odot$, the completeness is more than 30 per cent, out to beyond 500 kpc from the cluster core.  Most notably, the completeness for quiescent galaxies (classified from their rest-frame UVJ colours) above this mass limit is comparable to that of the sample as a whole. The drop in completeness at low stellar masses is partly driven by the fact that the 68th percentiles on the photometric redshift estimates get larger, so the parent sample to which we compare increases relative to the true underlying cluster population.

Based on this, we conclude that the spectroscopic catalogue provides a significant ($\sim 45$ per cent on average) sampling of the parent cluster population within $500$~kpc, and for masses $>10^{10.2}~\mathrm{M}_\odot$, that is not strongly biased with respect to galaxy type.  An exception to this might be strongly dust-reddened (likely massive, star-forming) galaxies, which would be biased against during target selection and also when determining spectroscopic redshifts.  

\section{Public Data Release Contents}\label{sec-DR}
Here we describe a summary of the public data release contents. 
Three main catalogues are included in this publication, and available via linked online resources.  These are the main cluster catalogue (\S~\ref{sec-cluscat}); the redshift catalogue (\S~\ref{sec-zcat}); and a photometric catalogue with a subset of available information expected to be most commonly useful (\S~\ref{sec-photcats}).  All catalogues, including the full $K_s$-selected photometric catalogues, and reduced data (images and spectroscopy) are available via the CADC (\url{https://www.cadc-ccda.hia-iha.nrc-cnrc.gc.ca/en/community/gogreen}), and NSF's NOIRLab (\url{https://datalab.noao.edu/gogreendr1/}).  Pointers to these catalogues and other information can currently be found on the GOGREEN web page at \url{http://gogreensurvey.ca/data-releases/data-packages/gogreen-and-gclass-first-data-release/}.

We provide two {\sc jupyter python3} notebooks with the data release.  The first, {\sc DR1\_Notebook}, provides examples for reading the data, displaying spectra and images, and reproducing many of the plots in this paper.  The second, {\sc build\_Table3}, is the notebook used to construct Table~\ref{tab-photo.fits} from the raw photometric and spectroscopic catalogues.

\subsection{Cluster Catalogue}\label{sec-cluscat}
We provide a {\sc fits} table {\sc clusters.fits} with information about each cluster in the GOGREEN and GCLASS samples.    Column names and descriptions are given in Table~\ref{tab-clusters.fits}.  This includes position, redshift and velocity dispersion measurements, as well as filenames for the corresponding images and photometric catalogues.
\begin{table*}
    \centering
    \begin{tabular}{|p{1cm}|p{2cm}|p{12cm}}
    \hline
        column & parameter name & description  \\
        \hline
        \hline
         1&cluster & Short name of each cluster.\\
         2&fullname & Longer format cluster name\\
         3&cluster\_id&An integer which is used to identify the corresponding photometry.  It is a unique number for each SpARCS and SPT cluster; it is 14 for all COSMOS clusters and 13 for those in the SXDF.\\  
         4-5& RA\_Best, DEC\_Best&Coordinates, in J2000 degrees, for the best estimate of the cluster centre.  For the SPT and SpARCS clusters, this is the location of the BCG.  For the COSMOS and SXDF clusters, it is the average position of members as described in \S\ref{sec-vdisp}.\\
         6-7&RA\_GMOS, DEC\_GMOS&Coordinates, in J2000 degrees, for the centre of the GMOS spectroscopic observations (GOGREEN only).\\
         8&PA\_GMOS&Position angle, in degrees, for the GMOS spectroscopic observations (GOGREEN only).\\
         9&Redshift&Best estimate of the cluster redshift, based on available spectroscopy, including publicly available spectra from other sources not included in this release.\\
         10-11&vdisp, vdisp\_err&Velocity dispersion and its uncertainty, in km/s, computed as described in \S\ref{sec-vdisp}.\\
         12-17&gogreen\_mN &Name of each GOGREEN GMOS mask, for N from 1 to 6, used to obtain spectra for this program. \\
         18-22&gclass\_mN &Name of each GCLASS GMOS mask, for N from 1 to 5, used to obtain spectra for this program. \\
         23&Kphot\_cat&Name of K-selected photometry catalogue\\
         24 & photoz\_cat & Name of photometric redshift catalogue\\
         25 & stelmass\_cat&Name of catalogue with stellar mass information\\
         26-37& IMAGE\_{X}&Name of image for filter X for SpARCS and SPT clusters.\\
         38& Preimage & Name of the GMOS z-band image, or Subaru pseudo-image, used for mask design.  Note the preimages were used for mask design but are not optimally reduced, specifically regarding sky subtraction and astrometry.  
    \end{tabular}
    \caption{A description of the contents of the {\sc clusters.fits} table, which contains information relevant to each cluster system in the GCLASS and GOGREEN surveys.}
    \label{tab-clusters.fits}
\end{table*}

\subsection{Images}
All GOGREEN images are resampled to $3000\times 3000$ pixels, projected onto the tangent plane with pixel scale $0.200$\arcsec/pix in the center ($\sim 10\arcmin$ on a side). Images in all filters are aligned in x and y position to aide in performing matched aperture photometry.  The five $z<1$ GCLASS clusters that are not part of GOGREEN are prepared in a similar manner, but are resampled to a pixel scale of $0.185$\arcsec/pix; furthermore the two northern clusters have a different size, of $5000\times 5000$ pixels.

For each cluster there are regular stacks, seeing-homogenised stacks (for aperture photometry), and (inverse-variance) weight maps.  The seeing-homogenized stacks are convolved with a kernel to give them Moffat-shaped PSFs with Beta-parameter of 3.0, with FWHM in arcseconds listed in the file psfsize\_target.dat.
There is a second set of psf-homogenized images *psf2\_*, for the $K_{s}$-band and the IRAC images (with sizes in the file psfsize\_target\_psf2.dat) that are used to construct homogeneous $K$-IRAC photometry as described in \citet{gogreen-smf}.

Conservative, manual mask images are provided.  These masks indicate pixels for which good data is available in all filters (excluding {\it HST}).  As this mask requires the presence of IRAC data, which covers a relatively small field of view compared to the optical/NIR data, these pixel masks are conservative.

\subsection{Photometry Catalogues}\label{sec-photcats}
Each SPT and SpARCS cluster in the GOGREEN and GCLASS sample has an associated set of catalogues based on the multiwavelength imaging; the exception is SpARCS1033, for which deep $K$-band imaging has not yet been obtained.  These catalogues are all {\sc ascii} format, and row-matched to the parent photometric catalogue. The detailed structure of these catalogues is described in the documentation distributed with the data release. 

The parent catalogue,  described in \citet{gogreen-smf}, is constructed from the original (unconvolved) $K_{s}$-band image as measured with \texttt{SExtractor}   \citep{sextractor}.
Each  detected object is required to have five adjacent pixels with at least 1.5 sigma significance.  All flux values have an AB magnitude zeropoint of 25 (equivalent to a flux scale of 0.3631 $\mu$Jy per count). Therefore m\_filter = $-2.5\times\mbox{log}_{10}(\mbox{flux\_filter}) + 25$.  An aggressive mask is applied, such that an object is unmasked (value {\sc totmask}$=0$) only if data exists in all available bands (and for all sources in the SPLASH and UltraVISTA catalogues). It is therefore very conservative and might not be appropriate for some analyses.  In particular, the IRAC data typically cover a limited area around the cluster; for science where IRAC coverage is not necessary, it can be appropriate to consider masked objects.

The COSMOS and SXDF photometry all comes from publicly available catalogues: \citet{Ultravista_Muzzin} and \citet{SPLASH}, respectively.  These are not described in detail here.  However, we provide {\sc jupyter notebooks} for reading these data in a consistent way with our own catalogues.

 For convenience, here we provide a single table {\sc photo.fits} with some of the most useful parameters gathered from these catalogues, for all objects with photometric measurements in all available filters.  
 The contents of this table are described in Table~\ref{tab-photo.fits}.  
 The descriptions about how each parameter is computed apply to the SpARCS and SPT systems.  For the COSMOS and SXDF, we use closely corresponding quantities from \citet{Ultravista_Muzzin} and \citet{SPLASH}, respectively.  For SXDF we calculate the rest-frame $UVJ$ colours ourselves, using the {\sc EAZY} code \citet{EAZY}, as these are not provided in the \citet{SPLASH} catalogue.

\begin{table*}
    \centering
    \begin{tabular}{|p{0.9cm}|p{2.8cm}|p{11.8cm}}
    \hline
        column & parameter name & description  \\
        \hline
        \hline
         1    &    Cluster    &    Name of the corresponding cluster, when there is an associated photometric catalogue.  Objects in the COSMOS or SXDF photometric catalogues are identified with those labels, unless there is a GOGREEN spectroscopic redshift, in which case we use the name of the associated target.  Note that SXDF49 and SXDF87 share a field, and are identified here only by SXDF49.  Similarly SXDF76a and SXDF76b are identified here as SXDF76.\\
         2    &    cPHOTID    &    This is a unique identifier for each object in this table.  The first digit identifies the source of the photometry (1: GOGREEN; 2: GCLASS; 3: UltraVISTA/COSMOS; 4:SPLASH/SXDF).  The next two digits are the cluster\_id column from Table~\ref{tab-clusters.fits}.  The remaining numbers are the PHOTID identifier in the main photometric catalogues.\\
         3    &    SPECID     &    The ID corresponding to Table~\ref{tab-zcat.fits} for objects with a corresponding GCLASS or GOGREEN spectrum.\\
         4,5  &    ra,dec     &    J2000 positions, in degrees.  Calibrated with SDSS DR7 or USNO-b whenever a cluster falls outside of the SDSS footprint.\\
         6,7  &    zspec,Redshift\_Quality    &    The spectroscopic redshift and quality flag for the associated spectrum, if any. Redshifts without a corresponding Redshift\_Quality are copied from the parent (UltraVISTA or SPLASH) catalogue. \\ 
         8,9,10    &    zphot,zphot\_l68,zphot\_u68      &    Photometric redshift, upper and lower uncertainties from the 68 per cent confidence region. Based on the {\it zpeak} output from {\sc EAZY} \citep{EAZY}, where for the GOGREEN galaxies a polynomial correction is applied to improve the correspondence with spectroscopy.  \\
         11,12 &    $U{-}V$,$V{-}J$     &    Rest-frame colours between Johnson $U$, $V$ and $J$, as measured with {\sc EAZY} \citep{EAZY}.  Small offsets, as described in \citet{gogreen-smf}, have been applied on a cluster-by-cluster basis to improve correspondence with UltraVISTA. For the COSMOS galaxies the rest-frame colours are from the UltraVISTA catalogue. \\
         13   &    Star       &    Star/galaxy classification based on colours, as described in \citet{gogreen-smf}.  Flag is 1 for a star, and 0 otherwise. \\
         14   &    K\_flag    &    \texttt{SExtractor} flag in the $K$-band.\\
         15   &    totmask    &    Manual mask at position of detection, where objects are masked (totmask$=1$) if they do not have an image in all available filters for that cluster.   Only sources with totmask=0 are included in this compilation catalogue.  Photometry for other sources must be obtained from the original catalogues. \\
         16   &    Mstellar    &    Total stellar masses, measured with the {\sc FAST} \citep{FAST} code and assuming the best redshift for the object (spectroscopic or {\it zphot}).  These assume $\tau-$model star formation histories, and are known to underestimate the stellar mass obtained with a non-parametric star formation history, by up to 0.3\,dex \citep{2019ApJ...876....3L}. For objects in COSMOS and SXDF the stellar masses are taken from their respective catalogues.  \\
        17-46   &    $X_i$\_tot    &     Total fluxes in each filter $X_i$.  These are derived from the Ks\_tot flux and the appropriate colour, computed in 2\arcsec\ diameter circular apertures from PSF-matched images. IRAC aperture fluxes have been measured in a two-step process, similar to the description in Appendix A of \citet{vdB+13}. The measurements within a 3\arcsec\ aperture are scaled by a factor determined by comparing the 2\arcsec\ aperture $K_s$ flux with that within a 3\arcsec\ aperture measured on an image convolved to match the IRAC point spread function.  This is done to avoid having to convolve all the high resolution ground-based data to the IRAC psf. For objects in COSMOS and SXDF the fluxes are taken from their respective catalogues, scaled by the corresponding Ks\_tot flux.
        Includes: $u, g, r, i, z, y, V, B, J, H, K_s, \mathrm{IRAC1}, \mathrm{IRAC2}, \mathrm{IRAC3}, \mathrm{IRAC4}, \mathrm{IA}484,\mathrm{IA}527,\mathrm{IA}624,\mathrm{IA}679,\mathrm{IA}738,$ $\mathrm{IA}767,\mathrm{IB}427,\mathrm{IB}464,\mathrm{IB}505,\mathrm{IB}574,\mathrm{IB}709,\mathrm{IB}827,{fuv},{nuv}$, and ${mips24}$. \\
        47-77    &     e$X_i$\_tot  &     Associated uncertainty estimates for filter $X_i$, assuming that the sole source of uncertainty is the background {\it rms}. It therefore depends on position on the stack (as the depth is not necessarily uniform), but does not depend on the source flux.\\        
        
    \end{tabular}
    \caption{A description of the contents of the {\sc photo.fits} table, which contains selected photometric data for each cluster system in the GCLASS, GOGREEN samples, COMOS UltraVISTA \citep[v1.4;][]{Ultravista_Muzzin} and SXDF SPLASH\citep[v1.6;][]{SPLASH} fields.}
    \label{tab-photo.fits}
\end{table*}

\subsection{Redshift Catalogue}\label{sec-zcat}
The redshift catalogue {\sc Redshift\_catalogue.fits}, described in Table~\ref{tab-zcat.fits}, includes an entry for every object with a GOGREEN or GCLASS spectrum\footnote{This excludes the stars deliberately observed on some masks for the purposes of telluric corrections.}.  There are no duplicates: if a galaxy has a spectrum in both surveys, only the GOGREEN entry is included here.  

\begin{table*}
    \centering
    \begin{tabular}{|p{1cm}|p{2cm}|p{12cm}}
    \hline
        column & parameter name & description  \\
        \hline
        \hline
         1&Cluster & Short name of each cluster; matches the entry in Table~\ref{tab-clusters.fits}\\
         2&SPECID & A unique identification number.  The first digit identifies the origin of the spectrum:  1 for GOGREEN and 2 for GCLASS.  The next two digits correspond to the {\bf cluster\_id} identifier in the Cluster catalogue, that specify the photometric field.   The remaining digits are the galaxy ID (only unique for a given field and source).\\
         3,4&RA(J2000), DEC(J2000)& Target coordinates, in J2000 degrees. For GOGREEN, these coordinates correspond to the $z^{\prime}$ image coordinates used for mask design.  These have been transformed to align with the $K_s$ images; however positions will not match exactly with coordinates in the photometric catalogues. \\ 
         5&OBJClass&This has a value of 1 for GOGREEN primary targets, i.e. those that match our photometric selection criteria. A value of 3 corresponds to a GOGREEN "mask filler" object, and 4 identifies a GCLASS spectrum.  (OBJClass=2 was reserved for stellar sources used for telluric correction, and these are not included in the catalogue).\\ 
         6&Redshift&The redshift measured from the spectrum\\ 
         7&Redshift\_Quality&The redshift quality flag.  Both quality 3 and 4 are secure galaxy redshifts and can be used for scientific analysis; the difference between them is subjective and not rigorously defined.  Quality 2 is a "best guess" but should be used with caution; this includes cases where there is plausible consistency with the photometric redshift, but no clearly identifiable spectral features. Quality 1 means no redshift is available.\\
         8&EXTVER&This is the science extension number in the {\sc fits} files with the 1D and 2D spectra (see \S~\ref{sec-1D2D}).\\
         9&Spec\_Flag&An integer used to identify spectra that have problems that might compromise the ability to measure a redshift or line indices of a spectrum.  Flags are assigned for the following:
    \begin{itemize}
        \item[1: ] Mild slit contamination or artefacts that should not strongly affect measurements
        \item[2:]  Non-galaxy-like spectrum and/or image
        \item[4:] Significant slit contamination from neighbouring objects.  Redshift and features may be compromised.
        \item[8:] Poor telluric correction or sky subtraction, due for example to inadequate correction for the stray light effect described in Appendix~\ref{sec-cdc}.
        \item[16:] Major artefacts or large masked regions that render the spectrum nearly useless.
    \end{itemize}
    Flags can be added.  So, for example, a flag of 12 means there is both contamination from neighbouring objects, and poor sky subtraction.  \\ 10&SNR\_8500\_VAR&The signal-to-noise ratio per pixel, measured in the range $7500<\lambda<9500$\AA.  The noise estimate is taken from the {\sc VAR} array associated with the spectrum. \\
    11&SNR\_8500\_RMS&The signal-to-noise ratio per pixel, measured in the range $7500<\lambda<9500$\AA.  The noise estimate is taken from the {\it rms} in the science spectrum over the same range.\\
    12,13&D4000, eD4000& The D$_{n}$4000 index as defined in \citet{PSG}, and its uncertainty.  See \S~\ref{sec-indices}\\
    14,15&EWOII, eEWOII& The equivalent width of the [O{\sc ii}] emission line and its uncertainty, in \AA, using the line index definitions in \citet{PSG}.  Positive values represent emission.  See \S~\ref{sec-indices}\\
    16,17&EWHdelta, eEWHdelta& The equivalent width  of the H$\delta$ absorption line and its uncertainty, in \AA, using the line index definitions in \citet{PSG}.  Positive values represent absorption.  See \S~\ref{sec-indices}\\
    18,19&EWOII\_model, eEWOII\_model&The equivalent width of the [O{\sc ii}] emission line and its uncertainty, in \AA, calculated from the Gaussian fitting model described in \citet{gogreen-sfr}.\\
    20,21&F\_OII,eF\_OII&The integrated flux of the [O{\sc ii}] emission line and its uncertainty, in ergs/s/cm$^2$/\AA,  calculated from the Gaussian fitting model described in \citet{gogreen-sfr}. \\
    22,23&SFR,eSFR&The star formation rate in solar masses per year, estimated from the [O{\sc ii}] emission line flux and the stellar mass, using the calibration of \citet{2010MNRAS.405.2594G}.  \\
    24&delta\_BIC& The difference in Bayesian Information Criterion used to identify the presence of [O{\sc ii}] emission ($\Delta{\rm BIC}>10$) or its absence ($\Delta{\rm BIC}<-10$).  See \citet{gogreen-sfr} for more details.\\
    25&member\_Clean&Applicable only to the 11 SPT and SpARCS clusters in GOGREEN, this indicates likely cluster membership based on the {\sc clean} algorithm of \citet{CLEAN}.  A value of 1 indicates a member, 0 is a non-member, and -1 indicates membership could not be determined.\\
    26&member\_EM&Applicable only to the 11 SPT and SpARCS clusters in GOGREEN, this indicates likely cluster membership based on the {\sc C.L.U.M.P.S.} algorithm of Munari et al. (in prep).  A value of 1 indicates a member, 0 is a non-member, and -1 indicates membership could not be determined.\\
    27&member&A flag that identifies likely cluster members (1) or nonmembers (0).  A value of $-1$ means membership could not be determined.  For SpARCS and SPT clusters in GOGREEN, this is the maximum of the member\_Clean and member\_EM flags.  For the five GCLASS clusters we use the membership given in \citet{GCLASS12}.  Finally, for the systems in COSMOS and SXDF we define members as those within 1~Mpc and 2.5$\sigma$ of the centre, as described in \S~\ref{sec-vdisp}. \\

     \end{tabular}
    \caption{A description of the contents of the spectroscopic redshift catalogue {\sc Redshift\_catalogue.fits}, which contains and entry for every unique object with a GOGREEN or GCLASS spectrum.}
    \label{tab-zcat.fits}
 \end{table*}

\subsection{1D and 2D reduced spectra}\label{sec-1D2D}
Each cluster has an associated multi-extension {\sc fits} file containing the final 1D spectra, and another with the 2D spectra, for every target.  Each object has up to three entries, labeled {\sc sci} for the science spectrum, {\sc var} for its estimated variance, and, for GOGREEN spectra only, {\sc dq} for a data quality flag.  The science spectrum for object $i$, for example, is accessed with the extension [SCI,i].  All spectra are 974 pixels long, with a linear dispersion of 3.906\,\AA\ per pixel, and starting at 6398.44\,\AA.  This information is provided in the header of each extension.  The header also includes keywords 'ORIG1', 'ORIG2' etc. containing the mask names and extensions of the spectra that were combined to make this final stack.  The associated {\sc var} array contains a first-principles estimate of the uncertainty on each pixel, as propagated through the data reduction pipeline.  This includes a systematic term: during extraction, if the positive and negative version of the same wavelength pixel in the spectrum differ by more than five times the square root of the estimated variance, that difference is added in quadrature to the variance.  The data quality ({\sc dq}) array (only present for GOGREEN spectra) is likely of limited use; it corresponds to the number of pixels that were combined (along a CCD column) during the 1D extraction.  Pixels and wavelength ranges that are flagged as bad\footnote{These wavelength ranges are also identified in the header as {\sc userbad}.} during the reduction or extraction process are excluded.  Thus, large numbers ($>10)$ represent good regions of the spectrum where most of the pixels in the slit were used;  lower values mean only part of the slit was used.  The latter happens most frequently when there is a neighbouring object contaminating the slit in one of the nod positions.  

The GCLASS spectra were extracted differently from GOGREEN, and do not have a comparable {\sc var} array.  We construct a similar array from the science and sky spectra provided and scale it so that, on average, the relation between {\sc var} and the {\it rms} of the science spectrum is the same as for GOGREEN.  

The 1D {\sc fits} file also contains an extension, labeled [MDF], which contains a binary table.  This table is derived from the mask definition file (hence the extension name), and includes the position and magnitude of the spectrum corresponding to each extension.  These and the other columns in this table are described in \ref{tab-1dmdf}.

\begin{table*}
    \centering
    \begin{tabular}{|p{1cm}|p{2cm}|p{12cm}}
    \hline
        column & parameter name & description  \\
        \hline
        \hline
        1&ID& Corresponds to {\sc SPECID} in Table~\ref{tab-zcat.fits}\\
         2&EXTVER & The extension number of each spectrum within this file. \\
         3,4&RA, DEC&J2000 coordinates, in degrees, as in Table~\ref{tab-zcat.fits}.\\
         5&MAG&Total magnitude in the $z^{\prime}$-band, from the preimages, used for target selection.  This should generally not be used; magnitudes in the $K$-selected photometric catalogues are better.\\
         6&priority&This is the same as OBJClass in Table~\ref{tab-zcat.fits}\\
         7&qop&This is the same as Redshift\_Quality in Table~\ref{tab-zcat.fits}\\
         8,9&SNR\_8500\_VAR,  SNR\_8500\_RMS & These are the same as the entries in Table~\ref{tab-zcat.fits}\\
         10&Etime&Total exposure time, in seconds.  \\
    \end{tabular}
    \caption{A description of the contents of the MDF table extension associated with each 1D spectrum. }
    \label{tab-1dmdf}
\end{table*}

The 2D spectra are provided in another MEF {\sc fits} file.  The format is similar to the 1D file, but only GOGREEN spectra are included, so there are fewer extensions.  There is no corresponding binary table extension in this file.  Every spectrum is 974 pixels long and either 19 or 21 pixels wide.  The dispersion axis is the same as for the 1D file; the spatial axis is in pixel units.  No relative or absolute flux calibration is applied to these spectra.

\section{Conclusions}
We have presented the first public Data Release of the GCLASS \citep{GCLASS12} and GOGREEN \citep{gogreen-survey} galaxy cluster surveys.  The main characteristics of the data sample are:
\begin{itemize}
    \item 26 clusters spanning the redshift range $0.8<z<1.5$, with dynamical halo masses spanning a wide range from the "massive group" scale of $\sim 10^{13}~{\rm  M}_\odot$ to the most massive clusters at these redshifts, $\sim 10^{15}~{\rm M}_\odot$.
    \item Homogeneous, $K_s$-selected photometric catalogues are available for all systems except SpARCS1033; these include PSF-matched photometry covering the full optical and near-infrared spectrum, at depths comparable to those of the UltraVISTA \citep{Ultravista_Muzzin} survey.
    \item Modest resolution ($R=440$) Gemini GMOS spectroscopy has been obtained covering the full virialized region of each cluster.  The redshift catalogue is $\gtrsim 30$ per cent complete and unbiased relative to spectral type  for stellar masses $M_\ast\gtrsim10^{10.2}~{\rm M}_\odot$, within $\sim 500$ kpc of the centre.
    \item The spectroscopic catalogue includes 2257 galaxies with good redshifts (Redshift\_Quality$>2$), including nearly 800 cluster members.
    \item Reduced \textit{HST}/WFC3 imaging is provided in the F160W (GOGREEN) and F140W (GCLASS) filters; the five clusters that overlap these surveys have images in both filters.  We also provide reduced archival ACS/F606W and ACS/F814W images for the three SPT clusters in GOGREEN.
    
\end{itemize}
This data release includes all reduced images and spectroscopy.  It represents hundreds of hours of 8-m class and space-based observations, dedicated to a comprehensive and homogeneous survey of dense galaxy systems that are rare in blank field surveys at this depth.  Catalogues of advanced data products including photometric and spectroscopic redshifts, rest-frame colours, stellar masses, spectral line indices and cluster membership are also provided.  The tabular data for most commonly used quantities are included in the tables within this paper.  
Future releases, science results and other updates will be announced via the GOGREEN website at  \url{http://gogreensurvey.ca/}.

\section*{Data Availability}
Access to the GOGREEN and GCLASS data release, including {\sc jupyter python3} notebooks for reading and using the data, is available at the CADC (\url{https://www.cadc-ccda.hia-iha.nrc-cnrc.gc.ca/en/community/gogreen}), and NSF's NOIRLab \url{(https://datalab.noao.edu/gogreendr1/}).  Future releases, science results and other updates will be announced via the GOGREEN website at  \url{http://gogreensurvey.ca/}.

\section*{Acknowledgments}
This work was enabled by observations made from the Gemini North, Subaru and CFHT telescopes, located within the Maunakea Science Reserve and adjacent to the summit of Maunakea. We are grateful for the privilege of observing the Universe from a place that is unique in both its astronomical quality and its cultural significance. 

We extend our thanks and appreciation to Gemini, and the Gemini communities, for supporting this Large Program.  The extension of the program from the original allocation of 3 years and 438 hours to the final, 5 years and 530 hours needed to complete the survey was critical for achieving our science goals.  Queue mode observations, and a small amount of Director's Discretionary time, was used to make up for time lost during Priority Visitor runs, ensuring the survey finished at nearly 100 per cent completeness.
We also thank Gemini for their financial support of junior observers on several of our observing runs.  
\par
The authors would like to acknowledge research grant funding that has enabled this research, including:  NSERC Discovery grants (MLB and AM) and CGS-D (KW);  National Science Foundation grants AST-1517815, AST-1517863 (GW), AST-1716690, and AST-1814159 (GHR), AST-1518257 and AST-1815475 (MC); NASA HST program grants GO-15294 (GW), AR-14289 (MC) and AR-14310 (GHR); by grant numbers 80NSSC17K0019 (GW) and 80NSSC19K0592 (GR) issued through the NASA Astrophysics Data Analysis Program (ADAP), PRIN MIUR 2017 n.20173ML3WW\_001 (PI Cimatti) and the INAF main-stream funding programme (BV); the Chilean Centro de Excelencia en Astrof\'isica y Tecnolog\'ias Afines (CATA) BASAL grant AFB-170002 (RD).
G.R. also acknowledges the support of an ESO visiting science fellowship.
 
This paper includes data gathered with the Gemini Observatory, which is operated by the Association of Universities for Research in Astronomy, Inc., under a cooperative agreement with the NSF on behalf of the Gemini partnership: the National Science Foundation (United States), the National Research Council (Canada), CONICYT (Chile), Ministerio de Ciencia, Tecnología e Innovación Productiva (Argentina), and Ministério da Ciência, Tecnologia e Inovação (Brazil); the 6.5 meter Magellan Telescopes located at Las Campanas Observatory, Chile, under programme IDs CN2015A-31, CN2016B-44, and CN2018B-14; the Canada-France-Hawaii Telescope (CFHT) which is operated by the National Research Council of Canada, the Institut National des Sciences de l'Univers of the Centre National de la Recherche Scientifique of France, and the University of Hawaii; MegaPrime/MegaCam, a joint project of CFHT and CEA/DAPNIA; Subaru Telescope, which is operated by the National Astronomical Observatory of Japan; the ESO Telescopes at the La Silla Paranal Observatory under programme IDs 097.A-0734 and 179.A-2005.  We make use of data products produced by CALET and the Cambridge Astronomy Survey Unit on behalf of the UltraVISTA consortium. Finally, this research has made use of the SVO Filter Profile Service (http://svo2.cab.inta-csic.es/theory/fps/) supported by the Spanish MINECO through grant AYA2017-84089. 

\par
\noindent
\section*{Affiliations}
$^{1}$Department of Physics and Astronomy, University of Waterloo, Waterloo, Ontario, N2L 3G1, Canada\\
$^{2}$Waterloo Centre for Astrophysics, University of Waterloo, Waterloo, Ontario, N2L 3G1, Canada\\
$^{3}$European Southern Observatory, Karl-Schwarzschild-Str. 2, 85748, Garching, Germany\\
$^{4}$Department of Physics and Astronomy, York University, 4700 Keele Street, Toronto, Ontario, ON MJ3 1P3, Canada\\
$^{5}$Department of Physics and Astronomy, The University of Kansas, Malott room 1082, 1251 Wescoe Hall Drive, Lawrence, KS 66045, USA\\
$^{6}$Department of Physics and Astronomy, University of California Riverside, 900 University Avenue, Riverside, CA 92521, USA\\ 
$^{7}$INAF-Osservatorio Astronomico di Trieste, via G. B. Tiepolo 11, 34143, Trieste, Italy\\
$^{8}$IFPU - Institute for Fundamental Physics of the Universe, via Beirut 2, 34014 Trieste, Italy\\
$^{9}$Departamento de Ingenier\'ia Inform\'atica y Ciencias de la Computaci\'on, Universidad de Concepci\'on, Concepci\'on, Chile\\
$^{10}$Department of Physics and Astronomy, University of California, Irvine, 4129 Frederick Reines Hall, Irvine, CA 92697, USA\\
$^{11}$South African Astronomical Observatory, PO Box 9, Observatory, 7935, South Africa\\
$^{12}$Centre for Space Research, North-West University, Potchefstroom 2520, South Africa\\
$^{13}$Canadian Astronomy Data Centre, NRC Herzberg, 5071 West Saanich Road, Victoria, BC, V9E 2E7, Canada\\
$^{14}$Research School of Astronomy and Astrophysics, The Australian National University, ACT 2601, Australia\\
$^{15}$ Centre for Gravitational Astrophysics, College of Science, The Australian National University, ACT 2601, Australia\\
$^{16}$Department of Physics and Astronomy, Texas A\&M University, College Station, TX, 77843-4242, USA\\
$^{17}$George P. and Cynthia Woods Mitchell Institute for Fundamental Physics and Astronomy, Texas A\&M University, College Station, TX, 77843-4242, USA\\
$^{18}$School of Physics and Astronomy, University of Birmingham, Edgbaston, Birmingham B15 2TT, England\\
$^{19}$European Space Agency (ESA), European Space Astronomy Centre, Villanueva de la Ca\~{n}ada, E-28691 Madrid, Spain\\
$^{20}$Department of Astronomy and Astrophysics, University of Toronto 50 St. George Street, Toronto, Ontario, M5S 3H4, Canada\\
$^{21}$Department of Physics, McGill University, 3600 rue University, Montr\'{e}al, Qu\'{e}bec, H3P 1T3, Canada\\
$^{22}$INAF - Osservatorio astronomico di Padova, Vicolo Osservatorio 5, IT-35122 Padova, Italy\\
$^{23}$Observatorio Astron\'omico de C\'ordoba (UNC) and Instituto de Astronom\'ia Te\'orica y Experimetal (CONICET-UNC). C\'ordoba, Argentina.\\
$^{24}$University of California at Santa Cruz, 1156 High Street, Santa Cruz, CA 95064, USA\\
$^{25}$Department of Astronomy, University of Washington, Box 351580, U.W., Seattle, WA 98195, USA\\
$^{26}$Department of Physics, University of Helsinki, Gustaf H\"allstr\"omin katu 2a, FI-00014 Helsinki, Finland\\ 
$^{27}$Laboratoire d'astrophysique, \'Ecole Polytechnique F\'ed\'erale de Lausanne (EPFL), 1290 Sauverny, Switzerland \\ 
$^{28}$Departamento de Ciencias F\'isicas, Universidad Andres Bello, Fernandez Concha 700, Las Condes 7591538, Santiago, Regi\'on Metropolitana, Chile\\
$^{29}$School of Earth and Space Exploration, Arizona State University, Tempe, AZ, 85287, USA\\
$^{30}$Department of Physics and Astronomy, McMaster University, Hamilton ON L8S 4M1, Canada\\
$^{31}$Steward Observatory and Department of Astronomy, University of Arizona, Tucson, AZ, 85721\\

\appendix
\section{Photometric calibration}\label{app-photcal}
\begin{figure*}{}
\includegraphics[clip=true,trim=0mm 0mm 0mm 0mm,width=3.4in,angle=0]{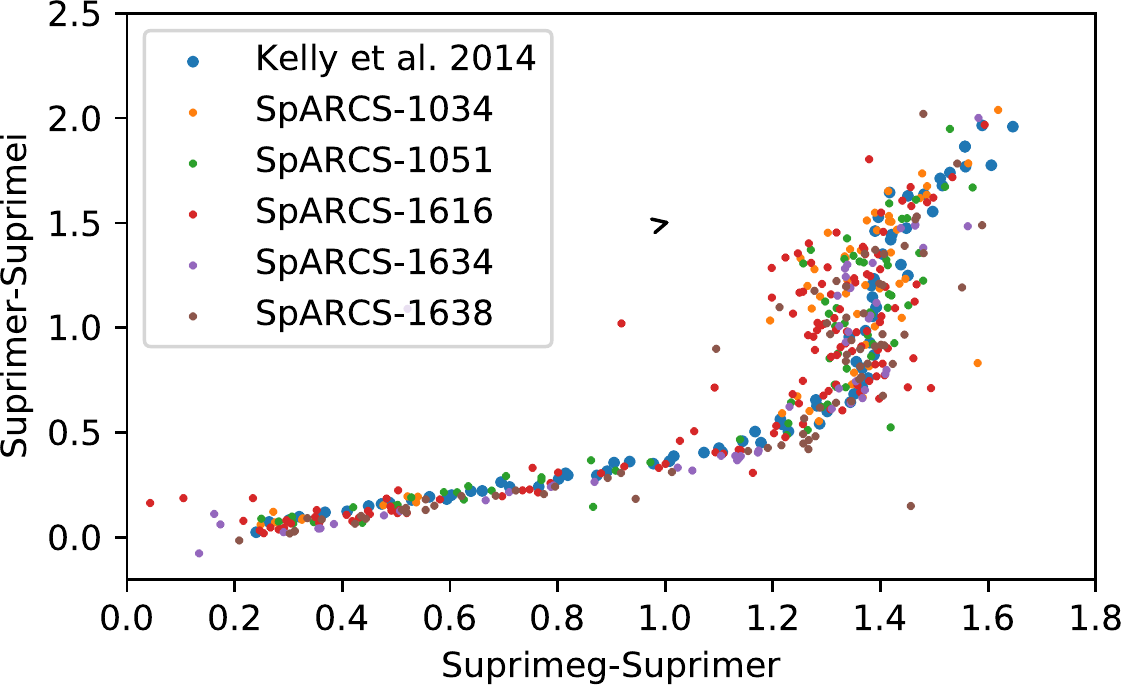}
\includegraphics[clip=true,trim=0mm 0mm 0mm 0mm,width=3.4in,angle=0]{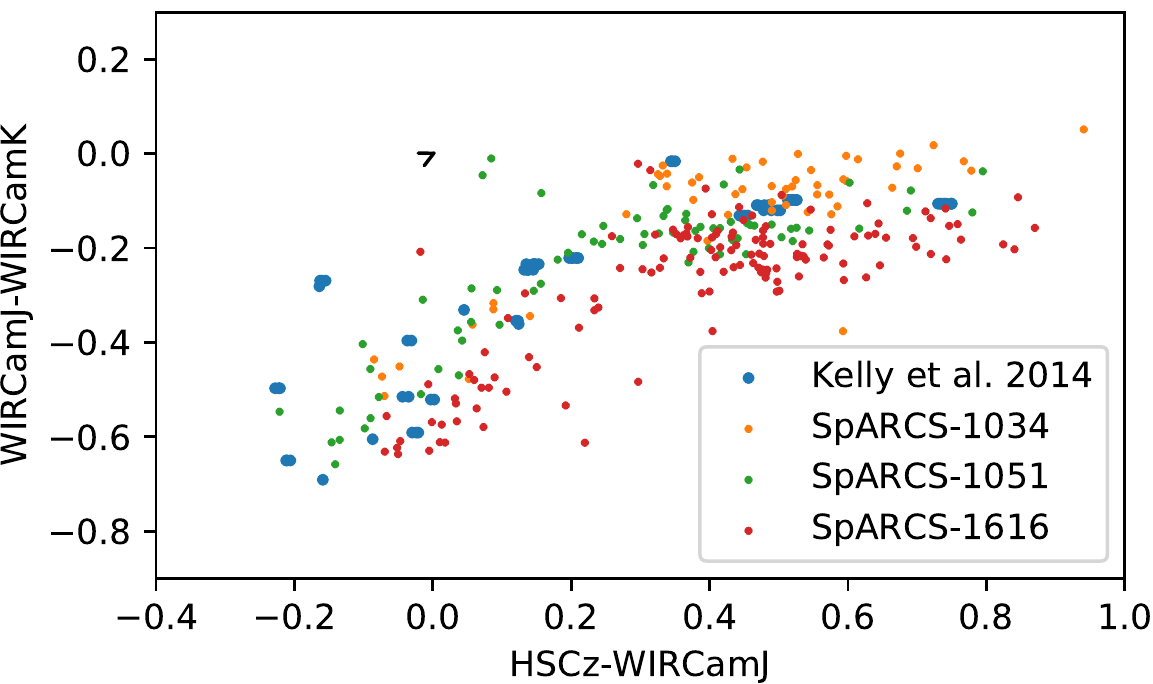}
\includegraphics[clip=true,trim=0mm 0mm 0mm 0mm,width=3.4in,angle=0]{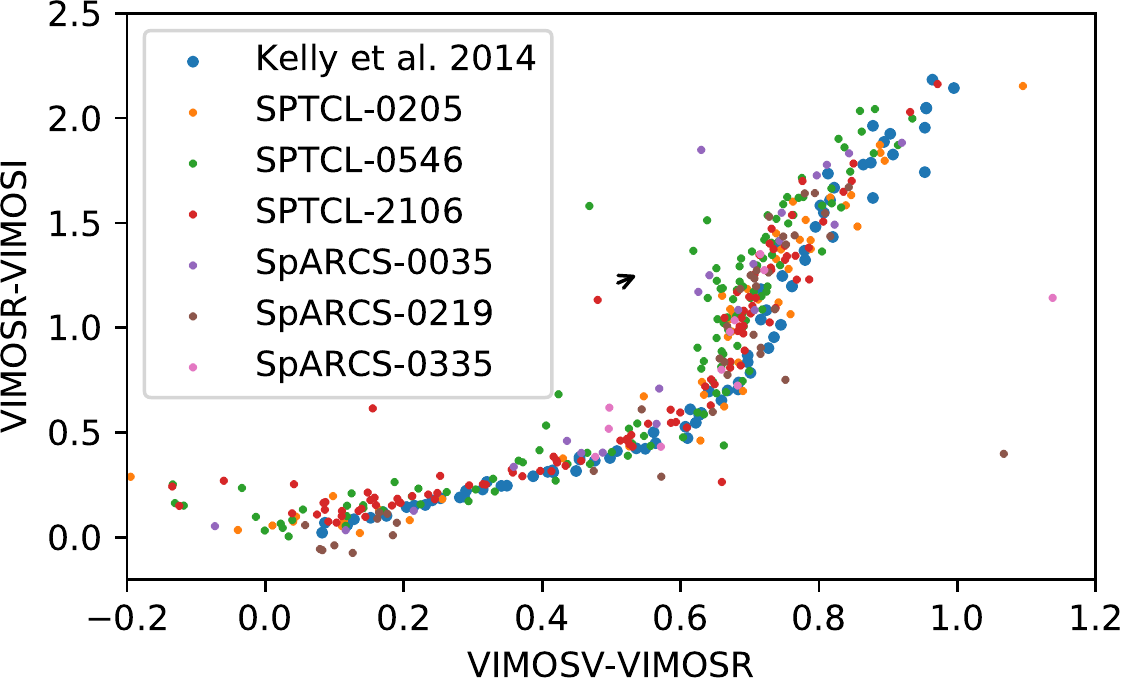}
\includegraphics[clip=true,trim=0mm 0mm 0mm 0mm,width=3.4in,angle=0]{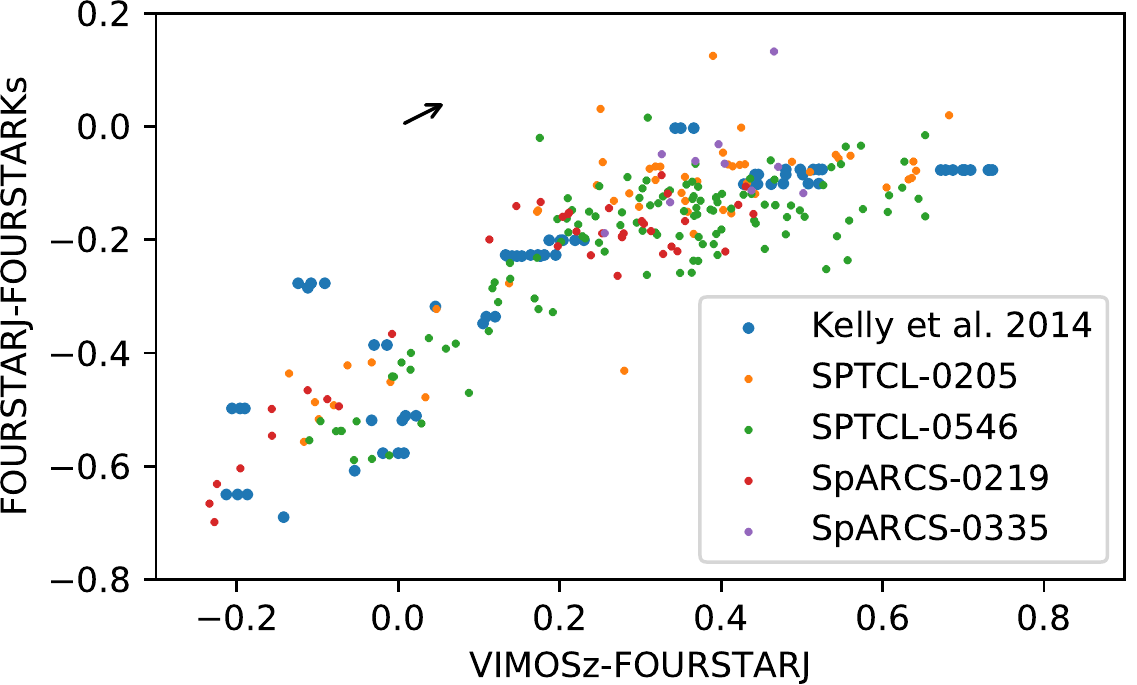}
	\caption{Different representations of the stellar locus that was used for calibration of the multi-band photometry. \textit{Thicker blue symbols:} stellar spectral library from \citet{Kelly2014}, convolved with the response function for different combinations of telescope optics/filter/detector/atmosphere. \textit{Top panels:} filters used for some of the Northern clusters. \textit{Bottom panels:} same for the Southern clusters. \textit{Left panels:} Stellar locus in optical/visible filters. \textit{Right panels:} Stellar locus in Near-IR. Different colours show calibrated photometry for the different clusters, after correcting for (the small amount of) galactic dust absorption \citep[][]{schlegel98dust}. Arrows show the maximum dust extinction that was corrected for within the cluster sample. \label{fig-stellarloci}}
\end{figure*}

As described in Sect.~\ref{sec-groundbasedimaging}, the multi-band photometry is calibrated with respect to the universal stellar locus. In Fig.~\ref{fig-stellarloci}, we show the stellar locus for several different filter combinations.  We note that the stars from the different clusters that are shown here, are selected based on their $u-J$ (or $g-J$) and $J-K_{\mathrm{s}}$ colours \citep[as in][]{gogreen-smf}. In contrast, the photometric calibration was originally done on stellar candidates selected by their magnitude and size (unresolved objects) as measured with \texttt{SExtractor}.

The reference stellar spectral library from \citet{Kelly2014} is primarily based on SDSS spectroscopy, which covers the wavelength range 4100\AA-9000\AA. It is extended both in the red and blue by matching with similar objects in the \citet{Pickles98} library, where similarity is judged based on the region where there is wavelength overlap between both samples. Some sources in the \citet{Pickles98} library are matched to multiple SDSS sources, which leads to a ``step-like'' pattern in Fig.~\ref{fig-stellarloci} in the Near-IR.

As one test to look for significant photometric calibration systematics in one particular filter,
we look for residuals relative to the best-fit EAZY templates used to calculate photometric redshifts \S~\ref{sec-photoz}.   Consistently large residuals in one band could indicate a calibration error.  

Specifically, we calculate $\sigma$=$\frac{f_{observed}-f_{model}}{err_{observed}}$ for each filter used in the EAZY photometric fitting for each galaxy with a spectroscopic redshift.  Then in each filter/cluster combination, we find the median residual across the entire redshift range spanned by the cluster's spectroscopic catalogue.  This includes both cluster members and field galaxies observed near the cluster, in an attempt to find any systematic effect as a function of redshift.  We list our residuals in Tables \ref{tab-eazyresidualsnorth} \& \ref{tab-eazyresidualssouth} for northern and southern cluster respectively.  We find that fits of all filter and cluster combinations are within 3$\sigma$ of the observed data, with the highest residuals residing at the blue end of the SED fits. It is not trivial to identify the underlying cause of these residuals, which in any case are small.  As one, illustrative test, we systematically increase the Suprime $g$-band input magnitudes by the median residual measured for the northern sample, and rerun EAZY.  No significant change is observed in the resulting residuals, relative to the new fits.  While this rules out a simple zeropoint shift in this one filter as the cause of the systematic, the actual cause remains unknown and may still be related to the photometric calibration.


\begin{table*}
\begin{tabular}{l|rrrrr}
         & SpARCS1034 & SpARCS1051 & SpARCS1616 & SpARCS1634 & SpARCS1638 \\
         \hline 
MegaCamu &            & 0.525    & 0.503    & 1.056    & 0.813    \\
\hline
Suprimeg & 1.400    & -1.373   & -1.892   & -1.805   & -2.327   \\
Suprimer & -0.824   & 0.783    & 0.926    & 0.407    & 0.632    \\
Suprimei & 0.739    & 0.871    & 0.923    & 1.094    & 1.563    \\
Suprimey &            &            &            &            & -1.163   \\
\hline
HSCz     & 0.443    & -0.086   & -0.047   &            &            \\
HSCy     & 0.384     & 0.299    & -0.565   &            &            \\
\hline
GMOSz    &            &            &            & 0.196    & 0.409    \\
\hline
WIRCamJ  & -0.861   & 0.007    & -0.46509   & -1.007   & -0.424   \\
WIRCamK & -0.376   & -0.608    & 0.63699    & 0.201     & -0.032     \\
\hline
IRAC1    & -0.970   & -0.413    & -0.58884   & -0.797   & -0.328   \\
IRAC2    & -0.511   & -0.331   & -0.26533   & -0.364   & -0.307   \\
IRAC3    & -0.054   & -0.013   & -0.04943   & -0.063     & -0.100   \\
IRAC4    & 0.233    & -0.025   & 0.09243    & 0.091    & 0.166  \\
\hline 
\end{tabular}
\caption{Residual ($\sigma$ = $\frac{f_{observed}-f_{model}}{err_{observed}}$) between input photometry of a given band and the modelled band flux using the EAZY best-fit template for northern clusters. Negative values indicate an overestimation during fitting and positive values indicate an underestimation.   \label{tab-eazyresidualsnorth}}
\end{table*}

\begin{table*}
\begin{tabular}{lrrrrrr}
\multicolumn{1}{l|}{}         & SPTCL0205 & SPTCL0546 & SPTCL2106 & SpARCS0035 & SpARCS0219 & SpARCS0335 \\
\hline
VIMOSU     & 1.702   & 2.046   & 1.451   & 2.326    & 1.082    & 2.043    \\
VIMOSB     & -0.796  & -0.100  & -0.413  & -0.681   & -0.814   & -0.507   \\
VIMOSV     & 0.121   & -0.368  & -0.028  & 0.157     & 0.051    & 0.32     \\
VIMOSR     & 0.229   & -0.062  & 0.075    & 0.217    & -0.007   & -0.294   \\
VIMOSI     & 0.275   & 0.061   & -0.447  & 0.197    & 0.762    & 0.078     \\
VIMOSz     & 0.672   & 0.517   & 0.285   &            & 1.095    & 0.805    \\
\hline
DECamz     &           &           &           & 0.093    &            &            \\
\hline
HAWKIJ     &           &           &           & 0.243    &            &            \\
HAWKIKs    &           &           & 0.284   & -1.216   &            &            \\
HAWKIY     &           &           &           &            &            & 0.047    \\
\hline
FOURSTARJ1 & -0.372  & -0.074  & -0.089  & 0.060    & -0.103   &            \\
FOURSTARJ  & -0.836  & -0.515  & 0.004   &            & -1.273   & -1.218   \\
FOURSTARKs & -0.350   & -0.003  &           &            & -0.863   & 0.808    \\
\hline
IRAC1      & -0.881  & -1.098  & -0.754  & -1.122   & -0.476   & -1.062    \\
IRAC2      & -0.412  & -0.214  & -1.286  & -0.063   & -0.319   & -0.062   \\
IRAC3      &           &           &           & -0.047   & 0.033    & -0.397   \\
IRAC4      &           &           &           & 0.647    & 0.576    & 0.469    \\
\hline
\end{tabular}
\caption{Residual ($\sigma$ = $\frac{f_{observed}-f_{model}}{err_{observed}}$) between input photometry of a given band and the modelled band flux using the EAZY best-fit template for southern clusters. Negative values indicate an overestimation during fitting and positive values indicate an underestimation. \label{tab-eazyresidualssouth}}
\end{table*}

\section{Correction for spectral contamination due to scattered light from nearby slits}\label{sec-cdc}
\begin{figure}{
		\includegraphics[clip=true,trim=0mm 0mm 0mm 0mm,width=3.5in,angle=0]{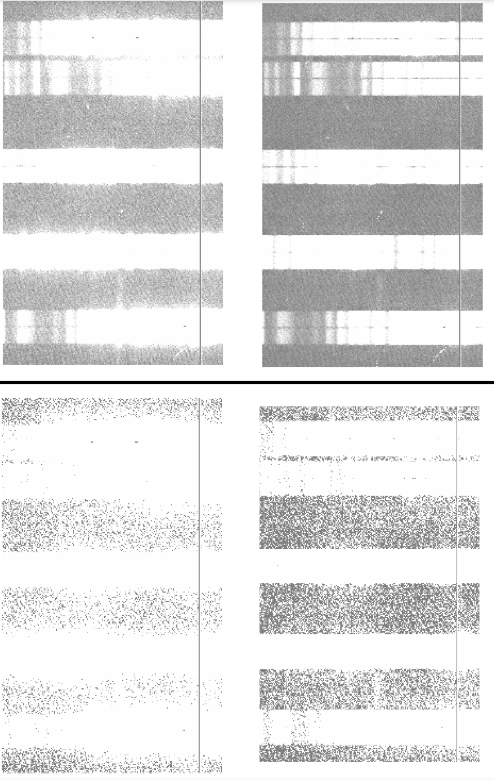}}
	\caption{Left:  An example region of a GMOS detector is shown in the top panel, and again in the bottom panel at higher stretch.  Five slits are visible; each has two adjacent spectra due to the nod-and-shuffle acquisition.  Light is observed to extend well outside the slit edges, particularly at the wavelengths of bright sky lines.  Right: the same regions are shown after applying our empirical correction (see text for details). \label{fig-CSex}}
\end{figure}

The standard data reduction steps resulted in spectra that were unusable beyond about 9000\AA, as the nod-and-shuffle background subtraction was failing to correctly remove the sky.  This was traced to a wavelength-dependent effect, where the amount of light extending beyond the slit edge increases with increasing wavelength.  This is catastrophic when slits are close together, as is the case for pairs of spectra taken in microshuffle mode, where the spectra from the two nod positions are adjacent to one another on the detector. An example is given in the left panels of Figure~\ref{fig-CSex}, which shows a region on the detector with five slits, at high stretch.  Light extends far beyond the slit edge, particularly noticeable at the location of bright sky lines.  This in particular blurs the distinction between the nod-and-shuffle pairs.

The same effect was noted by \citet{GDDS}, and attributed to charge diffusion in the detector.  However, we find the effect is quantitatively similar for  detectors used through the course of this project that have very different sized depletion zones, which makes this explanation unlikely. We considered scattering from the grating \citep[e.g.][]{Woods:94,2014SPIE.9151E..1VE}, but a quantitatively similar effect is seen in the undispersed through-mask images.   The origin remains unknown, therefore, but seems to be an optical effect independent of the grating and detector.

In \citet{gogreen-survey} we described a procedure for empirically correcting the effect.  However, this method was only applicable to nod-and-shuffle pairs of spectra that are of a fixed separation and aligned in wavelength.   Because of our high slit density, there are many other examples where light from a nearby slit contaminates another; an example is the two slits at the top of Figure~\ref{fig-CSex}.  As the amount of contamination is sensitive to wavelength and distance from the slit edge, this is more difficult to correct for.

Our approach is to build an empirical model that describes the amount of flux that "leaks" outside a slit, relative to the average intensity within the slit, as a function of wavelength, distance from the slit edge, and intensity.  This is made possible because we have so many different masks, obtained with nearly identical setup (same grism, slit length and nod parameters).  We are able to select a subset of slits which, over some wavelength range, are well separated from any other slit on the mask.  We then simply measure the amount of signal in the region outside the slit.  By combining results from many different slits, at different locations, this can be mapped as a function of wavelength and detector position.  

This measurement of "slit leakage" (for lack of a better term or understanding of its origin)
must be made on images that have been corrected for bias and large-scale scattered light, but before any sky subtraction or other processing has been done.  It also requires a wavelength solution for each slit.   Unfortunately, GMOS suffers from substantial scattered light effects\footnote{There are some good examples on the GMOS Data Reduction webpages, at \url{http://www.gemini.edu/instrumentation/gmos/data-reduction}}.  These are normally not a concern, because they vary over scales that are large compared with a slit length, so they are removed during sky subtraction.  It does, however, compromise our ability to measure the local "slit leakage" component of scattered light,
because we are measuring small flux residuals and combining measurements from spectra located at different locations in the focal plane.  We therefore have to first attempt to model and subtract off the large-scale component of scattered light.  Again this is made possible by the large volume of data.  We identify areas on a given mask that are far from any slit (farther than expected to be contaminated by the slit leakage effect), and assemble those to build a low resolution image of the scattered light.  This will not include any contribution from bright objects on the mask (usually alignment stars), but does capture the dominant large-scale component that appears to be fairly stable over time.  We then fit a polynomial surface to this, and subtract it from each image.

All the steps above were done separately for each amplifier on each different detector (three different detectors were used over the course of the survey).  This was done because we suspected the origin was due to charge diffusion.  This turned out not to be the case, and in fact the residual component is very similar for all three detectors considered, and independent of location on the chip.  

\begin{figure}{
		\includegraphics[clip=true,trim=0mm 0mm 0mm 0mm,width=3.5in,angle=0]{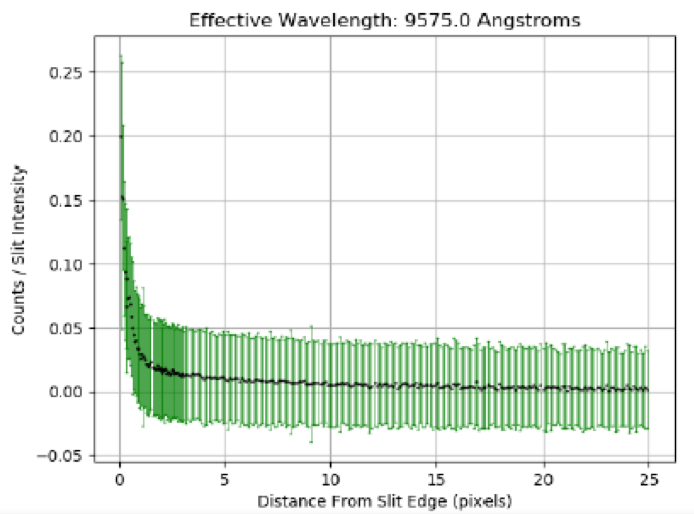}}
	\caption{The average number of counts per $0.081$\,\AA\ pixel, relative to the average slit intensity, as a function of distance away from the edge of a slit.  This is shown at fixed wavelength, of $9575$\AA.  The green vertical bars show the variance among the measured slits; they are not the error on the mean, and they are not independent. \label{fig-chargespread}}
\end{figure}

\begin{figure}{
		\includegraphics[clip=true,trim=0mm 0mm 0mm 0mm,width=3.5in,angle=0]{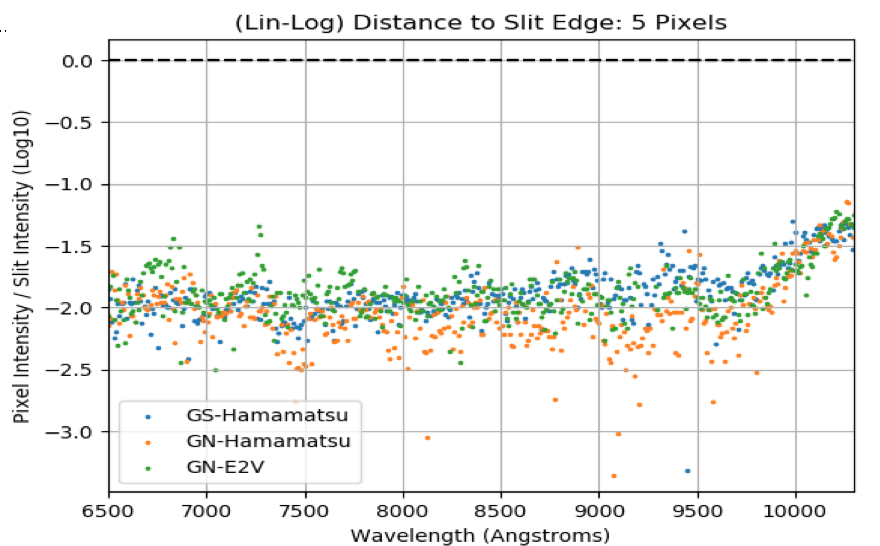}}
	\caption{Similarly to Figure~\ref{fig-chargespread}, but as a function of wavelength for a fixed distance of five pixels from the slit edge.  Different coloured points correspond to different detectors, as indicated in the legend; there is no evidence that the effect varies significantly between detectors.  While it is most problematic at the longest wavelengths, $>9600$\AA, there is at least a 1 per cent effect at all wavelengths. \label{fig-CSwave}}
\end{figure}

The measured average flux residual at $\lambda=9570$\AA\ is shown in Figure~\ref{fig-chargespread}, as a function of distance from the edge of the slit. It is non-negligible, at the level of $\sim 1$ per cent, even beyond 20 pixels (1.6 arcseconds).  At wavelengths corresponding to bright sky lines this corresponds to an amount that can be significantly larger than our source intensity.  The wavelength dependence is shown in Figure~\ref{fig-CSwave}, at a fixed distance of five pixels from the slit edge.  While the effect is most problematic at $\lambda>9600$\AA, there is a wavelength-independent, 1 per cent effect at all shorter wavelengths, as well.  A parametric model is fit to these data.  There is also a dependence on the average slit intensity $I$, such that the relative residual intensity it is actually stronger at lower intensities, roughly proportional to $I^{-1/3}$.    This is also included in our parametric fit.


We do not construct a similar model of light outside of alignment star boxes.  This requires some additional work and given that only a small subset of our data will be affected, we did not do it.  Another deficiency is that this model assumes the scattered light is dominated by sky  emission within the slit, as is generally the case, at least at the location of bright sky lines.  If the source itself is bright, there is an additional contribution, that may be asymmetric between the two pairs of slits because the object is nodded along the spatial direction.  This has a particularly notable effect on band-shuffle masks, which normally should not be affected by this charge leakage.  In the band-shuffled case, one set of spectra (corresponding to the A position) are at one end of the detector, and the other set are at the opposite end.  The scattered light pattern is the same in both, and generally subtracts off without problem.  This isn't true if the science target contributes significantly to the flux.  In particular it is a problem for the mask alignment boxes; these contain bright stars in the A position, but they are mostly nodded out of the box in the B position.  Where these contribute to the scattered light, they lead to an asymmetry in the two sets of spectra.  The problem becomes acute because, compared with microshuffle observations, the same number (3--4) of alignment stars are packed into a third of the area; thus there is a greater contamination of neighbouring slits.  Our procedure does not correct for this, and it means that spectra near alignment star boxes in band-shuffle masks still suffer from this effect.

\begin{figure}{
		\includegraphics[clip=true,trim=0mm 0mm 0mm 0mm,width=3.5in,angle=0]{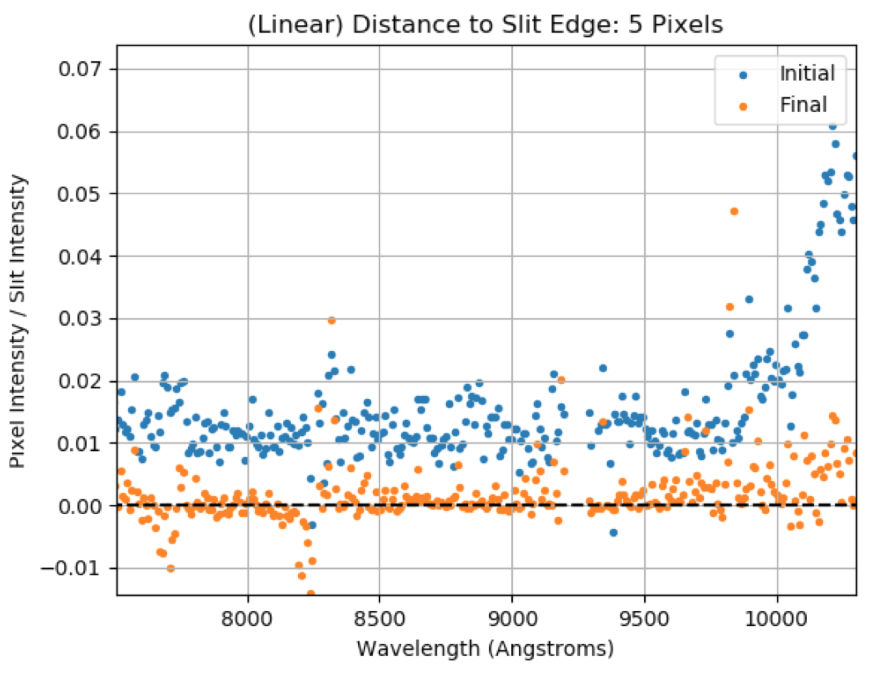}}
	\caption{The residual intensity relative to the average slit intensity is shown as a function of wavelength, similar to that shown in Figure~\ref{fig-CSwave}.  The blue points show the original data, and the orange points are the result after applying our empirical correction.  For $\lambda<10,000$\AA, the residuals are reduced to much less than 1 per cent. \label{fig-CSfix}}
\end{figure}

Having constructed the parametric model, we then apply it to both sides of every spectrum on every GOGREEN mask, and subtract off the estimated contribution.  The result of our correction is quantitatively quite excellent for most spectra.  The right panels of Figure~\ref{fig-CSex} show the same detector regions as before, at the same contrast, but after the correction has been applied. A quantitative comparison is shown in Figure~\ref{fig-CSfix}, which shows the residual intensity at a distance of five pixels from the slit edge, as a function of wavelength.  One set of points corresponds to the original data, and the other is after we have applied our correction. This empirical correction is not perfect, and fails for a small percentage of slits, particularly those near alignment star boxes.   However, the improvement for most of the data is significant and employing the correction enables redshift and line index measurements for hundreds of additional galaxies.

Uncertainties related to this correction are estimated from the variance in residuals and added to the variance vectors propagated through the data reduction procedure.  
\section{GOGREEN Telluric corrections}\label{sec-telluric}
\begin{figure}{}
\includegraphics[clip=true,trim=0mm 0mm 0mm 0mm,width=3in,angle=0]{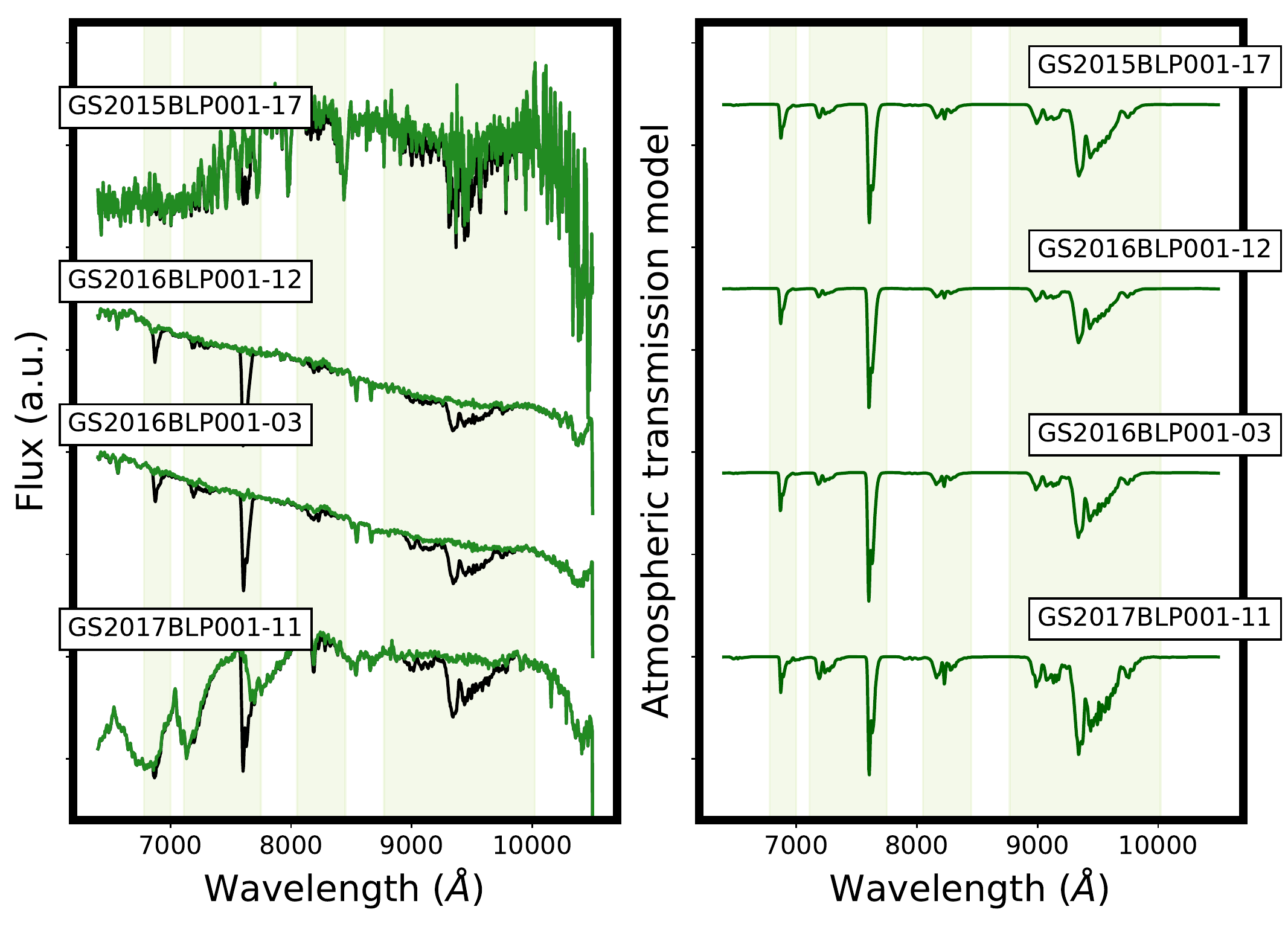}
 	\caption{Example of the telluric corrections performed by \texttt{molecfit} on the four masks of SpARCS0219 cluster. Left panel shows the brightest spectra in each mask before (black) and after (green) telluric correction. Right panel shows the atmospheric transmission model used for telluric corrections. The shaded regions indicate the spectral regions used in the fit. \label{fig-molecfit}}
\end{figure}

To model and correct the telluric absorption in the GOGREEN spectra we use the ESO code \texttt{molecfit} \citep{Kausch2014,Smette2015}. First, this code reads the science spectrum and a set of ambient input parameters. Then, it creates a single profile of the Earth’s atmosphere at the time of observation by gathering data from three sources: ENVISAT, GDAS, and the corresponding ground-based ESO Meteo Monitor measurements. The next step is the construction of a synthetic atmospheric absorption model by fitting user-defined spectral regions dominated by telluric absorption, for which \texttt{molecfit} relies on the radiative transfer code LBLRTM. In particular, we fit the spectral regions $\lambda\lambda\,$6780,7000; $\lambda\lambda\,$7110,7750; $\lambda\lambda\,$8050,8450; and, $\lambda\lambda\,$8770, 10020\,\AA, where strong H$_{2}$O, O$_2$, and O$_3$ telluric features are found, as shown in Figure \ref{fig-molecfit}. We exclude regions from the fit if they coincide with a strong spectral feature in the spectrum, such as an [O\textsc{ii}] emission line. The code modifies the model spectrum, using a polynomial fit of the continuum and the wavelength grid of each spectral region and convolving it with a kernel mimicking the instrumental profile, to match the science spectrum. The fitting of the science and the model spectrum is performed by the \texttt{mpfit} package, which uses a $\chi^2$ minimization procedure based on the Levenberg-Marquard iterative technique. If the desired fit quality is not reached, \texttt{mpfit} changes the fit parameters (e.g., molecular abundances) to search for a better $\chi^2$. Finally, the atmospheric transmission model for the full wavelength range of the input science spectrum is calculated.

\section{Spectral flux calibration}\label{sec-fluxcal}
As described in \S~\ref{sec-spectroscopy}, the wavelength dependence of the spectral flux calibration is calibrated using standard star observations.  However this does not account for slit losses and atmospheric effects that reduce the overall amplitude of the final spectrum.   
To obtain an absolute flux calibration for the GOGREEN and GCLASS spectra, we use $i$-band photometric data described in Section~\ref{sec-photcats}. We use the Suprime-Cam $i$-band for northern clusters and VIMOS $i$-band for the southern clusters.  The filter response curve, $R$, is interpolated using a cubic-spline to match the wavelength sampling of the spectroscopy.  
We then integrate over the interpolated filter response curve multiplied by the spectral flux density to give the average spectral flux density, in $\rm{erg\;cm^{-2}s^{-1}}$\AA$^{-1}$:
    \begin{align}
    f_{\rm \lambda, tot, spec} = \int_{\lambda_{\rm min}}^{\lambda_{\rm max}}(R \,f_{\lambda} \,\lambda\, d\lambda)/\int_{\lambda_{\rm min}}^{\lambda_{\rm max}}(R \,\lambda \,d\lambda).
    \end{align}
In this calculation, regions of the spectroscopy flagged as bad data are omitted.  The synthetic photometric flux in Jy is then determined as 
    \begin{align}
    f_{\rm \nu, tot, spec} =  3.34\times 10^{4} \; \lambda_{\rm pivot}^{2}  \; f_{\rm \lambda, tot, spec}\;,
    \end{align}
where the pivot wavelength 
    \begin{align}
        & \lambda_{\rm pivot} = \sqrt{ \int_{\lambda_{\rm min}}^{\lambda_{\rm max}}{R\,\lambda \,d\lambda}/\int_{\lambda_{\rm min}}^{\lambda_{\rm max}}{R/\lambda \, d\lambda}} 
    \end{align}
    
Finally, we compare this flux with the total i-band flux from the photometric catalogues $f_{\rm \nu, tot, phot}$, to obtain a flux calibration ratio 
        \begin{align}
        f_{\rm cal} = f_{\rm \nu, tot, spec}/f_{\rm \nu, tot, phot}
    \end{align}

The flux calibrated spectrum and variance are then calculated as: 
     \begin{align}
     & f_{\rm \lambda, cal} = f_{\rm \lambda}/f_{\rm cal},\\
     & Var(f_{\rm \lambda,cal}) = Var(f_{\rm \lambda})/(f_{\rm cal})^{2}.
      \end{align}

\section{Spectroscopic completeness}\label{sec-obscomp}
Here we describe in more detail our method for estimating the  spectroscopic incompleteness of the GOGREEN sample, as a function of the galaxy position in the color magnitude diagram and of the galaxy distance from the cluster center.

To compute the incompleteness as a function of the position on the color-magnitude diagram, we first subdivided the diagram of each field into cells and then we compared  the number of objects in the spectroscopic catalog with the number in the parent photometric catalog.  The parent catalog included all entries in the GOGREEN photometric catalog that were retained as targets for spectroscopy. The ratio of these two numbers yielded a weight as a function of galaxy apparent magnitude and color (W$_{\rm mag}$). As the target selection in the $z^\prime{-[}3.6$] vs [3.6] plane was slightly different from cluster to cluster, depending on the redshift of the cluster, we first computed these weights field by field instead of binning all fields together.

\begin{figure*}
\includegraphics[clip=true,trim=0mm 0mm 0mm 0mm,width=7.5in,angle=0]{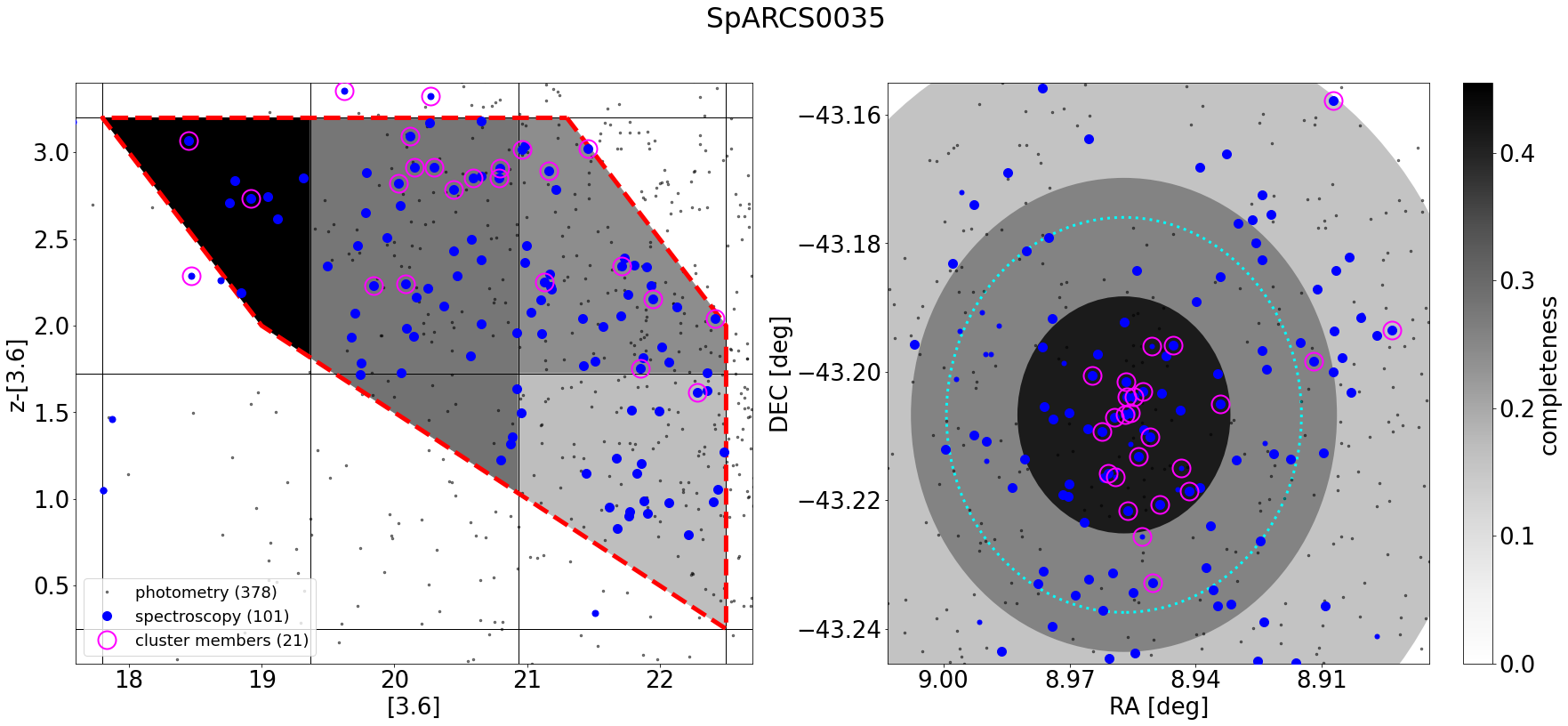}
\caption{These images show the spectroscopic sampling rate in SpARCS0035.  It is analagous to Figure~\ref{fig-zIRACcompleteness}, but for this single cluster.  \label{fig-SpARCS0035}}
\end{figure*}

\begin{figure*}
\includegraphics[clip=true,trim=0mm 0mm 0mm 0mm,width=7.5in,angle=0]{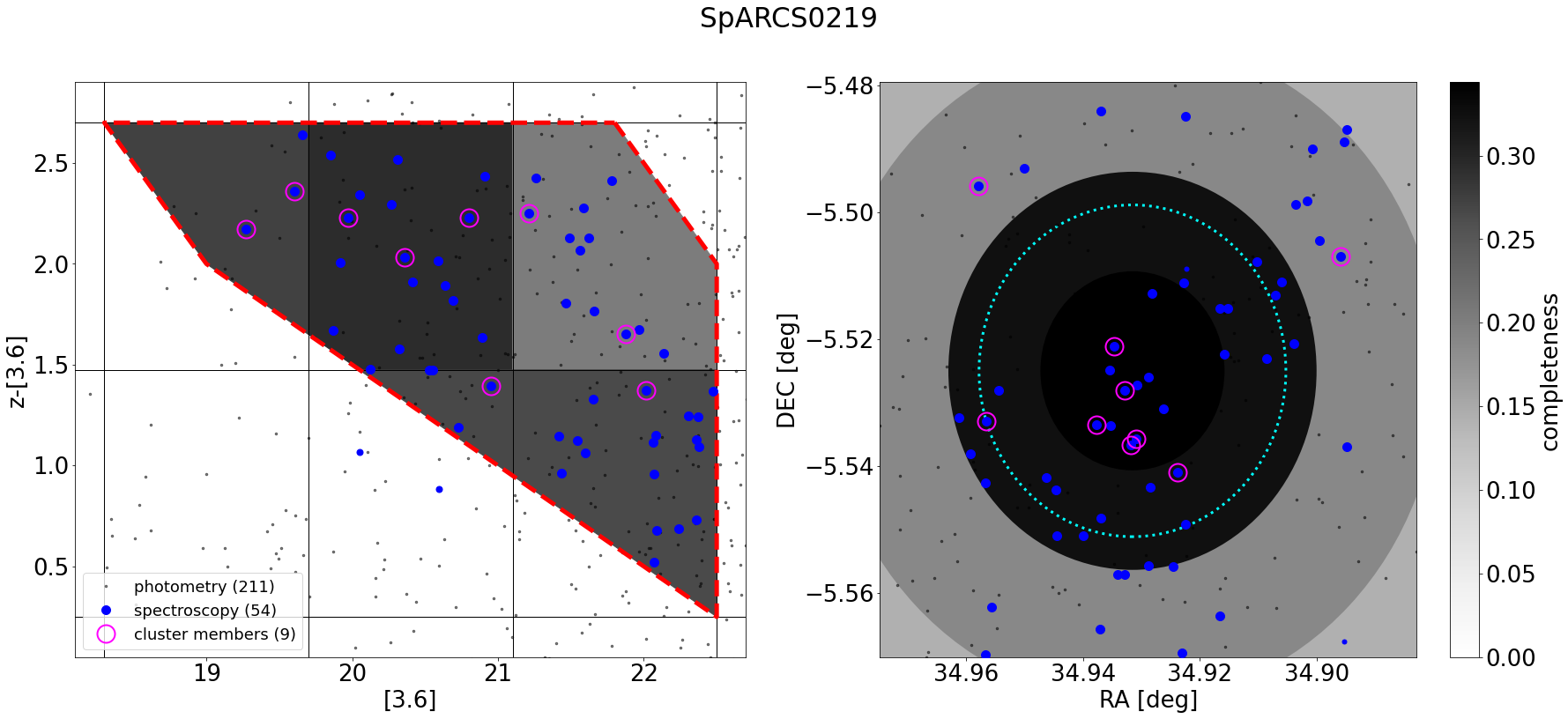}
\caption{As Figure~\ref{fig-SpARCS0035}.  \label{fig-SpARCS0219}}
\end{figure*}

\begin{figure*}
\includegraphics[clip=true,trim=0mm 0mm 0mm 0mm,width=7.5in,angle=0]{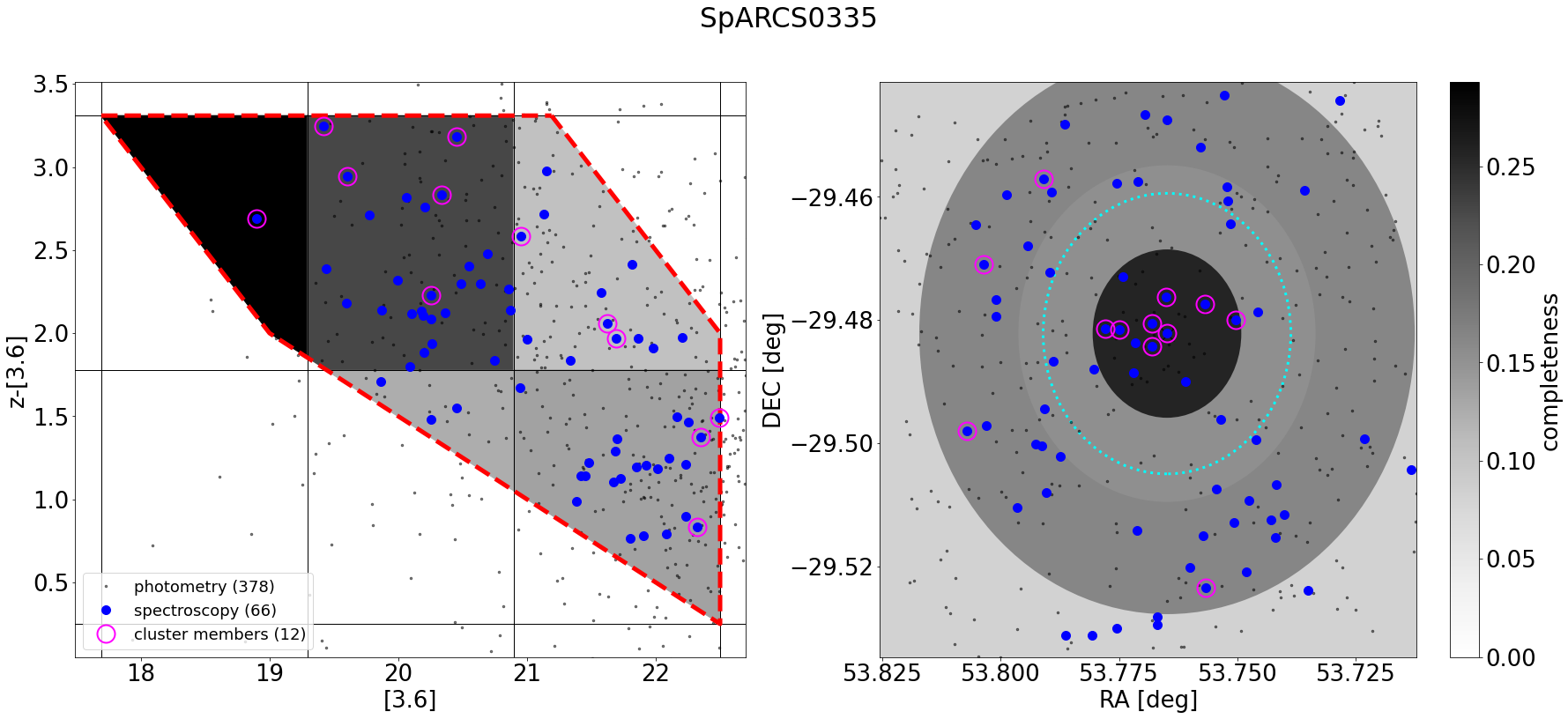}
\caption{As Figure~\ref{fig-SpARCS0035}.  \label{fig-SpARCS0335}}
\end{figure*}

\begin{figure*}
\includegraphics[clip=true,trim=0mm 0mm 0mm 0mm,width=7.5in,angle=0]{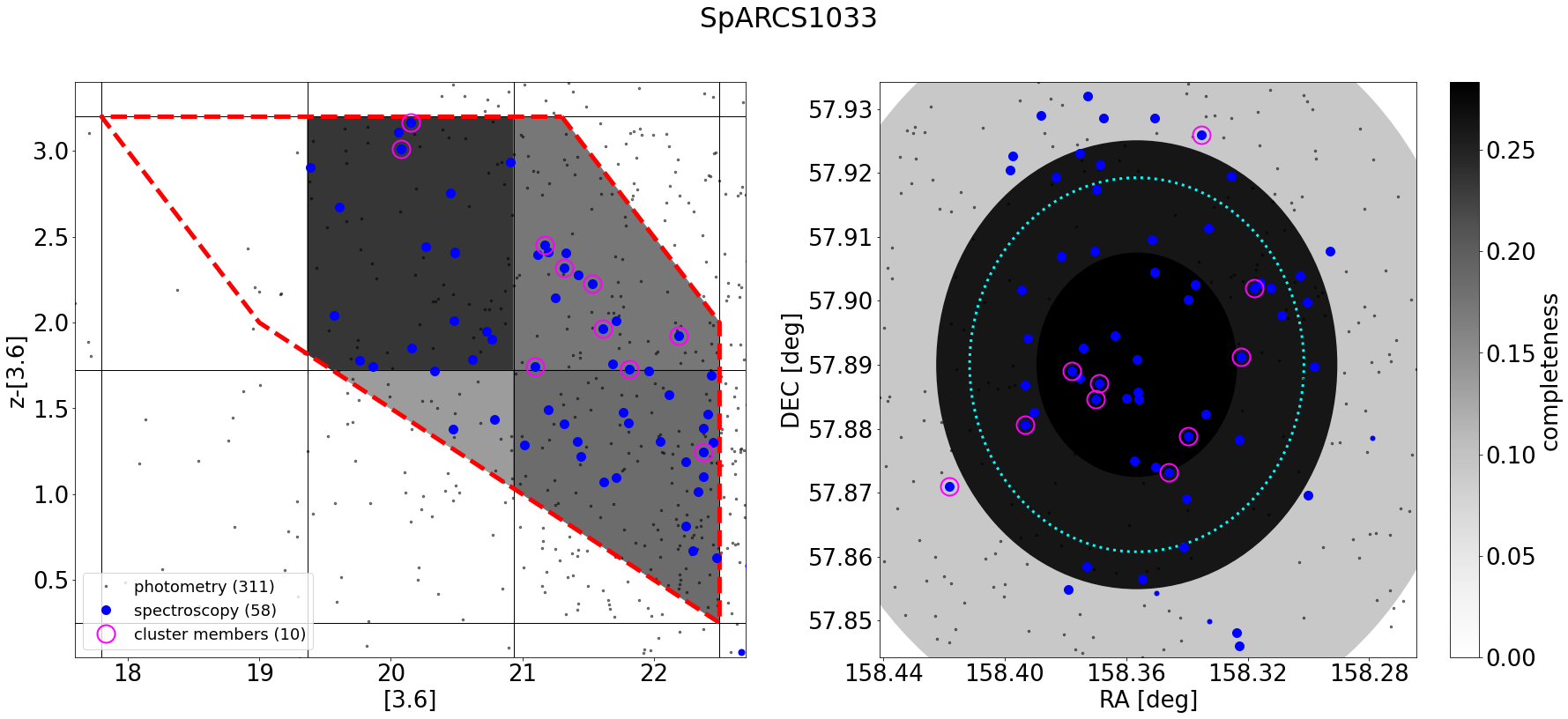}
\caption{As Figure~\ref{fig-SpARCS0035}.  \label{fig-SpARCS1033}}
\end{figure*}

\begin{figure*}
\includegraphics[clip=true,trim=0mm 0mm 0mm 0mm,width=7.5in,angle=0]{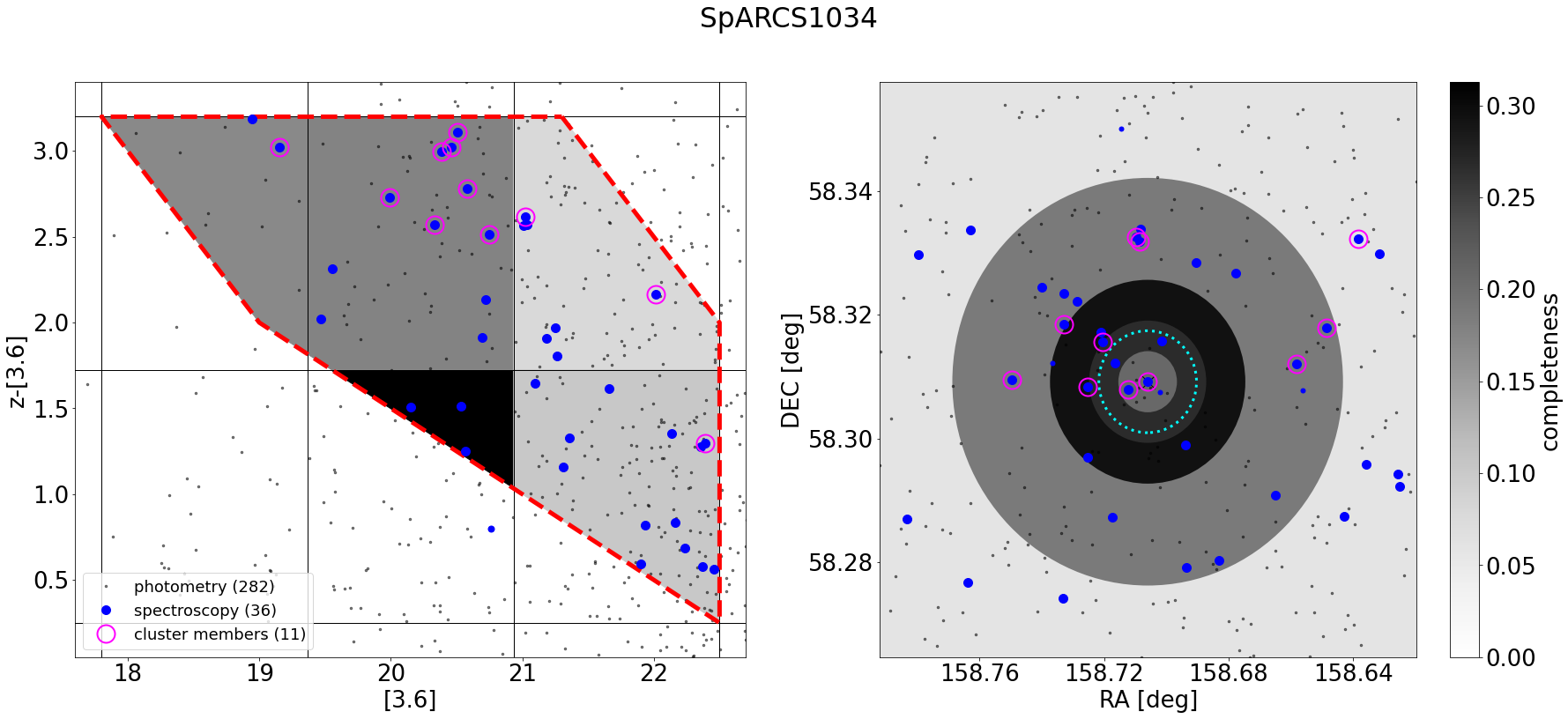}
\caption{As Figure~\ref{fig-SpARCS0035}.  \label{fig-SpARCS1034}}
\end{figure*}

\begin{figure*}
\includegraphics[clip=true,trim=0mm 0mm 0mm 0mm,width=7.5in,angle=0]{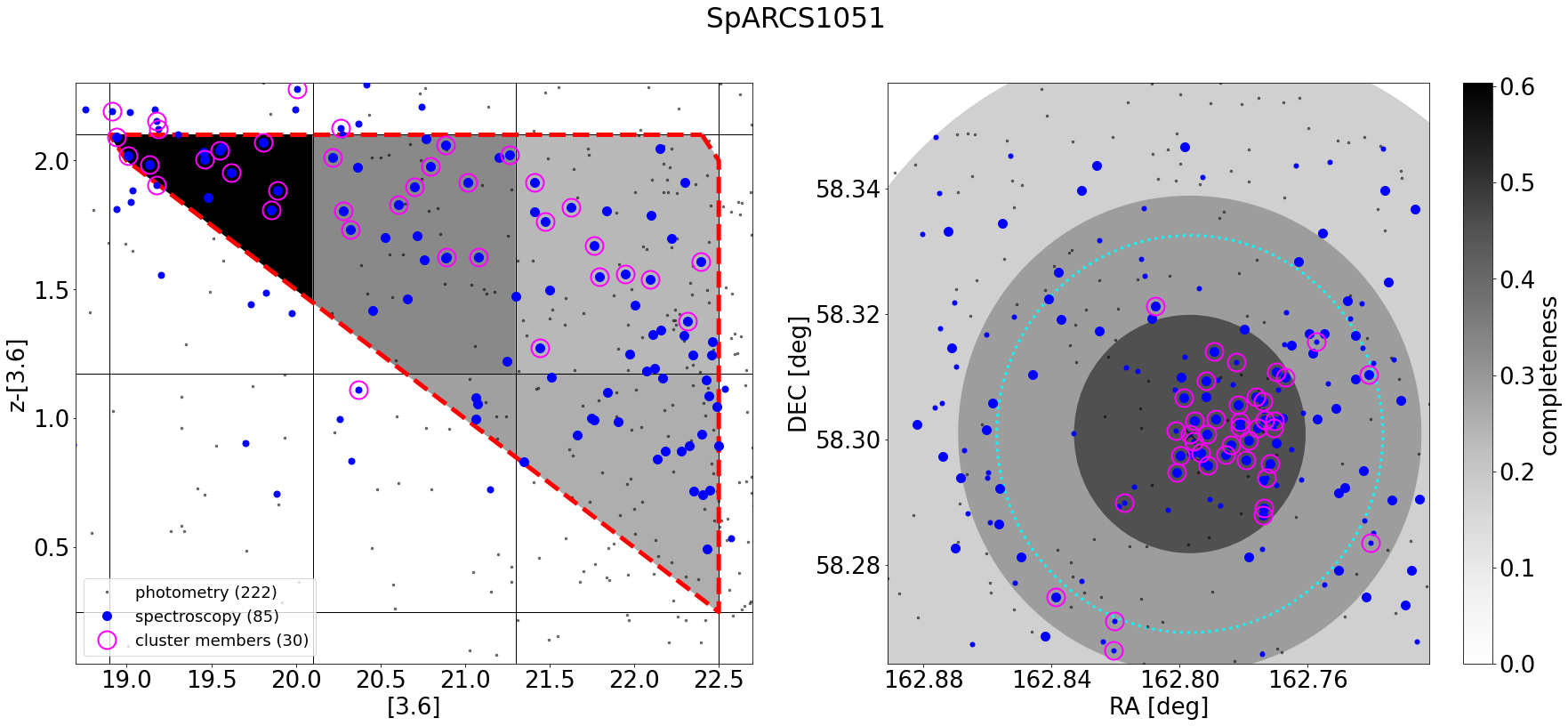}
\caption{As Figure~\ref{fig-SpARCS1051}.  \label{fig-SpARCS1051}}
\end{figure*}

\begin{figure*}
\includegraphics[clip=true,trim=0mm 0mm 0mm 0mm,width=7.5in,angle=0]{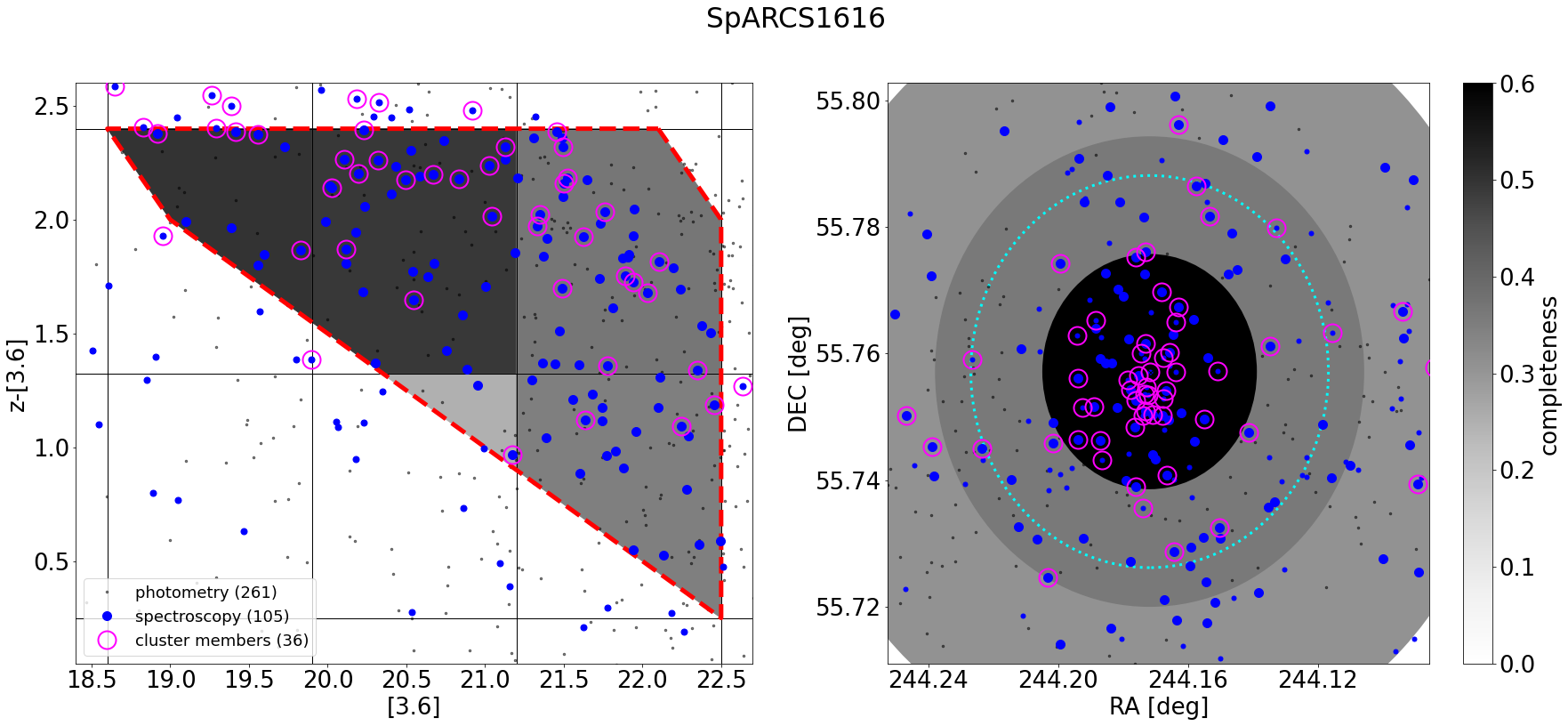}
\caption{As Figure~\ref{fig-SpARCS0035}.  \label{fig-SpARCS1616}}
\end{figure*}

\begin{figure*}
\includegraphics[clip=true,trim=0mm 0mm 0mm 0mm,width=7.5in,angle=0]{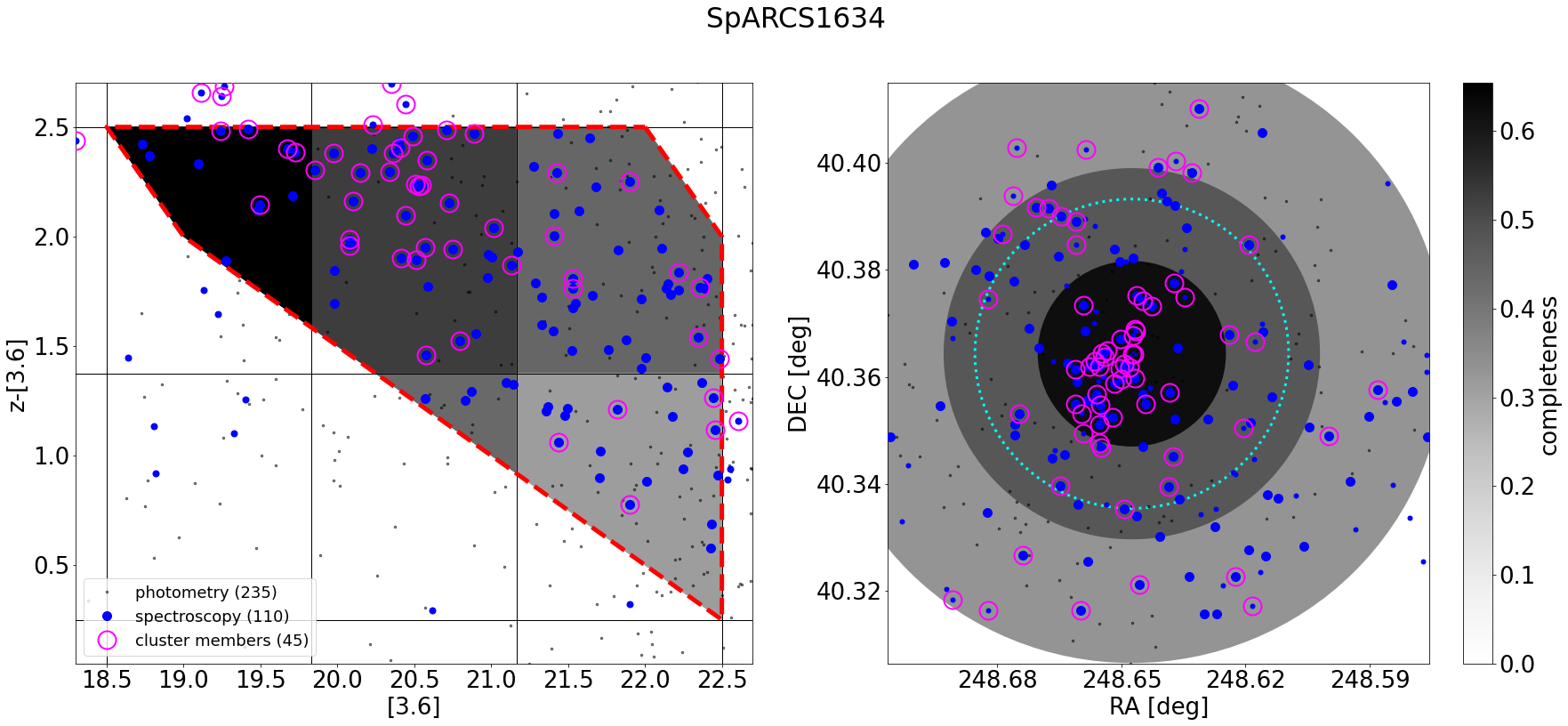}
\caption{As Figure~\ref{fig-SpARCS0035}.  \label{fig-SpARCS1634}}
\end{figure*}

\begin{figure*}
\includegraphics[clip=true,trim=0mm 0mm 0mm 0mm,width=7.5in,angle=0]{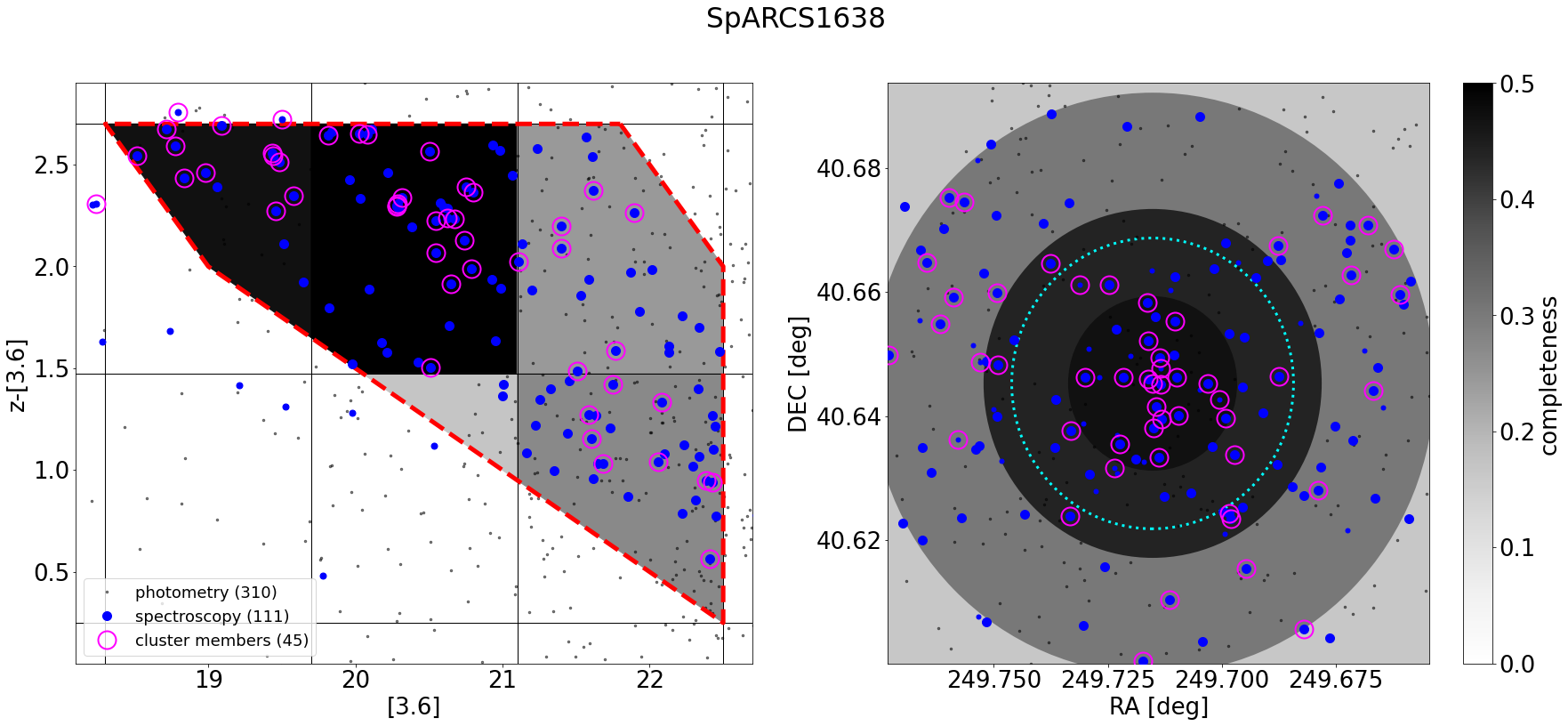}
\caption{As Figure~\ref{fig-SpARCS0035}.  \label{fig-SpARCS1638}}
\end{figure*}

\begin{figure*}
\includegraphics[clip=true,trim=0mm 0mm 0mm 0mm,width=7.5in,angle=0]{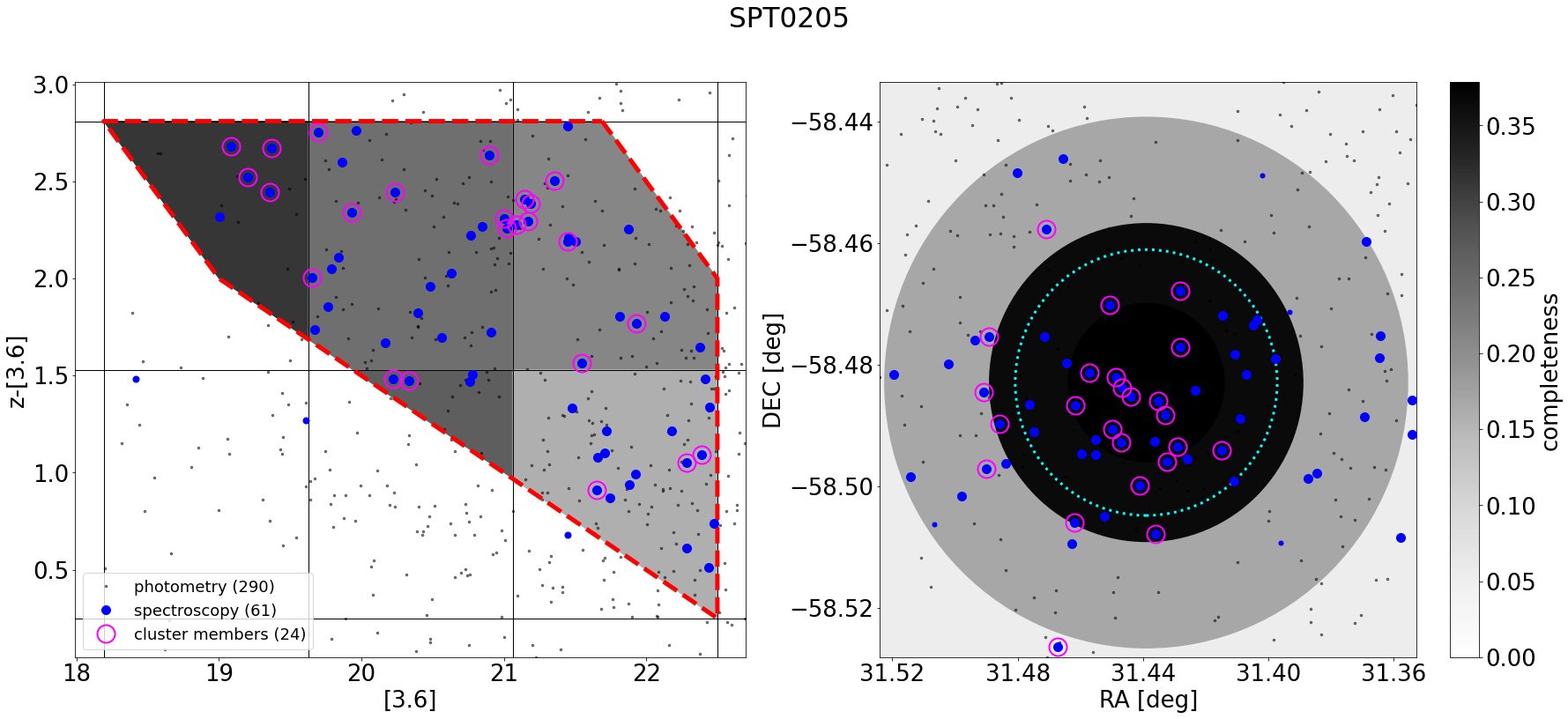}
\caption{As Figure~\ref{fig-SpARCS0035}.  \label{fig-SPT0205}}
\end{figure*}

\begin{figure*}
\includegraphics[clip=true,trim=0mm 0mm 0mm 0mm,width=7.5in,angle=0]{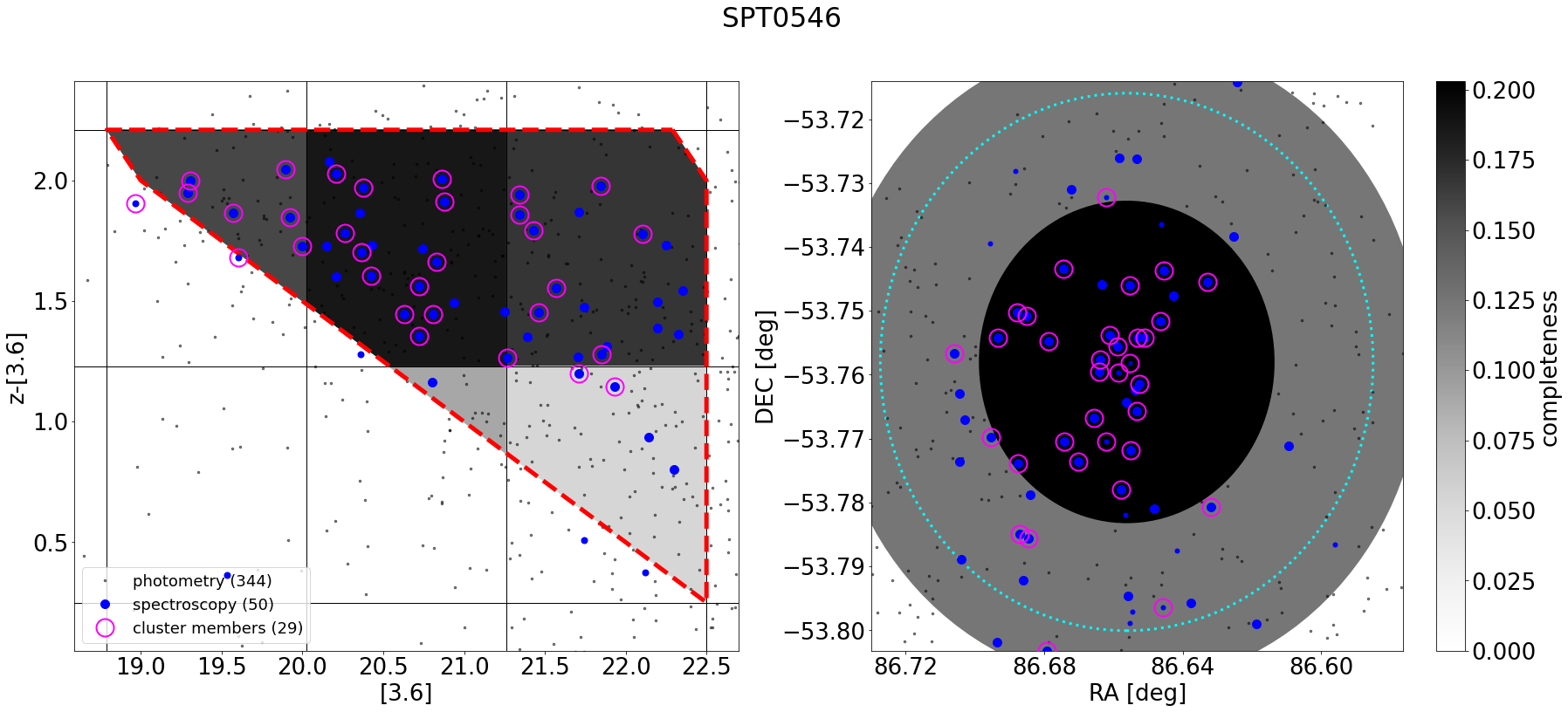}
\caption{As Figure~\ref{fig-SpARCS0035}.  \label{fig-SPT0546}}
\end{figure*}

\begin{figure*}
\includegraphics[clip=true,trim=0mm 0mm 0mm 0mm,width=7.5in,angle=0]{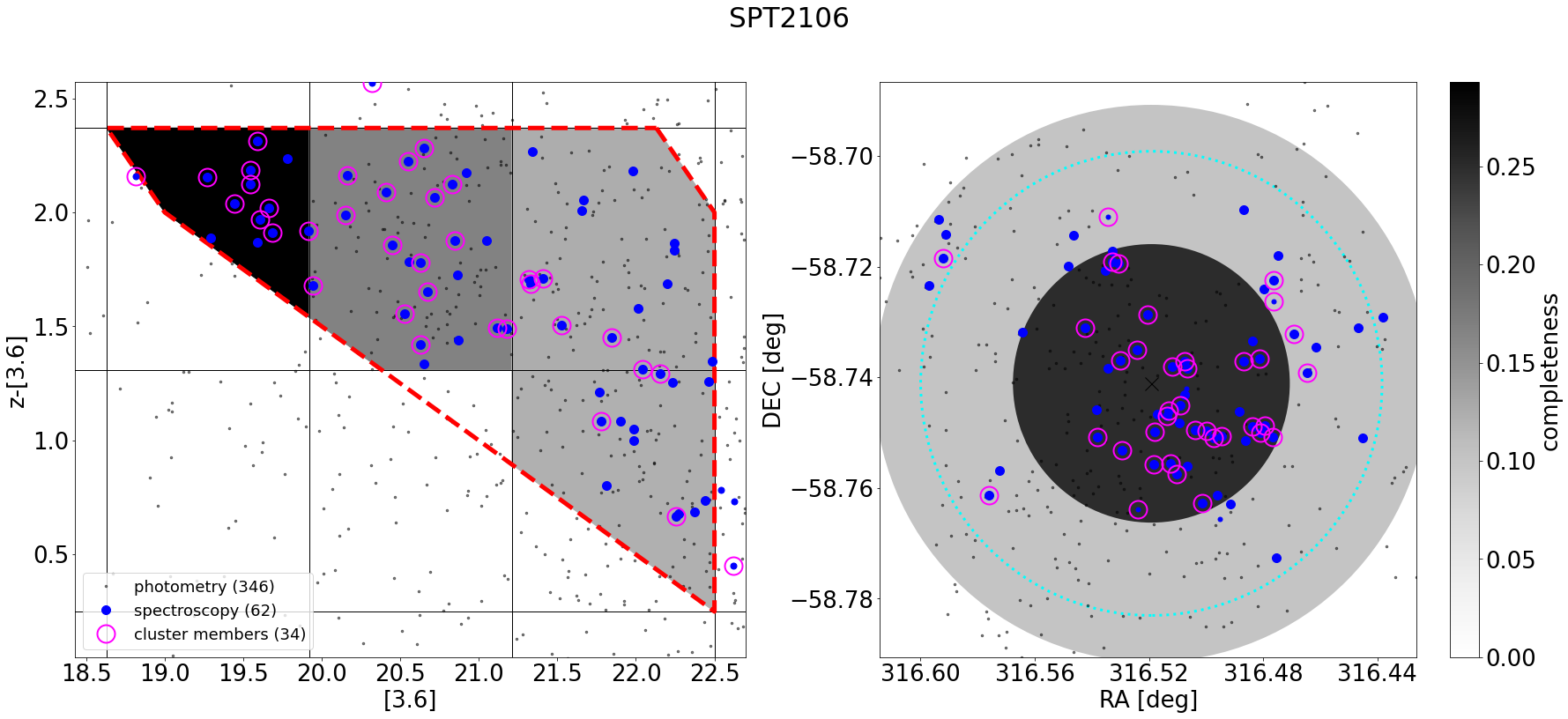}
\caption{As Figure~\ref{fig-SpARCS0035}.  \label{fig-SPT2106}}
\end{figure*}

We also quantified the presence of geometrical effects due to possible variations in the sampling as a function of the clustercentric radius. Geometrical effects can affect a spectroscopic sample of a cluster due to the fact that cluster galaxies are indeed more concentrated toward the cluster center, so observational constraints on the minimum distance between slits typically result in a lower sampling of these central regions. 
The geometrical completeness W$_{\rm geo}$ was computed comparing the number of galaxies in the spectroscopic and in the parent photometric catalogs in four annuli with R < 0.6, 0.6 < R < 1.2, 1.2 < R < 2, and R > 2 in units of R$_{200}$. 
Figures~\ref{fig-SpARCS0035}-\ref{fig-SPT2106} show the completeness diagrams for the clusters in the sample. The results for the combined sample are given in the main body of the paper, in Figure~\ref{fig-zIRACcompleteness}.

\section{Spectroscopy Log}\label{sec-speclog}
Table~\ref{tab-specobs} provides a log of all GOGREEN spectroscopy obtained as part of this program.  We identify the dates of observation, mask name, nod-and-shuffle mode (microshuffle or bandshuffle), telescope/detector combination, total integration time, and whether the data were taken in Queue mode.

\clearpage
\onecolumn
\begin{longtable}{lllllll} 
\hline
Target &Date &Mask &Band/Micro &Telescope/ &Integration &Notes\\
		&       &     &           &Detector   & time (ks)  &           \\
\hline
		\hline
		\endfirsthead
\multicolumn{7}{l}
{\tablename\ \thetable\ -- \textit{Continued from previous page}} \\
\hline
Target &Date &Mask &Band/Micro &Telescope/ &Integration &Notes\\
		&       &     &           &Detector   & time (ks)  &           \\
\hline
\endhead
\hline \multicolumn{7}{r}{\textit{Continued on next page}} \\
\endfoot
\hline
\endlastfoot
		SPT0205 & Nov 16,18, 2014 & GS2014BLP001-06 & Microshuffle & GS/Ham & 6.48&\\ 
		& Oct 29-30, Nov 3 2016& GS2016BLP001-02 & Microshuffle & GS/Ham &10.8 &\\ 
		& Oct 28-29, 2016	& GS2016BLP001-09 & Microshuffle & GS/Ham & 9.36&\\ 
		& Nov 13, 2017 & GS2017BLP001-03 & Microshuffle & GS/Ham & 8.64&GS2017BDD10\\
		& Nov 14, 2017 & GS2017BLP001-04 & Microshuffle & GS/Ham & 10.8&\\
		& Jan 10, 16, 18, 2018 & GS2017BLP001-03 & Microshuffle & GS/Ham & 10.8&\\
		\hline
		SPT0546 & Nov 15-16, 2014 & GS2014BLP001-09 & Microshuffle & GS/Ham & 5.76&\\ 
		& Nov 17,19, 2014 & GS2014BLP001-10 & Microshuffle & GS/Ham & 7.2&\\ 
		& Nov 20, 2015 & GS2015BLP001-15 & Microshuffle & GS/Ham & 7.92 & \\ 
		& Nov 21, 2015 & GS2015BLP001-16 & Microshuffle & GS/Ham & 2.16 & \\ 
		& Feb 10, 2016 & GS2015BLP001-16 & Microshuffle & GS/Ham & 14.4 & \\
		& Nov 13-14, 16, 2017 & GS2017BLP001-12 & Microshuffle & GS/Ham & 9.36\\
		& Nov 15-16, 2017 & GS2017BLP001-13 & Microshuffle & GS/Ham & 10.08\\
		\hline
		SPT2106 & June 15, 2018 & GS2018ALP001-01 & Microshuffle & GS/Ham & 10.8&Queue\\ 
		         & June 16-17, 2018 & GS2018ALP001-02 & Microshuffle & GS/Ham & 10.8&Queue\\ 
		           & Sept 6, 2018 & GS2018BLP001-04 & Microshuffle & GS/Ham & 10.8&\\
		           & Sept 7, 2018 & GS2018BLP001-05 & Microshuffle & GS/Ham & 10.8&\\
\hline
		SpARCS0035 & Nov 21, 2015 & GS2015BLP001-05 & Bandshuffle & GS/Ham & 9.36 & \\
		& Nov 20, 2015 & GS2015BLP001-06 & Microshuffle & GS/Ham & 7.2 & \\ 
		& Oct 28, 2016 & GS2016BLP001-01 & Microshuffle & GS/Ham & 7.9& \\
		& Oct 27, 2016 & GS2016BLP001-07 & Microshuffle & GS/Ham & 10.8& \\ 
		& Nov 11, 2017 & GS2017BLP001-01 & Microshuffle & GS/Ham & 7.9\\
		& Nov 12, 2017 & GS2017BLP001-02 & Microshuffle & GS/Ham & 10.8\\
\hline
		SpARCS0219 & Nov 20, 2015 & GS2015BLP001-17 & Microshuffle & GS/Ham & 10.8 & \\ 
		& Oct 30, 2016& GS2016BLP001-03 & Microshuffle & GS/Ham &8.64 & \\ 
		& Oct 27-28, 2016	& GS2016BLP001-12 & Microshuffle & GS/Ham & 9.36& \\ 
		& Nov 15, 2017 & GS2017BLP001-11 & Microshuffle & GS/Ham & 9.36\\
\hline
		SpARCS0335 & Nov 18-19, 2014 & GS2014BLP001-01 & Bandshuffle & GS/Ham & 7.2&\\ 
		& Feb 1, 2017	& GS2016BLP001-13 & Bandshuffle & GS/Ham & 9.36&\\ 
		& Oct 26-29, 2016	& GS2016BLP001-14 & Bandshuffle & GS/Ham &10.8 &\\ 
		& Nov 11, 2017 & GS2017BLP001-07 & Microshuffle & GS/Ham & 7.92\\
		& Nov 12-13, 2017 & GS2017BLP001-08 & Microshuffle & GS/Ham & 8.64\\
		\hline
		SpARCS1051 & Feb 18\&29, 2016 & GN2016ALP004-03 & Microshuffle & GN/EEV & 18.0 & Queue\\
				   & April 25, 2017 & GN2017ALP004-08 & Microshuffle & GN/Ham & 12.0&\\
				   & April 26, 2017 & GN2017ALP004-07 & Microshuffle & GN/Ham & 13.8&\\
		            & Feb 12, 2018 & GN2018ALP004-07 & Microshuffle & GN/Ham & 13.68&\\
				   \hline
		SpARCS1033 & April 18, 2017 & GN2017ALP004-01 & Bandshuffle & GN/Ham & 7.2&\\
		& April 19, 2017 & GN2017ALP004-02 & Microshuffle & GN/Ham & 10.08&\\
		& April 20, 2017 & GN2017ALP004-03 & Microshuffle & GN/Ham & 10.08&\\
		& Feb 11, 2018 & GN2018ALP004-01 & Microshuffle & GN/Ham & 7.92&\\
		& Feb 11, 13 2018 & GN2018ALP004-02 & Microshuffle & GN/Ham & 7.92&\\
		\hline
		SpARCS1034 & April 24, 2017 & GN2017ALP004-04 & Bandshuffle & GN/Ham & 4.3&\\
					& April 12\&27, 2017 & GN2017ALP004-05 & Bandshuffle & GN/Ham & 10.08 & \\
					& May 21,22, 29, 31, June 3, 2017 & GN2017ALP004-06 & Microshuffle & GN/Ham & 10.8 &Queue \\
		            & Feb 13 2018 & GN2018ALP004-04 & Microshuffle & GN/Ham & 9.36&\\
		            & Feb 17, June 12, 2018 & GN2018ALP004-05 & Microshuffle & GN/Ham & 7.2&\\
					\hline
		SpARCS1616 & June 1, 2016 & GN2016ALP004-06 & Microshuffle & GN/EEV & 14.4 & \\
				   & June 2, 2016 & GN2016ALP004-07 & Microshuffle & GN/EEV & 18.0 & \\
				   & April 18\&27, 2017 & GN2017ALP004-09 & Microshuffle & GN/Ham & 17.28 & \\
		            &June 10, 12, 2018 & GN2018ALP004-08 & Microshuffle & GN/Ham & 17.28&\\
				   \hline
		SpARCS1634 & May 30, 2016 & GN2016ALP004-04 & Microshuffle & GN/EEV & 10.8 & \\
				   & May 30-31, 2016 & GN2016ALP004-05 & Microshuffle & GN/EEV & 18.0 & \\
				   & April 19/26, 2017 & GN2017ALP004-10 & Microshuffle & GN/Ham & 18.0 & \\
		            &June 12-13, 2018 & GN2018ALP004-09 & Microshuffle & GN/Ham & 17.28&\\
				   \hline
		SpARCS1638 & May 28-20, 2016 & GN2016ALP004-01 & Microshuffle & GN/EEV & 10.8 & \\
					& May 29/June2, 2016 & GN2016ALP004-02 & Microshuffle & GN/EEV & 18.0 & \\
					& April 20/28, 2017 & GN2017ALP004-11 & Microshuffle & GN/Ham & 18.0 & \\
		            &June 21-23, 2018 & GN2018ALP004-10 & Microshuffle & GN/Ham & 19.44&\\
					\hline
		COSMOS-28 & Jan 30, 2016 & GN2015BLP004-03 & Microshuffle & GN/EEV & 18.0&Queue \\ 
		        & Mar 4,6, 2019 & GN2019ALP004-01 & Microshuffle & GN/Ham & 20.88&Queue \\ 
		        & April 7, 24-26, 2019 & GN2019ALP004-02 & Microshuffle & GN/Ham & 19.44&Queue \\ 
		        & April 27, May 1, 8, 10-11, 2019 & GN2019ALP004-03 & Microshuffle & GN/Ham & 19.44&Queue \\ 
		\hline
		COSMOS-63 & Jan 31, 2016 & GN2015BLP004-02 & Microshuffle & GN/EEV & 18.0&Queue\\ 
		\hline
		COSMOS-125 & Jan 31, 2016 & GS2016ALP001-02 & Microshuffle & GS/Ham & 15.12 &\\
		& Feb 25, 2015 & GS2015ALP001-02 & Microshuffle & GS/Ham & 12.25&\\ 
		\hline
		COSMOS-221 & Feb 24, 2015 &  GS2015ALP001-01 & Microshuffle & GS/Ham & 10.08&\\ 
		& Feb 23, 2015 & GS2014BLP001-05 & Microshuffle & GS/Ham & 5.04&\\ 
		& Feb 13, 2016 & GS2016ALP001-01 & Microshuffle & GS/Ham & 10.8 & \\
		\hline
		SXDF49/87 & Oct 9, 2015 & GN2015BLP004-01 & Microshuffle & GN/EEV & 18.0&Queue\\ 
		           & Nov 15, 2014 & GS2014BLP001-07 & Microshuffle & GS/Ham & 8.64&\\
		           & Nov 1, 2018 & GN2018BLP004-01 & Microshuffle & GN/Ham & 10.8&\\
		           & Sept 6, 2018 & GS2018BLP001-01 & Microshuffle & GS/Ham & 12.96&\\
		           & Sept 8, 2018 & GS2018BLP001-02 & Microshuffle & GS/Ham & 12.96&\\
		           & Sept 7, 9, 2018 & GS2018BLP001-03 & Microshuffle & GS/Ham & 12.24&\\
		\hline
		SXDF64 & Nov 17, 2014 & GS2014BLP001-08 & Microshuffle & GS/Ham & 7.2&\\
		\hline
		SXDF76 & Nov 15, 2014 & GS2014BLP001-02 & Microshuffle & GS/Ham & 5.76&\\
		           & Nov 5, 2018 & GN2018BLP004-02 & Microshuffle & GN/Ham & 15.12&\\
		           & Nov 5, 2018 & GN2018BLP004-03 & Microshuffle & GN/Ham & 15.12&\\
		           & Nov 6, 12, 2018 & GN2018BLP004-04 & Microshuffle & GN/Ham & 15.12&\\
		\hline
	\caption{A log of all spectroscopic data obtained for GOGREEN.  All data were acquired in Priority Visitor mode unless otherwise indicated in the Final column.  \label{tab-specobs}}
\end{longtable}
\twocolumn


\bsp	
\label{lastpage}
\bibliographystyle{mnras} 
\bibliography{ms2}
\end{document}